%% file: 0-main.tex
\documentclass[a4paper,11pt]{article}
\pdfoutput=1 

\usepackage{jheppub} 

\usepackage[T1]{fontenc} 

\usepackage{etoolbox}

\makeatletter
\patchcmd{\maketitle}{\@fpheader}{Published
in \href{https://doi.org/10.1007/JHEP03(2023)038}{JHEP 03 (2023) 038}}{}{}
\makeatother

\usepackage{longtable}
\usepackage{afterpage}
\usepackage{ifthen} 
\usepackage{bm}
\newboolean{pdflatex}
\setboolean{pdflatex}{true} 
\newboolean{articletitles}
\setboolean{articletitles}{true} 
\newboolean{uprightparticles}
\setboolean{uprightparticles}{false}
\usepackage{microtype}
\usepackage{lineno}  
\usepackage{xspace}
\usepackage{colortbl}
\usepackage{amsfonts}
\usepackage{upgreek}

\usepackage{mciteplus}

\usepackage{pifont}

\usepackage{booktabs}

\usepackage{rotating}

\usepackage{relsize}
\usepackage{units}

\usepackage{amssymb}

\usepackage[section]{placeins}

\usepackage{color}
\usepackage{graphicx}
\usepackage{subfigure}
\usepackage{amsmath}

\usepackage{cancel}
\usepackage{enumitem}
\newlist{todolist}{itemize}{2}
\setlist[todolist]{label=$\square$}
\usepackage{arydshln}

\usepackage{url}
\usepackage[dvipsnames,usenames]{xcolor}

\usepackage[english]{babel}

\usepackage[all]{hypcap}

\usepackage[hyperref]{backref}
\usepackage[noabbrev]{cleveref}
\renewcommand{\figurename}{figure}

\usepackage{footnotebackref}
\renewcommand*{\cite}{\citep}
\usepackage{hypernat} 

\input{inputs/X-newcommands.tex}

\newcommand*\patchAmsMathEnvironmentForLineno[1]{
  \expandafter\let\csname old#1\expandafter\endcsname\csname #1\endcsname
  \expandafter\let\csname oldend#1\expandafter\endcsname\csname end#1\endcsname
  \renewenvironment{#1}
     {\linenomath\csname old#1\endcsname}
     {\csname oldend#1\endcsname\endlinenomath}}
\newcommand*\patchBothAmsMathEnvironmentsForLineno[1]{
  \patchAmsMathEnvironmentForLineno{#1}
  \patchAmsMathEnvironmentForLineno{#1*}}
\AtBeginDocument{%
\patchBothAmsMathEnvironmentsForLineno{equation}
\patchBothAmsMathEnvironmentsForLineno{align}
\patchBothAmsMathEnvironmentsForLineno{flalign}
\patchBothAmsMathEnvironmentsForLineno{alignat}
\patchBothAmsMathEnvironmentsForLineno{gather}
\patchBothAmsMathEnvironmentsForLineno{multline}
\patchBothAmsMathEnvironmentsForLineno{eqnarray}
} 

\newif\ifbackrefshowonlyfirst
\backrefshowonlyfirsttrue
\makeatletter
\let\BR@direct@old@hyper@natlinkstart\hyper@natlinkstart
\renewcommand*{\hyper@natlinkstart}{\phantomsection\BR@direct@old@hyper@natlinkstart}
\let\BR@direct@oldBR@citex\BR@citex
\renewcommand*{\BR@citex}{\phantomsection\BR@direct@oldBR@citex}

\long\def\hyper@page@BR@direct@ref#1#2#3{\hyperlink{#3}{#1}}

\ifx\backrefxxx\hyper@page@backref
    \let\backrefxxx\hyper@page@BR@direct@ref
    \ifbackrefshowonlyfirst
    \fi
\else
    \ifbackrefshowonlyfirst
    \fi
\fi

\RequirePackage{etoolbox}
\patchcmd{\Hy@backout}{Doc-Start}{\@currentHref}{}{\errmessage{I can't seem to patch backref}}
\makeatother

\renewcommand*{\backref}[1]{}
\renewcommand*{\backrefalt}[4]{
  \ifcase #1 
  (Not cited.)
  \or
  (Cited in section~#2.)
  \else
  (Cited in sections~#2.)
  \fi}

\makeatletter
\def\NAT@citexnum[#1][#2]#3{%
  \NAT@reset@parser
  \NAT@sort@cites{#3}%
  \NAT@reset@citea
  \@cite{\def\NAT@num{-1}\let\NAT@last@yr\relax\let\NAT@nm\@empty
    \@for\@citeb:=\NAT@cite@list\do
    {\@safe@activestrue
     \edef\@citeb{\expandafter\@firstofone\@citeb\@empty}%
     \@safe@activesfalse
     \@ifundefined{b@\@citeb\@extra@b@citeb}{%
       {\reset@font\bfseries?}
        \NAT@citeundefined\PackageWarning{natbib}%
       {Citation `\@citeb' on page \thepage \space undefined}}%
     {\let\NAT@last@num\NAT@num\let\NAT@last@nm\NAT@nm
      \NAT@parse{\@citeb}%
      \ifNAT@longnames\@ifundefined{bv@\@citeb\@extra@b@citeb}{%
        \let\NAT@name=\NAT@all@names
        \global\@namedef{bv@\@citeb\@extra@b@citeb}{}}{}%
      \fi
      \ifNAT@full\let\NAT@nm\NAT@all@names\else
        \let\NAT@nm\NAT@name\fi
      \ifNAT@swa
       \@ifnum{\NAT@ctype>\@ne}{%
        \@citea
        \NAT@hyper@{\@ifnum{\NAT@ctype=\tw@}{\NAT@test{\NAT@ctype}}{\NAT@alias}}%
       }{%
        \@ifnum{\NAT@cmprs>\z@}{%
         \NAT@ifcat@num\NAT@num
          {\let\NAT@nm=\NAT@num}%
          {\def\NAT@nm{-2}}%
         \NAT@ifcat@num\NAT@last@num
          {\@tempcnta=\NAT@last@num\relax}%
          {\@tempcnta\m@ne}%
         \@ifnum{\NAT@nm=\@tempcnta}{%
          \@ifnum{\NAT@merge>\@ne}{}{\NAT@last@yr@mbox}%
         }{%
            \Hy@backout{\@citeb\@extra@b@citeb}
           \advance\@tempcnta by\@ne
           \@ifnum{\NAT@nm=\@tempcnta}{%
             \ifx\NAT@last@yr\relax
               \def@NAT@last@yr{\@citea}%
             \else
               \def@NAT@last@yr{--\NAT@penalty}%
             \fi
           }{%
             \NAT@last@yr@mbox
           }%
         }%
        }{%
         \@tempswatrue
         \@ifnum{\NAT@merge>\@ne}{\@ifnum{\NAT@last@num=\NAT@num\relax}{\@tempswafalse}{}}{}%
         \if@tempswa\NAT@citea@mbox\fi
        }%
       }%
       \NAT@def@citea
      \else
        \ifcase\NAT@ctype
          \ifx\NAT@last@nm\NAT@nm \NAT@yrsep\NAT@penalty\NAT@space\else
            \@citea \NAT@test{\@ne}\NAT@spacechar\NAT@mbox{\NAT@super@kern\NAT@@open}%
          \fi
          \if*#1*\else#1\NAT@spacechar\fi
          \NAT@mbox{\NAT@hyper@{{\citenumfont{\NAT@num}}}}%
          \NAT@def@citea@box
        \or
          \NAT@hyper@citea@space{\NAT@test{\NAT@ctype}}%
        \or
          \NAT@hyper@citea@space{\NAT@test{\NAT@ctype}}%
        \or
          \NAT@hyper@citea@space\NAT@alias
        \fi
      \fi
     }%
    }%
      \@ifnum{\NAT@cmprs>\z@}{\NAT@last@yr}{}%
      \ifNAT@swa\else
        \@ifnum{\NAT@ctype=\z@}{%
          \if*#2*\else\NAT@cmt#2\fi
        }{}%
        \NAT@mbox{\NAT@@close}%
      \fi
  }{#1}{#2}%
}%
\makeatother

\expandafter\def\expandafter\UrlBreaks\expandafter{\UrlBreaks\do\a\do\b\do\c\do\d\do\e\do\f\do\g\do\h\do\i\do\j\do\k\do\l\do\m\do\n\do\o\do\p\do\q\do\r\do\s\do\t\do\u\do\v\do\w\do\x\do\y\do\z\do\&}

\newcommand{\T}{\ensuremath{T}\xspace}

\newcommand{\CPV}{\ensuremath{CP}~violation\xspace}
\newcommand{\DDb}{{\Dz\xspace\Dzb\xspace}}

\newcommand{\DDbSys}{\DDb}

\newcommand{\DDbP}{{\Dz\xspace\Dzb\xspace\pi^0\xspace}}
\newcommand{\DDbG}{{\Dz\xspace\Dzb\xspace\gamma\xspace}}
\newcommand{\DbDP}{{\Dzb\xspace\Dz\xspace\pi^0\xspace}}

\newcommand{\XtoDDbP}{\X \to \DDbP}
\newcommand{\XtoDDbG}{\X \to \DDbG}

\newcommand{\panda}{$\rm\overline{P}$ANDA\xspace}
\newcommand{\MeVm}{{MeV/$c^2$}\xspace}
\newcommand{\keVm}{{keV/$c^2$}\xspace}
\newcommand{\GeVm}{{GeV/$c^2$}\xspace}

\newcommand{\MeVp}{{MeV/$c$}\xspace}

\newcommand{\MeVE}{{MeV}\xspace}

\newcommand{\GeVE}{{GeV}\xspace}
\newcommand{\TeVE}{{TeV}\xspace}
\newcommand{\DCPp}{{D_{+}}}

\newcommand{\DCPm}{{D_{-}}}
\newcommand{\mb}{\boldmath}

\newcommand{\BptoXKp}{\Bp \to \X K^+}
\newcommand{\BtoXK}{B \to \X K}
\newcommand{\BptoPsiKp}{\Bp \to \psix K^+}

\title{\boldmath Novel 
correlated $\DDbSys$ systems for 
$c/b$
physics \\and tests of $T/CPT$}

\author[]{Paras Naik%
\footnote{Now at Oliver Lodge Laboratory, University of Liverpool, Liverpool,
United Kingdom}}

\affiliation{H.H. Wills Physics Laboratory, University of Bristol, \\Bristol,
United Kingdom}

\collaborationImg{\vspace{2pt}\includegraphics[height=62.5pt,width=!]{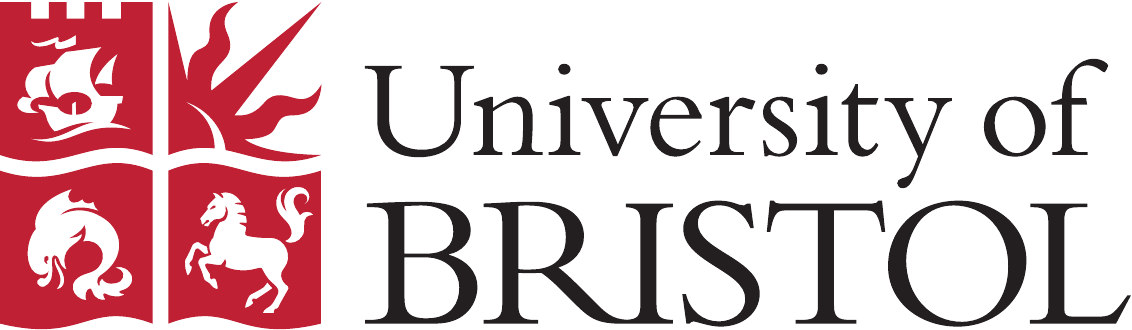}\vspace{24pt}}

\emailAdd{Paras.Naik@cern.ch}

\abstract{Decays
of charmonia(-like) particles with definite $J^{PC}$ ({e.g.} $\X$)
to a $\DDbSys$ system and any combination of $C$-definite decay particles,
are sources of quantum-correlated $\DDbSys$ systems with $C = P = \pm 1$.
Several $b$-hadron decays also produce quantum-correlated $\DDbSys$ systems.
Advantages of isolating these systems in their $C = +1$ components for 
amplitude analyses and studies of lineshapes are discussed. Methods to separate
the $C = \pm 1$ $\DDbSys$ components from $\X$ decay samples are presented. Studies of $T$ and $CPT$ conservation in 
$C = +1$ $\DDbSys$ systems can be performed with
more easily reconstructible final states, when compared to $C = -1$ $\DDbSys$
systems. 
Experimental mechanisms 
that can produce 
$C = \pm 1$ $\DDbSys$ systems are described in an appendix.
}

\arxivnumber{2102.07729}

\begin{document}
\maketitle
\flushbottom

\input{inputs/1-introduction}

\input{inputs/2-formalism} 
\input{inputs/3-simulationSeparation}

\input{inputs/4-timeReversal}
\input{inputs/5-conclusion}

\acknowledgments
I thank Jeremy Dalseno for detailed discussions about spin in decay amplitudes
and the use of {\sc EvtGen}~\cite{Ryd:2005zz} as I established the 
$\XtoDDbP$ decay model used in my simulations.
I thank Drew Silcock, Jennifer Smallwood (and the Ogden Trust), Toby Twigger and
Tom Williams for exploratory contributions that inspired this note.
I thank Jonas Rademacker and Konstantinos Petridis for several useful
discussions and their ongoing support.
Along with Jeremy, Jonas, and Konstantinos, I thank Daniel O'Hanlon,
Peter Onyisi, Mark Whitehead, and Patr\'{i}cia Magalh\~{a}es for their careful
reading and useful comments. 
I also acknowledge
the support of the UK Science and Technology Facilities Council (STFC).
\appendix
\input{inputs/A-appendices}
\bibliographystyle{JHEP}
\bibliography{bibliography}
\end{document}

%% file: inputs/X-newcommands.tex
\newcommand{\prt}[1]{\ensuremath{\mathit{#1}}}

\newcommand{\zeroCharge}{\ensuremath{\mathrm{0}}}

\newcommand{\DbarLHCb}    {\kern 0.2em\overline{\kern -0.2em \ensuremath{D}}{}}
\newcommand{\D}       {\prt{D}}
\newcommand{\Db}      {\prt{\DbarLHCb}}
\newcommand{\Dz}      {\prt{\D^{\zeroCharge}}}
\newcommand{\Dzb}     {\prt{\DbarLHCb^{\zeroCharge}}}
\newcommand{\Dst}     {\Dstar}
\newcommand{\Dstz}    {\prt{D^{*\zeroCharge}}}
\newcommand{\Dstzb}   {\prt{\DbarLHCb^{*\zeroCharge}}}

\newcommand{\DmDb}    {\prt{(\Dz - \Dzb)}}

\newcommand{\Dstar}{\prt{D^{*}}}
\newcommand{\DstarPlus}{\prt{D^{*+}}}

\newcommand{\B}{\prt{B}}

\newcommand{\Bp}{\prt{\B^{+}}}
\newcommand{\Bm}{\prt{\B^{-}}}

\newcommand{\Kp}{\prt{K^{+}}}
\newcommand{\Km}{\prt{K^{-}}}

\newcommand{\pip}{\prt{\pi^{+}}}
\newcommand{\pim}{\prt{\pi^{-}}}

\newcommand{\piz}{\prt{\pi^{\zeroCharge}\xspace}}

\newcommand{\cleoc}{\mbox{{C}{L}{E}{O}{-}{c}}\xspace}

\newcommand{\besiii}{\mbox{{B}{E}{S}{I}{I}{I}}\xspace}

\newcommand{\lhcb}{{L}{H}{C}{b}\xspace}
\newcommand{\babar}{{B}{a}{B}{a}{r}\xspace}

\newcommand{\belleII}{{B}{e}{l}{l}{e}{ }{I}{I}\xspace}
\newcommand{\cms}{{C}{M}{S}\xspace}
\newcommand{\atlas}{{A}{T}{L}{A}{S}\xspace}

\newcommand{\CP}{\ensuremath{\mathit{CP}}}
\newcommand{\CPT}{\ensuremath{\mathit{CPT}}}

\newcommand{\f}{\ensuremath{f}}

\newcommand{\bra}[1]{\ensuremath{\langle #1 |}}
\newcommand{\ket}[1]{\ensuremath{| #1 \rangle}}

\newcommand{\Xx}{{\prt{X}(3872)}\xspace}
\newcommand{\Xs}{{\prt{X}}\xspace}
\newcommand{\Zc}{{\prt{Z_{c}}(3900)}\xspace}
\newcommand{\X}{{\prt{\chi_{c{\rm 1}}}(3872)}\xspace}
\newcommand{\chicJ}{{\prt{\chi_{cJ}}(3930)}\xspace}
\newcommand{\Yx}{{\prt{\psi}(4230)}\xspace}
\newcommand{\psix}{{\prt{\psi}(3770)}\xspace}
\newcommand{\psiff}{{\prt{\psi}(4040)}\xspace}

%% file: inputs/1-introduction.tex
\FloatBarrier\section{Introduction}
\label{Introduction}

The study of quantum-correlated charmed mesons\footnote{A pair of
charmed mesons in a definite eigenstate of charge conjugation ($C$),
where as a consequence the two mesons must have either the same or the opposite
$\CP$ eigenvalues, depending on the $C$ eigenvalue of the pair.
} has been a fantastic success for constraining the parameters of $\Dz$$-$$\Dzb$ mixing, 
and as powerful input in constraints of the CKM phase $\gamma$, 
helping to push the total uncertainty on the Standard Model benchmark 
value of $\gamma$ below
$5^\circ$~\cite{LHCb:2021dcr,HFLAV:2022pwe, 
Rosner:2008fq,Asner:2008ft,Asner:2012xb,Insler:2012pm,Libby:2010nu,Ablikim:2014gvw,Nayak:2014tea,Malde:2015mha,Evans:2016tlp,Harnew:2017tlp,Resmi:2017fuo,Ablikim:2020lpk,Ablikim:2020cfp,BESIII:2020hlg,BESIII:2021eud,
ADS, GLW1, GLW2, GGSZ, GLS,  
PhysRevD.15.1254, Bigi:1986dp, Xing:1996pn,
Gronau:2001nr, Atwood:2002ak, Asner:2005wf, 
Atwood:2000ck,  Atwood:2003mj,
Grossman:2005rp,  Bondar:2005ki, Bondar:2008hh}. 
Improving these constraints will require 
further quantum-correlated samples
to be included~\cite{BESIII:2022qkh,BESIII:2022wqs,BESIII:2022ebh} and
studied~\cite{Charles:2009ig, Poluektov:2017zxp, Bediaga:2018lhg}.
Further constraints on the Standard Model benchmark 
value of $\gamma$ are crucial, to 
identify potential deviations from
measurements of $\gamma$ where new physics effects may be introduced through
quantum loops~\cite{Bhattacharya:2018pee}. 

To date, the quantum-correlated $\DDbSys$ systems
used for these analyses are obtained from $e^+ e^-$ collisions at $\psix$
threshold.
However, such correlations exist independent of the production mechanism used to
create the initial $\psix$ state. 
Moreover, such correlations 
also exist for $\DDbSys$ systems
from the decays of
other initial particle
states,
so long as the $\DDbSys$ systems produced are
formed in a eigenstate of charge conjugation ($C$).

The $\X$ (previously known as $\Xx$) exotic meson was
first discovered by the Belle collaboration in 2003~\cite{Choi:2003ue}, rapidly confirmed by 
CDF~\cite{Acosta:2003zx},
D0~\cite{Abazov:2004kp}, and BaBar~\cite{Aubert:2004ns}, and established as a
state that can decay to systems of open charm, as near-threshold $\Dstz\Dzb$ and $\Dz\Dzb\pi^0$ 
enhancements corresponding to the $\X$ have been
observed~\cite{Gokhroo:2006bt,Aubert:2007rva,Adachi:2008sua}.
The observation of the decay mode $\X \to
J/\psi \gamma$ in 2011 established that the $\X$ meson
has a charge-conjugation
eigenvalue of $C = +1$~\cite{Bhardwaj:2011dj}. In 2013, the LHCb
collaboration 
established the full quantum numbers $J^{PC}={1}^{++}$ for the
$\X$ state~\cite{Aaij:2013zoa}.
The most precise measurements of the $\X$ mass and decay width are
$3871.695 \pm 0.096$~\MeVm\xspace and $1.39 \pm 0.26$~\MeVm,
respectively~\cite{Aaij:2020qga, LHCb:2020fvo}. The $\X$ mass coincides with
the sum of the $\Dz$ and $\Dstz$ meson masses~\cite{PDG2020}. The branching fraction of $\X \to \Dz
\Dstzb + \Dzb \Dstz$ has been determined to be 
$(52.4_{-14.3}^{+25.3})\%$~\cite{Li:2019kpj}.

This knowledge allows us to now consider utilizing $\DDbSys$ systems 
from $\X$
decays, through a natural extension of existing work, which considers
$\DDbSys$ systems from $e^+e^-$
interactions~\cite{PhysRevD.15.1254,Bigi:1986dp,Xing:1996pn,Gronau:2001nr,
Atwood:2002ak,Asner:2005wf}.
Perhaps because the $\X$ is a $J^{PC}={1}^{++}$ state, and resonant
production of $C = +1$ states is 
very highly suppressed
at $e^+e^-$
colliders~\cite{BESIII:2022mtl}, directly or through initial state radiation
(ISR), the idea of using
these decays for quantum-correlated analyses has not previously been discussed.
However, as is the case for $\psix$ decays, substantial
numbers of 
$\X$ decays could be obtained through several
alternative production mechanisms. 

The necessary formalism is presented in \cref{Formalism}, with some
applications discussed. In \cref{simulationSeparation},   
methods are proposed that 
enable the separation of $C = -1$ and $C = +1$
$\DDbSys$ systems from $\X$ decays.
In
\cref{TV}, a new proposal to use 
$C = +1$ $\DDbSys$ systems ({e.g.} from $\X$ decays) to quantify
time-reversal violation and \CPT\xspace violation in the charm system is
outlined.
Additionally, \cref{Laboratories} 
details 
possible correlated $\DDbSys$ production
mechanisms, highlighting potential advantages of exploiting $\X$, compared to
exploiting $\psix$, in some cases.

%% file: inputs/2-formalism.tex
\FloatBarrier\section{Quantum correlations formalism}
\label{Formalism}

A meson-antimeson system such as $\DDbSys$ exists in an eigenstate of parity
($P$) and an eigenstate of $C$, with both eigenvalues defined by:
\begin{equation}
C_{{\DDbSys}} =
P_{{\DDbSys}}  
= (-1)^{L_{\DDbSys}},
\end{equation}
where $L_{\DDbSys}$ is the relative orbital angular momentum of the $\Dz$ meson 
and $\Dzb$ meson system. 
In order to form an eigenstate of $C$, a $\DDbSys$ 
system in a state of definite $L_{\DDbSys}$ is limited to the 
following 
quantum state configurations~\cite{PhysRevD.15.1254}: 
\begin{equation}
\frac{\ket{\Dz\Dzb}+\ket{\Dzb\Dz}}{\sqrt{2}}
~{\rm when}~ C_{{\DDbSys}} = P_{{\DDbSys}} = +1,~L_{{\DDbSys}}~{\rm is~even}
\label{plus}
\end{equation}
and
\begin{equation}
\frac{\ket{\Dz\Dzb}-\ket{\Dzb\Dz}}{\sqrt{2}}
~{\rm when}~ C_{{\DDbSys}} = P_{{\DDbSys}} = -1,~L_{{\DDbSys}}~{\rm is~odd}.
\label{minus}
\end{equation}

\FloatBarrier\subsection{Scenarios leading to the production of \texorpdfstring{$\DDbSys$}{D0
anti-D0 systems} in a \texorpdfstring{$C$}{C} eigenstate}
\label{subsec:C}

In $e^+ e^-$ colliders the production 
mechanism for the resulting
decay particles is electromagnetic annihilation through a virtual photon
($\gamma^*$); 
this is specifically true when the collision
energy is such that the weak $Z$ boson does not contribute. 
The charge conjugation eigenvalue of the initial state is:
\begin{equation}
C_{e^+ e^-} = C_{\gamma^*} = -1.
\end{equation}

For a 
final state composed of 
states that do not
combine themselves to form any particle-antiparticle systems, 
the $C$ parity is the product of $C$ parities
for each particle system.
Recall that a photon has $J^{PC} = 1^{--}$, and a $\piz$ meson has 
$J^{PC} = 0^{-+}$.
Thus, for the example decay 
$e^+ e^- \xrightarrow[]{\gamma^*} \DDbSys + m \gamma + n
\piz$:
\begin{equation}
C_{\gamma^*} = 
C_{{\DDbSys}} \times (-1)^m \times 1
\label{classic}
\end{equation}
thus
\begin{equation}
C_{{\DDbSys}} 
= (-1)^{1-m} 
= (-1)^{m+1},
\label{Cfrompsi}
\end{equation}
where $m$ and $n$ are positive integers.
If other $C$-definite particles or particle systems were also produced 
({e.g.} 
$J/\psi$, $\rho^0$, $(K^+ K^-)_{C = +1}$) in the
final state, additional multiplicative terms could be included in
the two equations above. 

In $e^+e^-$ colliders, the virtual photon energy 
can be tuned
for resonant production of $c\overline{c}$ states, at or 
above $\DDbSys$ threshold. In the
specific case of the $C$-conserving 
decay $e^+e^- \to \psix \to \DDbSys$, $m
= n \equiv 0$; thus for the resulting $\DDbSys$ system, $C = -1$. 
There is a weak current contribution to $\psix \to \DDbSys$, thus
charge-parity conservation in principle could be relaxed;
however, angular
momentum conservation requires the
$\DDbSys$ system to be in an antisymmetric state (see
\cref{minus}), ensuring the $\DDbSys$ system has the same quantum numbers as
the initial resonance~\cite{Asner:2008nq}.

Note that
$C_{\psix} = -1$, so if running an $e^+e^-$ collider at
a resonance decaying to open neutral charm, we could substitute ${\psix}$
for ${\gamma^*}$ in the above equations and obtain the same physics. 
More energetic $C = -1$ resonance states such as $\psiff$ could be considered;
then both $C=+1$ and $C=-1$ $\DDbSys$ can be produced depending on the number of associated photons
or neutral pions. In these cases,
the production mechanism of the 
$C = -1$ resonance
is irrelevant and may be dropped.

The principles above may also be applied for the strong/electromagnetic 
decay of a particle or state with 
$C = +1$, such as the $\X$ exotic meson,
to the $\DDbSys + m \gamma + n \piz$ final state. 
For the $C$-conserving 
decay $\X \to \DDbSys + m
\gamma + n \piz$:
\begin{equation}
C_{\X} 
= C_{{\DDbSys}} \times (-1)^m \times 1
\end{equation}
thus
\begin{equation}
\label{CfromX}
C_{{\DDbSys}} 
= (-1)^{-m} =
(-1)^{m}.
\end{equation}

In $\X$ decays with $\DDbSys$
in the final state
$m = n = 0$ is not allowed by parity conservation.
The $n = 1$, $m = 0$ decay $\X \to \DDbP$ has been observed, is expected to have
a large branching fraction, and is dominantly produced 
via 
$\X \to \Dstzb \Dz$ and $\X \to \Dstz \Dzb$ resonant decays~\cite{PDG2020},
which may have varying degrees of coherence depending on the nature of 
binding between the charm and anti-charm components of the $\X$
exotic meson~\cite{Voloshin:2003nt}.
A direct $\X \to \DDbP$ non-resonant
component has not yet been ruled out~\cite{PDG2020}.
Thus the decay $\X \to \DDbG$ is also expected; 
for this decay $n = 0$, $m = 1$. 
The 
$\Dstzb$ is expected to decay twice as much to $\Dzb \piz$ as to $\Dzb
\gamma$, but due to the smaller phase space of $\X \to \DDbP$,
$\X \to \DDbG$ is expected to be slightly enhanced
\cite{Hanhart:2007yq}. Decays with $ n > 1 $ are not possible, and
 decays with $m > 1$ are henceforth not considered as they are likely
to be suppressed.

To obtain a pure $C = +1 (-1)$ $\DDbSys$ system, the $\X$ decays would
have to be 
cleanly
reconstructed in the $\DDbP (\DDbG)$ decay mode. 
Separation techniques 
addressing the reconstruction of these $\DDbSys$ systems, without
reconstructing the photon or neutral pion,
are discussed in 
\cref{simulationSeparation};
purity of the $C$ identification is briefly discussed in
\cref{subsec:purity}.

\Cref{Cfrompsi,CfromX} can be fully generalized to all $C$-conserving
charmonium(-like) decays of particles with fixed $J^{PC}$ to the 
$\DDbSys + m \gamma + n \piz$ final state:
\begin{equation}
C_{{\DDbSys}} = 
(-1)^{m} C_{J^{PC}_{c\overline{c}}}.
\end{equation}

This may be useful in the case that existing, new, or
exotic charmonia-like particles are determined to produce correlated $\DDbSys$
systems in quantity. The existence of the neutral $\Zc$ 
with
$J^{PC} =  1^{+-}$~\cite{Ablikim:2015gda,Collaboration:2017njt} should motivate
such searches for such alternative sources of correlated $\DDbSys$.

Lastly, it should be noted that $\DDbSys$ systems from weak decays such as
$B_{(s)}^0 \to \DDbSys$ must also exist in fixed states of $C = +1$ due to
angular momentum conservation~\cite{TimG}. 
In this case, the relative orbital angular momentum of the $\DDbSys$ system must
be $L = 0$, 
and a boson pair with $L = 0$ must be
in a symmetric state.

\FloatBarrier\subsection{Quantum state of \texorpdfstring{$\X \to \Dz\Dstzb$}{X(3872)
decays to D0 anti-D*0}}
\label{subsec:quantumStateOfX}

The $\Dz\Dstzb$ system is not an eigenstate of $C$, so the strong decay $\X \to
\Dz\Dstzb$ cannot occur in isolation~\cite{Voloshin:2003nt}.
It is only possible to produce $\X \to \Dz \Dstzb$, and conserve $C$, via
a symmetric (and henceforth implied)
superposition with $\X \to  \Dzb \Dstz$ decays:
\begin{equation}
\frac{\ket{\Dz\Dstzb}+\ket{\Dzb\Dstz}}{\sqrt{2}}.
\label{superplus}
\end{equation}

Thus in $\X \to \Dz\Dstzb$ decays, the full quantum state must evolve as either:
\begin{equation} 
\label{eq:evolve1}
\ket{\X} \to
\frac{\ket{\Dz\Dstzb}+\ket{\Dzb\Dstz}}{\sqrt{2}}
\to 
\frac{\ket{\Dz\Dzb} + \ket{\Dzb\Dz}}{\sqrt{2}}
\ket{\piz}
\end{equation}
or
\begin{equation} 
\label{eq:evolve2}
\ket{\X} \to
\frac{\ket{\Dz\Dstzb}+\ket{\Dzb\Dstz}}{\sqrt{2}}
\to 
\frac{\ket{\Dz\Dzb} - \ket{\Dzb\Dz}}{\sqrt{2}}
\ket{\gamma}.
\end{equation}
This is 
verified, for example, by following the arguments of Bondar and
Gershon~\cite{Bondar:2004bi}. 

\FloatBarrier\subsection{Configurations of
\texorpdfstring{$\DDbSys$}{D0 anti-D0} systems
in
\texorpdfstring{$C$}{C} eigenstates}
\label{subsec:Configurations}

In the case of $\X \to \Dz \Dstzb$ decay, angular momentum
relationships between the $\Dz$ and $\Dstzb$ mesons, and also the $\Dzb$
meson and light neutral $\piz$ meson or photon, 
can be determined
easily, but recall from \cref{plus,minus} that 
to have a correlated $\DDbSys$ system, an 
angular momentum relationship must exist
in the system. This relationship limits the possible $\DDbSys$ states to:
\begin{equation}
J^{PC}_{\DDbSys} = {0^{++}, 1^{--}, 2^{++}, \ldots}
\end{equation}
In
general, we may use strong conservation laws to determine the possible
relationships, demonstrating that the types of decays that we are interested in
are not forbidden.
In this section, the $J^{PC}_{\DDbSys}$
possibilities for the cases of $\X \to \DDbP$ and $\X \to \DDbG$ are explicitly
determined.

The $\X$ exotic meson is known to
have large isospin-breaking 
effects~\cite{Kusunoki:2005qs}; this is explained by the possibility that the
$\X$ is a superposition of $\Dz\Dstzb$, $\D^{+}\D^{*-}$, and
their charge conjugate states, while the $\X$ is below threshold to
produce a $\D^{+}\D^{*-}$ state.
The $\X$ exotic meson may be in as much as an equal superposition 
of 
isospin $0$ and isospin $1$~\cite{Close:2003sg}.
It is assumed that 
the $\X$ contains an
isospin component that does not forbid the following transitions.

Addressing the other quantum numbers, the
decays and their spin schematics may be written as follows:
\begin{eqnarray}
\X \to (\DDbSys)_{L_{\DDbSys}} \piz \qquad &::& \qquad 1^{++} \to
\underbrace{J^{PC}_{\DDbSys} \oplus 0^{-+}}_{L_R}\\
\X \to (\DDbSys)_{L'_{\DDbSys}} \gamma \qquad &::& \qquad 1^{++} \to
\underbrace{J'^{P'C'}_{\DDbSys} \oplus 1^{--}}_{L_R'}
\end{eqnarray}

For $\XtoDDbP$, $C$ conservation dictates that $C_{\DDbSys} = 1$, thus
$P_{\DDbSys} = 1$, and $J_{\DDbSys}$ can only be even. $P$ conservation then
also dictates that $L_R$ is odd.
For example, since $J_\X = L_R \oplus J_{\DDbSys}$, if $J_{\DDbSys} = 0$ then
$L_R = 1$.\footnote{One could consider initial state particles of other $J^{PC}$:
{e.g.} for a $J^{PC} = 1^{+-}$ initial state decaying to $\DDbP$,
$C_{\DDbSys}$ and $P_{\DDbSys}$ change sign, $J_{\DDbSys}$ is odd, and $L_R$ is even.}

For $\XtoDDbG$, $C$ conservation dictates that $C'_{\DDbSys} = -1$, thus
$P'_{\DDbSys} = -1$, and $J'_{\DDbSys}$ is odd. $P$ conservation then also
dictates that $L'_R$ is even. For example, since
$J_\X = L'_R \oplus (J'_{\DDbSys} \oplus 1)$, 
if $J'_{\DDbSys} = 1$ then $L'_R = 0$ or $2$.

The possibilities where ${J} {{}^{(}} ' {{}^{)}}_{\DDbSys} \leq 2$ are shown
in \cref{table:angmomtable}. Since the $\X$ mass is nearly the same as
the sum of the $\D$ meson and $\Dst$ meson masses,
it is reasonable to presume that the modes with higher angular momenta will be
suppressed.
\begin{table}[t]
\begin{center}
\begin{tabular}{c|c|c} 
\hline\hline
Decay &  $J_{\DDbSys}$ & $L_R$ \\
\hline\hline
$\XtoDDbP$ & 0 & 1 \\
$\XtoDDbP$ & 2 & 1 or 3\\
$\XtoDDbG$ & 1 & 0 or 2 \\
\end{tabular}
\caption{Allowed angular momentum configurations for $J_{\DDbSys} \leq 2$}
\label{table:angmomtable}
\end{center}
\end{table}

\FloatBarrier\subsection{Comment on effect of \texorpdfstring{$\Dz\Dstzb$}{D0 anti-D*0}
correlations}
\label{subsec:effectOfCorrs}

It is worthwhile to briefly comment further on the decay $\XtoDDbP$, 
which should be produced in copious quantites from the $\X$ state. The
decay has a small phase space, yet still has structure due to the
possibility of $\D^*$ resonances. 

The correlations of both
\cref{plus,superplus} should appear
in the full $\XtoDDbP$ amplitude model. In order for this to be
the case, the full 
amplitude for the decay 
must be:
\begin{equation}
\frac{
(\bra{\DDbP}+ \bra{\DbDP})
(\ket{\Dz\Dstzb}+\ket{\Dzb\Dstz})
(\bra{\Dz\Dstzb}+\bra{\Dzb\Dstz})
(\ket{\X})}{2\sqrt{2}}
\label{eq:amp}\end{equation}

Note that there are no arbitrary phases between the
individual components, besides a global phase that is irrelevant to the decay
dynamics. This doubly-correlated amplitude is effectively the sum of 
two decay amplitudes that are
identical, 
thus the
$\DDbSys$ correlation 
does not change
the probability
distribution of $\XtoDDbP$ decays over the phase space. 
A similar argument
applies to the $\XtoDDbG$ decay.

\FloatBarrier\subsection{Exploiting correlations in the \texorpdfstring{$\DDbSys$}{D0
anti-D0} system}
\label{subsec:exploitingCorrs}

The production of $\Dz$ and $\Dzb$ in the configurations of
\cref{plus,minus} allow several charm mixing and relative strong
phase parameters to be extracted for neutral $\D$ decays, via decay rates 
of the resulting $\D$ mesons. 
It is possible to
determine the parameters of interest via a 
time-integrated analysis as discussed
in refs.~\cite{PhysRevD.15.1254,Bigi:1986dp,Xing:1996pn,Gronau:2001nr,
Atwood:2002ak,Asner:2005wf, Asner:2012xb}.

These parameters were determined in refs.
\cite{Rosner:2008fq,Asner:2008ft,Asner:2012xb,Ablikim:2014gvw,BESIII:2022qkh}, 
but no $C$-even $\DDbSys$ systems were
used. 
In the
simple case of using only two-body hadronic and semileptonic $\D$ decays,
$C$-even $\DDbSys$ systems offer 
linear sensitivity to the charm mixing parameter $x$,
a quantity that is directly proportional to the mass difference between 
the two neutral $\D$ mass eigenstates, that $C$-odd $\DDbSys$ systems do
not (these are only sensitive to $x^2$). $C$-even $\DDbSys$ also offer
considerably improved sensitivity (by approximately a factor of two) 
to the charm mixing parameter $y$, a quantity directly proportional 
to the difference in the widths of the neutral $\D$ mass 
eigenstates~\cite{Asner:2005wf}, assuming  
that the same number of $C$-even and
$C$-odd decays are obtained.

Of equal importance, $C$-even and $C$-odd $\DDbSys$ systems also provide 
complementary 
sensitivity to strong phases that exist in $\D$ decays, for example $\Dz \to
K^-\pi^+$. Both $C$-even and $C$-odd $\DDbSys$ systems provide sensitivity
to the cosine of the $\Dz \to K^-\pi^+$ strong phase difference,
$\delta_{K\pi}$, whereas only $C$-even decays are sensitive to
$\sin(\delta_{K\pi})$, in the case where no all-hadronic multi-body $\D$ decay modes are used.\footnote{It is possible 
to gain direct sensitivity to $x$ and $\sin(\delta_{K\pi})$, by
including multi-body $\D$-decay final states of mixed \CP\xspace and exploiting
their interference with two-body decays (see ref.~\cite{Asner:2012xb}).}

\backrefsetup{disable}\begin{table}[t]
\begin{center}
\begin{tabular}{c|c|c|c} 
\hline\hline
$\D$ decay mode & $\bar{\D}$ decay mode & {$\mathcal B$} (naive)
 & {$\mathcal B$} (including correlations) 
\\
\hline\hline
$K^-\pi^+$  & $K^+\pi^-$ & $1.60 \times 10^{-3}$ & $1.60 \times 10^{-3}$  \\
$K^+K^-$, $\pi^+\pi^-$  & $\pi^+\pi^-$, $K^+K^-$ & $1.23 \times 10^{-5}$ & $2.45
\times 10^{-5}$ \\
$K^+K^-$  & $K^+K^-$ & $1.69 \times 10^{-5}$ & $3.38 \times 10^{-5}$ \\
$\pi^+\pi^-$  & $\pi^+\pi^-$ & $2.22 \times 10^{-6}$ & $4.44 \times 10^{-6}$ \\
\end{tabular} 
\caption{Approximate product branching fractions {$\mathcal B$} of a $C =
+1$ $\DDbSys$ pair reconstructed in the corresponding $\D\bar{\D}$ decay mode,
both under the naive expectation~\cite{Amhis:2019ckw} and after the effects of quantum
correlation~\cite{Asner:2005wf,Asner:2008ft}, excluding small effects due to charm mixing and ignoring doubly-Cabibbo suppressed decays. }
\label{table:Cdefinite}
\end{center}
\end{table}
\backrefsetup{enable}
The $(\DDbSys)_{C=+1}$ branching fraction to 
positive $\CP$-definite final states (such as those
described in \cref{TV}) for both $\D$ mesons will 
be approximately a factor
of two larger than the naive expectation, 
due to the quantum correlations 
described by refs.~\cite{Asner:2005wf,Asner:2008ft}. 
This is advantageous, as the positive $\CP$-definite final states include fully
charged-track decay modes that are easier to reconstruct experimentally. Representative
effects on the branching fractions of selected doubly
two-body $\D$ decay modes are shown in \cref{table:Cdefinite}.

The current best suggestions for obtaining these $C$-even decays would
be to extend the time-independent analysis to $\DDbSys$ from a more energetic
resonance {e.g.} $e^+e^- \to \psi(4040)$~\cite{Bondar:2010qs}, or to
simply collect $e^+e^- \to \DDbSys + m \gamma + n
\piz$ above the charm threshold~\cite{Asner:2005wf}. However, with existing
samples these are much less competitive in statistical power. 

If it is possible to separate the $C = +1$ and $C = -1$ $\DDbSys$ in
$\X$ decays, then this would provide an alternative source. 
A large number of prompt $\X$ candidates producing $\DDbSys$ systems in the
final state have recently been observed at \lhcb~\cite{Aaij:2019evc}, but no attempt was made to separate the $\XtoDDbP$ and $\XtoDDbG$ decay samples.
Potential separation techniques are demonstrated in
\cref{simulationSeparation}, and  
potential advantages of using $\X$ decays are discussed 
in
\cref{Laboratories}.

\FloatBarrier\subsection{\texorpdfstring{$C$}{C}-definite \texorpdfstring{$\DDbSys$}{D0
anti-D0} in
\texorpdfstring{$\XtoDDbP$ and $\B \to \DDbSys X$}{X(3872) to D0 anti-D0 pi0
and B to D0 anti-D0 X} decays}
\label{subsec:CdefiniteBdecays}

As a consequence of quantum correlations, the two $\D$ mesons in a $\DDbSys$
system must have either the same or the opposite $\CP$ eigenvalues, depending on
the $C$ eigenvalue of the pair. By selecting both $\D$ mesons by
their decays to positive $\CP$-definite final states, $\XtoDDbG$ decays, where
the $\DDbSys$ systems are $C = -1$ correlated, will not be observed,
leaving only the $\XtoDDbP$ decay mode for study. This would, for example, allow
the lineshape of $\XtoDDbP$ to be studied in isolation.

This consequence of quantum correlations also allows explicit selections of $\B
\to \DDbSys X$ decays, where $X$ is any set of additional final state particles,
in 
$\B \to (\DDbSys)_{C=-1} X$ and $\B \to
(\DDbSys)_{C=+1} X$ components. For example, by selecting both $\D$ mesons by
their decays to positive $\CP$-definite final states, the resulting $\B \to
\DDbSys X$ decays will only have $C=+1$ $\DDbSys$ components. An amplitude
analysis of such a sample could help to better constrain relative strong phases
between $\D K$  and $C = +1$ $\DDbSys$ resonances, 
within the decays $B^+ \to \DDbSys K^+$ and $B^0 \to \DDbSys K^0_S$. 
Note that if a $\B \to
\DDbSys X$ decay is high in $\B \to
(\DDbSys)_{C=+1} X$ content, then selecting both $D$ mesons
in positive $\CP$-definite final states 
would have enhanced branching fractions relative to selecting $D$ mesons in
flavor final states. More discussion on this topic may be found in
\cref{subsec:QCfrom-B}.
%

\FloatBarrier\subsection{Purity of the initial \texorpdfstring{$C$}{C}-definite
\texorpdfstring{$\DDbSys$}{D0 anti-D0} state, and using
\texorpdfstring{$\DDbSys$}{D0 anti-D0 systems} of mixed \texorpdfstring{$C$}{C}}
\label{subsec:purity}

A $C$-definite initial $\DDbSys$ state may be diluted by
radiated photons, which would flip the $C$ eigenvalue of the $\DDbSys$ system 
and thus result in an uncertainty that would need to be accounted for. However,
photon radiation is problematic only if it alters the relative angular momentum 
between the $\Dz$ meson and $\Dzb$ meson in the $\DDbSys$ system. Thus initial
and final state radiation, and bremstrahlung
photon emission, do not affect the $C$ eigenvalue because these processes all
occur either before the formation of the $\DDbSys$ system or after the
constituent $\Dz$ and $\Dzb$ mesons have decayed. There are virtual
processes that would cause a $C$ eigenvalue flip {e.g. $\psix \to 0^+_{\rm virtual}
\gamma \to \DDbG$, $\psix \to {(\Dz \Dstzb)}_{\rm virtual}
\to \DDbG$}, however such processes are predicted to  
occur at the order of $10^{-8}$ relative to the dominant
processes~\cite{PetrovPrivateCommunication}. 
It should be reasonably expected that for \X decay we should 
expect similar suppresion of such virtual states.

It is certainly possible for coherent $\DDbSys$ systems to have their $C$
eigenstate misidentified. However, 
the size of the admixture can be determined
from the data if both $\D$ mesons are reconstructed, and 
the effect of the
admixture can be accounted for in an analysis of such data~\cite{Asner:2005wf}.
There is a dilution of sensitivity if the sample is not pure, but the
analysis can still be performed. If the number of $C$-even and $C$-odd
decays in the admixture are exactly equal though, 
then some key charm mixing parameters
cancel and cannot be measured. 

%% file: inputs/3-simulationSeparation.tex
\FloatBarrier\section{Simulation and variables of separation}
\label{simulationSeparation}

In this section, the possibility of separating the correlated $C=+1$
and $C=-1$ $\DDbSys$ systems from $\X$ decays is discussed. 
The techniques are demonstrated with representative decays 
simulating specific detector effects, however they could be used 
in alternative laboratory environments as well.

\FloatBarrier\subsection{\texorpdfstring{{\scshape{RapidSim}} and
{\scshape{EvtGen}}}{RapidSim and EvtGen}}

\label{subsec:simulation}

%
Simulations of 
$\XtoDDbP$ and $\XtoDDbG$ 
are generated, which take into account the 
momentum and impact parameter (IP) resolution representative of that of the
\lhcb detector and experiment~\cite{Alves:2008zz,Aaij:2014jba}, and the effects of
final state radiation (FSR).
Such representative decays are generated with {\sc
RapidSim}~\cite{Cowan:2016tnm}, which can generate its own decays or utilize
{\sc EvtGen}~\cite{Ryd:2005zz}, and 
FSR is generated with {\sc PHOTOS++ v3.61}~\cite{Davidson:2010ew}. 

The simulations generated by {\sc RapidSim} presume production
profiles based on proton-proton collisions occuring at a center-of-mass energy
of 14~\TeVE.
Prompt $\X$ 
decays are generated under the 
assumption that the low momentum release 
in the decay dictates, that in the measurement frame, the particles 
in the $\DDbSys$ system will move approximately colinearly; 
thus the included 
``{\sc FONLL}'' 
charmed meson transverse
momentum and pseudorapidity distributions~\cite{Cacciari:2001td,
Cacciari:1998it} are
extended to represent low-mass charmonium by simply doubling the transverse
momenta generated, while maintaining the pseudorapidity. 
Decays producing $\X$ 
from $B$ mesons are simulated with the $B$
mesons having {\sc FONLL}
beauty meson transverse momentum and pseudorapidity distributions~\cite{Cacciari:2001td,
Cacciari:1998it}. 

In addition to particle decay models already contained within {\sc
EvtGen}, for $\XtoDDbP$ decays a specific amplitude model to fully describe
\cref{eq:amp} as the coherent sum of 
$\X \to \Dz\Dstzb$ and 
$\X \to \Dstz\Dzb$ amplitudes is also created. 
In this model, the masses of the $\Dst$ mesons, neutral $\D$ mesons, $\piz$
meson, and $\X$ exotic meson are fixed to their PDG values~\cite{PDG2020}.
The $\Dstz \to \Dz \piz$ decay width is fixed to 43~\keVm, as estimated by
Voloshin in ref.~\cite{Voloshin:2003nt}. 
The amplitude model, described 
in \cref{sec:Amplitude}, 
has
been implemented using {\sc EvtGen}~\cite{Ryd:2005zz}, and has the ability to
produce the intermediate state $\Dz\Dstzb$ and charge conjugate amplitudes 
in both $L=0$ (``Model S-Wave'') and $L=2$ (``Model D-Wave'') angular momentum
states. 

Since the $\X$ mass is nearly the same as the sum of the $\Dz$ and
$\Dstzb$ mass, it is reasonable to presume that the $L=2$ $\Dz\Dstzb$ mode 
is suppressed. However, the possibility of $L=2$ $\Dz\Dstzb$ is included in the
following sections to demonstrate the robustness of the separation techniques.

Using an amplitude model for $\XtoDDbP$ in the following sections
instead of the particle decay models for $\X \to \D\Dst \to \DDbP$ already
contained within {\sc EvtGen} had a mild effect on the separability of the
$\DDbP$ and $\DDbG$ samples in the simulation.
The construction of an analogous 
$\XtoDDbG$ amplitude model was not pursued, as
the use of a $\XtoDDbG$ model 
instead of the particle decay models for $\X \to \D\Dst \to \DDbG$
is expected to have approximately the same mild effect.

\FloatBarrier\subsection{Simulations of \texorpdfstring{$\XtoDDbP$, 
$\XtoDDbG$, \\and prompt $\DDbSys$ background}{X(3872) to D0
anti-D0 pi0, X(3872) to D0 anti-D0 gamma, and prompt
D0 anti-D0 background}
}
\label{sec:simulatedDecays}

The following simulated data samples are generated for the purposes of
demonstrating distributions of the variables proposed in 
\cref{subsec:DmDb_m2,subsec:angleDDb}, 
thus the number of events generated for each
sample is arbitrary. 

{\sc RapidSim} produces standard
decays with representative effects of detector simulation, and also ``TRUE''
decays without detector simulation effects; for every sample described both
standard and TRUE events are produced. All $\Dz$ mesons decay to $K^-\pi^+$, and
all $\Dzb$ mesons decay to $K^+\pi^-$.
Note that no mass constraint is 
applied to either neutral $\D$ meson in {\sc RapidSim}.

In {\sc RapidSim}, $2 \times 10^5$ events each of Model S-Wave and
Model D-Wave $\XtoDDbP$ decays 
are produced.
The non-interfering sum of the Model S-Wave and
Model D-Wave samples is called the ``Model'' sample.

In addition,  
events are produced in which 50\% are $\X \to \Dz\Dstzb \to \DDbP$ and 50\% are
$\X \to \Dstz\Dzb \to \DDbP$ to simulate a possible non-interfering decay mode; 
$2 \times 10^5$ events are produced in both $(\Dst\D)_{L=0}$
(``50/50 S-Wave'') and $(\Dst\D)_{L=2}$ (``50/50 D-Wave'') modes. The
non-interfering sum of these samples is called the ``50/50'' sample. In all
50/50 samples, the same mass and width parameters are
used as in the Model samples for the particles involved in the decay.
Equivalent samples are produced for $\XtoDDbG$ decays; the $\Dstz \to \Dz
\gamma$ decay width is fixed to 26~\keVm, as estimated by 
Voloshin in ref.~\cite{Voloshin:2003nt}.

In the case of both the Model and the 50/50 samples, the $\X \to \D\Dst$
decay amplitude considers the $\Dst$ to be on-shell; any conclusions drawn from
these samples depend on this assumption.
However, because the $\X$ mass is so close to $\Dz\Dstzb$ threshold, neutral $\Dst$ decays
may also be produced off-shell~\cite{Dubnicka:2010kz}, and thus the decays
$\XtoDDbP$ and $\XtoDDbG$ are expected to have non-trivial lineshapes.

In order to mitigate the assumption that the $\Dst$ is on-shell, $2 \times
10^5$ events are produced decaying uniformly throughout the $\XtoDDbP$ decay phase space (``PHSP'' events) 
to simulate a possible non-resonant / off-shell decay mode. 
An equivalent sample is produced for $\XtoDDbG$ decays.
The ``PHSP'' events are meant to be representative; 
real decays are not uniform in the phase space. As discussed in
\cref{subsec:Configurations}, there are various possible angular momentum
relationships among the final state particles in $\XtoDDbP$ 
and $\XtoDDbG$ 
decays.
However, this simplified representation is only for
demonstration, as decays which do not occur through a $\Dst\D$ intermediate
state are more easily separable --- this will be discussed in
\cref{monochromatic}.
%
%
%
%

All 
$\X$ decays are generated
with prompt proton-proton collision profiles and also with  
$\BptoXKp$ production profiles. Note that the prompt $\X$ decays are
produced ``unpolarized'', 
i.e. 
with equal weight for each polarization.
All $\Bp$ mesons are generated with the mass
and lifetime properties listed in ref.~\cite{PDG2014}. Events produced were only
accepted if all decay daughters were within the LHCb detector's nominal
acceptance,
though the ``light neutral'' 
(i.e. $\gamma$ or
$\piz$) may be treated as visible or invisible to the detector for this
acceptance requirement. 
Samples with invisible light neutral particles are shown in most cases;
however when angular distributions are presented, samples where the light
neutral particles are required to be visible are substituted.
Note that for the {\sc RapidSim} prompt and ``secondary'' $\BptoXKp$
samples, $\X$ were generated only at the $\X$ mass (no width), to ensure
conservation of energy in $\X \to \Dst\Db$ when
generating $\BptoXKp$ samples, and consistency between the $\X$ width in
the prompt and secondary samples; on the scales of interest this choice is not
expected to have a significant impact.

Separately, prompt $\DDbSys$
background samples are generated in {\sc RapidSim} with uniform
density in allowed values of $m_{\DDbSys}$.
For these samples the ``{\sc TGenPhaseSpace}'' generator of {\sc ROOT}
\cite{ROOT} is used, instead of {\sc EvtGen}; the effects of FSR are also not
included.
The background
distribution is not meant to be completely representative of real background,
but has a similar profile to the $m_{\DDbSys}$ distribution of the background 
found under the $\X$ and $\psix$ decay peaks in
Fig. 5 of ref.~\cite{Aaij:2019evc}. 
A ``Flat Background Sample'' of 
approximately $2 \times 10^6$
events 
were produced
that fall within the
acceptance of the LHCb detector and within 
a range of $m_{\DDbSys} < 3.76$~\GeVm after representative detector simulation. 
%
%
%
%

\FloatBarrier\subsection{The monochromatic \texorpdfstring{$\piz$}{neutral
pion} and \texorpdfstring{$\gamma$}{photon}} \label{monochromatic}

For the
case of PHSP decays, the energy of the $\piz$ meson ($E_\piz$) in the
$\XtoDDbP$ decay rest frame has a relatively narrow distribution compared to the energy of the
photon ($E_\gamma$) in the $\XtoDDbG$ rest frame. 
If the
decay $\XtoDDbG$ was dominated by a PHSP-like amplitude (such as the case where
$\XtoDDbP$ and $\XtoDDbG$ decay off-shell), then separating $\X \to \DDbP$
(where $\DDbSys$ has $C=+1$) and $\XtoDDbG$ (where $\DDbSys$ has $C=-1$) could
be straightforward 
if an experiment has 
adequate detector resolution.


However, the decays $\XtoDDbP$ and $\XtoDDbG$ are likely
dominated by $\X \to \Dz\Dstzb$ and $\X \to \Dstz\Dzb$ amplitudes. 
Note that
with the above parameters for particle masses, which are in agreement with the most precise
measurements~\cite{Aaij:2020qga},
the $\X$ and neutral $\Dst$ rest
frames effectively coincide. 

Thus
in the $\X$ rest frame the $\piz$ and $\gamma$ energy and momentum will 
be observed to be nearly the same as 
their breakup energy and
momenta from $\Dst$ decays, 
listed in \cref{table:breakupTable}.
%
%
%
\begin{table}[t]
\begin{center}
\begin{tabular}{c|c|c} 
\hline\hline
Decay & $E_{\piz/\gamma}$~(\MeVE) & $\vert p_{\piz/\gamma} \vert$~(\MeVp) \\
\hline\hline
$\Dstzb \to \Dzb\piz$ & 141.5 & ~\,42.6  \\
$\Dstzb \to \Dzb\gamma$ & 137.0 & 137.0 \\
\end{tabular} 
\caption{$\Dstzb$ breakup energy and momenta, resulting in the 
monochromatic
$\piz$ or $\gamma$ in $\X \to \Dz\Dstzb$ decays.}\label{table:breakupTable}
\end{center}
\end{table}
It is possible to
simulate the $\piz$ (or $\gamma$) energy distribution without reconstructing it.
In the $\X$ rest frame we can define the
variable:
\begin{equation}
\Delta{E}
= m_\X c^2 - E_{\DDbSys},
\end{equation}
where $E_{\DDbSys} = E_{\Dz} + E_{\Dzb}$.
Since the decay $\XtoDDbP$ occurs so close to threshold, the
invariant mass-squared variable $m_{\DDbSys}c^2$ produces 
a near-identical distribution to $E_{\DDbSys}$.
This is because:
\begin{equation}
m_{\DDbSys}^2 c^4 = E_{\DDbSys}^2 - \vec{p}_{\DDbSys}^{\,\,2} c^2,
\end{equation}
where $\vec{p}_{\DDbSys} = \vec{p}_{\Dz} + \vec{p}_{\Dzb}$. 
In the 
$\XtoDDbP$ rest frame, 
$\vec{p}_{\DDbSys}^{\,\,2}c^2$ is negligible compared to $E_{\DDbSys}^2$, so
$m_{\DDbSys}^2 c^4$ is almost identical. For $\XtoDDbG$ decays, 
the $\gamma$
momentum is still small but no longer negligible, and 
$m_{\DDbSys}c^2$ shifts higher from $E_{\DDbSys}$ for these decays by
$\sim$$2.5$ MeV.
This variable has the advantage of being Lorentz-invariant, and relies 
only on information from experimental tracking.

Thus, one could use the invariant mass difference:
\begin{equation}
\Delta{m} = m_\X - m_{\DDbSys},
\end{equation}
as a discriminating variable similar to $\Delta{E}$.  For PHSP decays, the $\Delta m$ distributions will
allow for a similar separation between $\XtoDDbP$ and $\XtoDDbG$ decays as 
$E_\piz / E_\gamma$.
%
However in the case where $\Dz\Dstzb$ is an intermediate state,
$\Delta m$ alone would be
insufficient to separate $\XtoDDbP$ and $\XtoDDbG$ decays if detector
resolution on $\Delta{m}$ is much greater than 1~\MeVm.
\FloatBarrier\subsection{\texorpdfstring{Invariant $\DmDb$}{Invariant 
(D0 - anti-D0)} mass-squared}
\label{subsec:DmDb_m2}

Fortunately, it is possible to determine other variables that
have better separation power, if a high-resolution measurement of the light
neutral particle's momentum is not available, or the light
neutral particle's momentum cannot be inferred by an
external constraint (e.g. $e^+ e^- \to \Upsilon(4S) \to \Bp\Bm$, 
where $\Bm$ decays to a fully reconstructed state, and $\BptoXKp$ is
reconstructed without the light neutral).

In $\XtoDDbP$ and $\XtoDDbG$ decays, the $\DDbSys$ momentum (|$\vec{p}_\Dz +
\vec{p}_\Dzb$|) in the $\X$ decay rest frame has the
property that it should be equal and opposite to the momentum of the light neutral particle 
in the $\X$ rest frame. However, boosting the $\DDbSys$ momentum into the 
available $\DDbSys$ rest frame 
removes access to this
information. 

Here it is proposed to instead examine |$\vec{p}_\Dz - \vec{p}_\Dzb$|. Using
momentum variables from the $\X$ frame, we know that:
\begin{eqnarray}
|\vec{p}_\Dz + \vec{p}_\Dzb| &=&  \sqrt{ |\vec{p}_\Dz|^2 + |\vec{p}_\Dzb|^2 +
2(\vec{p}_\Dz\cdot\vec{p}_\Dzb)} = p_{\rm \piz/\gamma},
 \nonumber \\ 
|\vec{p}_\Dz - \vec{p}_\Dzb| &=&  \sqrt{ |\vec{p}_\Dz|^2 + |\vec{p}_\Dzb|^2 -
2(\vec{p}_\Dz\cdot\vec{p}_\Dzb) } ,
\label{dotrule}
\end{eqnarray}

and thus
\begin{equation}
|\vec{p}_\Dz - \vec{p}_\Dzb| =   \sqrt{ 
p_{\rm \piz/\gamma}^2-4 (\vec{p}_\Dz
\cdot \vec{p}_\Dzb) } .
\label{parallelogram2}
\end{equation}
Effectively, this is the light neutral momentum ($p_{\rm \piz/\gamma}$) 
smeared 
by the collinearity and similarity in magnitude of the $\Dz$ and $\Dzb$
momenta.
Recall that in $\X \to \Dst\D$ decay, the $\D$ meson not associated with a
$\Dst$ is near rest in the $\X$ rest frame, in which case $|\vec{p}_\Dz -
\vec{p}_\Dzb| \simeq  p_{\rm \piz/\gamma }$.
As shown in \cref{table:breakupTable}, the light neutral particle's
momentum in the $\X$ rest frame can be used to separate $\DDbP$ and $\DDbG$
decays, but using $|\vec{p}_\Dz -
\vec{p}_\Dzb|$ we obtain it without using any information from the light
neutral particle.

As we conventionally interchange the sum of the $\Dz$ and $\Dzb$ four-momenta 
with the concept of ``$\DDbSys$ four-momentum'', 
hereafter interchanged is the difference of the $\Dz$ and $\Dzb$
four-momenta with the concept of ``$\DmDb$ four-momentum'' (and the
implied corresponding three-momentum when relevant). 
For $\X$ decays the $\DmDb$ momentum
is preserved to an excellent approximation when it is boosted into the $\DDbSys$
rest frame (see~\cref{sec:DmDbPreservation}).

Invariant mass-squared for four-momenta is always defined in terms of the energy
and momentum components of that four-momenta in any chosen frame. We can
choose to write the $\DmDb$ invariant mass-squared using the $\DmDb$ four-momentum in the
$\DDbSys$ frame:
\begin{equation}
{m^2_{\DmDb}} c^4 
= 
{\left(E_\Dz^{\D\Db} - E_\Dzb^{\D\Db}\right)}^2
- {\left({\vec{p}_\Dz}^{\,\D\Db} -
{\vec{p}_\Dzb}^{\,\D\Db}\right)}^2 c^2. 
\label{boosterEnergy-DmDb-2}
\end{equation}

In the $\DDbSys$ frame, the $\Dz$ and $\Dzb$ momenta are by definition equal and
opposite, and since both particles have the same mass, the
$\Dz$ and $\Dzb$ energies are by definition equal.
Thus 
we conclude:
\begin{equation}
{m^2_{\DmDb}} 
= 
- {\left({\vec{p}_\Dz}^{\,\D\Db} -
{\vec{p}_\Dzb}^{\,\D\Db}\right)}^2 / c^2 = -
{\left(2\,p_\Dz^{\,\D\Db}/c\right)}^2.
\label{boosterEnergy-DmDb-3}
\end{equation}
\Cref{boosterEnergy-DmDb-3} of course is generally true for any
two-particle system where the masses of the particles are identical ---
for example there are applications of a similar variable when studying spin
correlations of $\Lambda$ hyperon pairs~\cite{Cheng:2265715}.

Applying the conclusion from 
\cref{sec:DmDbPreservation} 
that the $(\Dz -
\Dzb)$ momenta in the $\X$ rest frame and the $\DDbSys$ rest frame 
are essentially the same, under the condition that the $\Dz$ and
$\Dzb$ momenta in the $\X$ rest frame are small: 
\begin{equation}
-{m^2_{\DmDb}} 
\simeq 
 {\left|\vec{p}_\Dz -
\vec{p}_\Dzb \right|}^2/c^2 \simeq 
{\left(p_{\rm \piz/\gamma }\,/c\right)}^2.
\label{equalsEnergy-DmDb}
\end{equation}

The quantity $-m^2_{\DmDb}$ has the advantage of being Lorentz invariant, and
thus can be calculated in the measurement frame. As no mass
constraints are applied in this study, the quantity $-m^2_{\DmDb}$ may be
negative for some events when calculated from the measurement frame
four-vectors.

\begin{figure}[th]
\flushright
\includegraphics[page=6,scale=0.35]{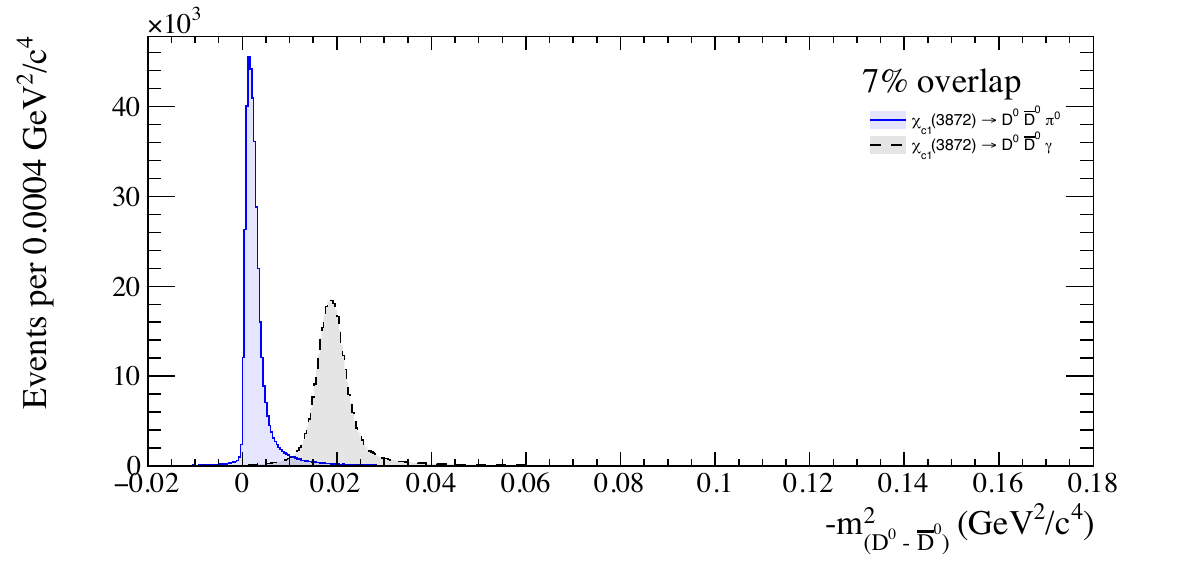}(a)
\includegraphics[page=6,scale=0.35]{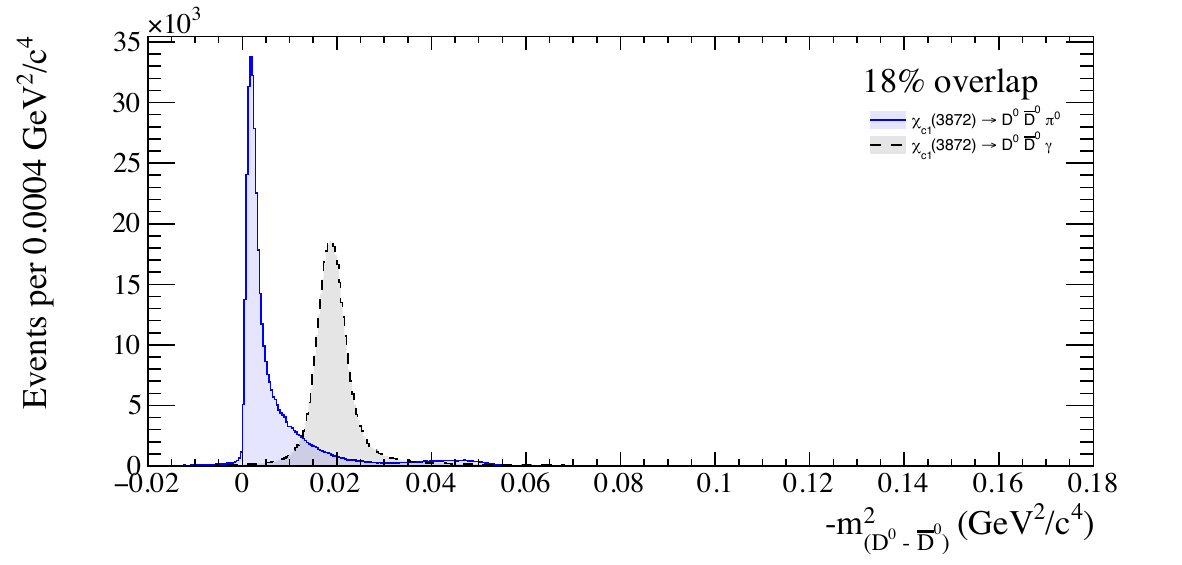}(b)
\includegraphics[page=6,scale=0.35]{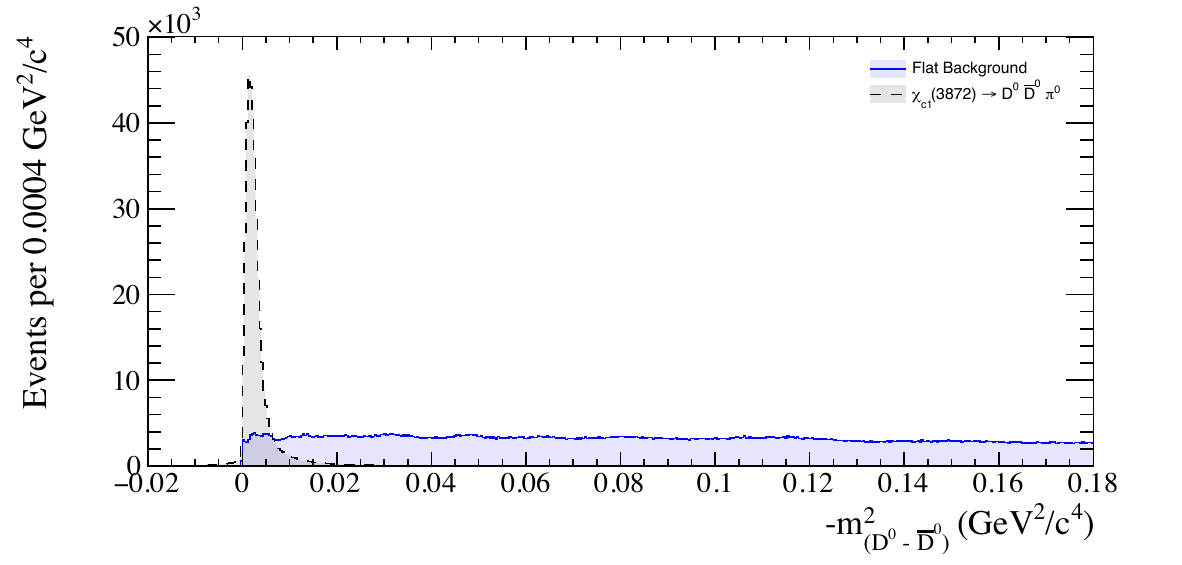}(c)
\includegraphics[page=6,scale=0.35]{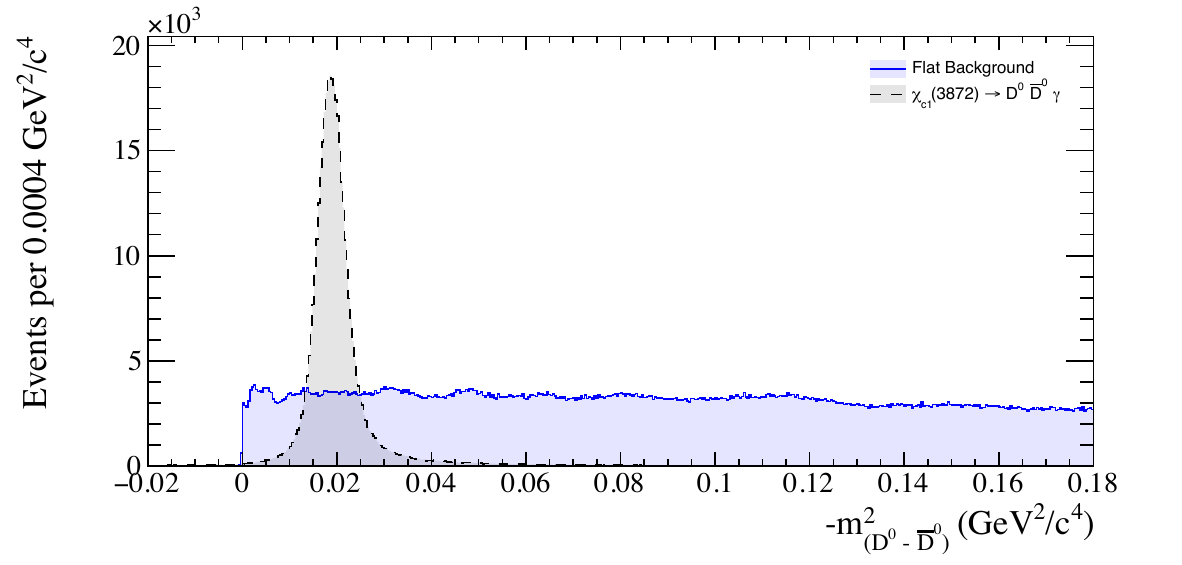}(d)
\includegraphics[page=6,scale=0.35]{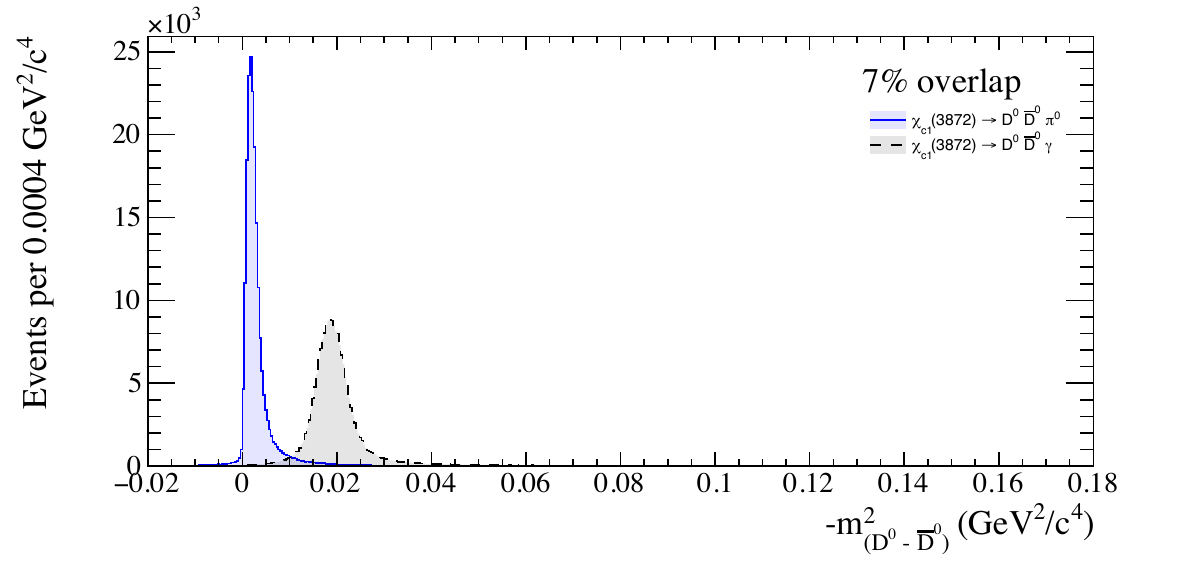}(e)
\includegraphics[page=6,scale=0.35]{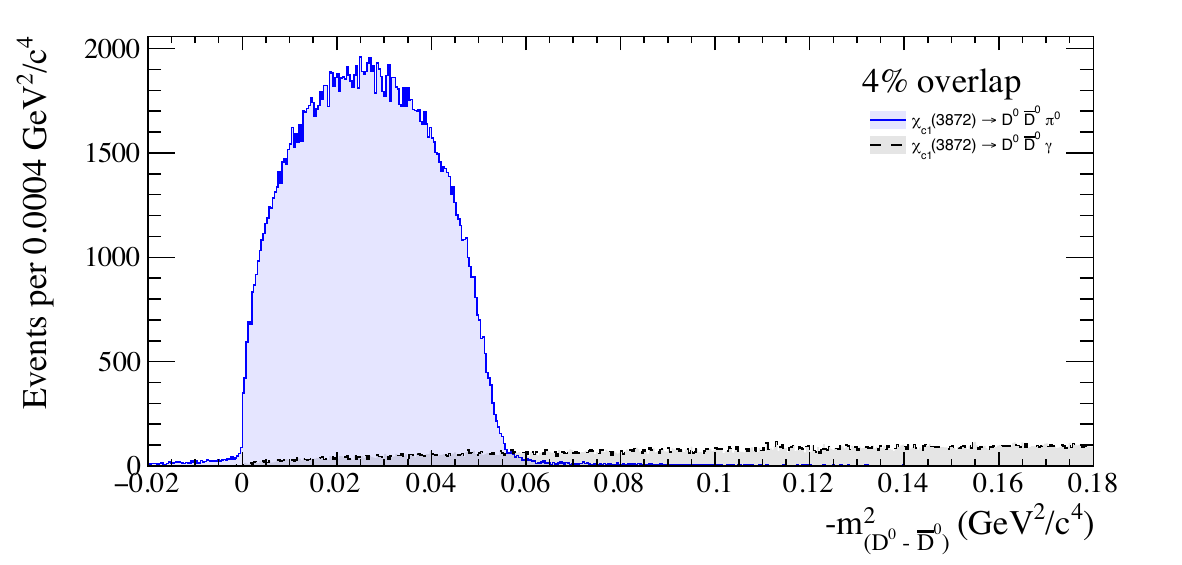}(f)
\caption{
$-m^2_{\DmDb}$
compared for prompt 
(a) $\XtoDDbP$ 50/50 (light blue) vs. $\XtoDDbG$ 50/50 (grey, dashed line), 
(b) $\XtoDDbP$ Model (light blue) vs.  $\XtoDDbG$ 50/50 (grey, dashed line), 
(c) $\DDbSys$ Flat Background Sample (light blue) vs. $\XtoDDbP$ 50/50 (grey,
dashed line), (d) $\DDbSys$ Flat Background Sample (light blue) vs. $\XtoDDbG$
50/50 (grey, dashed line), (e) $\XtoDDbP$ 50/50 S-Wave only (light blue) vs. $\XtoDDbG$ 50/50 S-Wave
only (grey, dashed line), and (f) $\XtoDDbP$ PHSP (light blue) vs $\XtoDDbG$ PHSP (grey, dashed line). 
Where the sample sizes are 
equal, the fraction of $\XtoDDbP$ (or equivalently $\XtoDDbG$) events in the
overlap region is displayed on each plot. 
\label{fig:compare-minus-DmDb-m2-DDbFrame}}
\end{figure}

\Cref{fig:compare-minus-DmDb-m2-DDbFrame}  shows comparisons of 
$-m^2_{\DmDb}$, in simulated 
prompt events (corresponding to final-state 
kinematics with \lhcb detector effects applied, as produced by {\sc RapidSim}),
for both $\XtoDDbP$ and $\XtoDDbG$ decays; plots of this variable for secondary 
decays are qualitatively identical. 
Note that in this simulation there is clear
separation between $\XtoDDbP$ and $\XtoDDbG$ decays in this
variable.

\FloatBarrier\subsection{Angle between \texorpdfstring{$K$}{K} and
\texorpdfstring{$\Dz$}{D0} in the
\texorpdfstring{$\DDbSys$ rest frame}{D0 anti-D0 rest frame}}
\label{subsec:angleDDb}

In the case of secondary $\X$ decay, e.g. $\BtoXK$ decay, where 
$\X \to \Dstz\Dzb$, 
it is possible to take advantage of additional kinematics.
A particularly interesting variable is ${\theta_K}^{\X}$, 
the angle between the
$K$ meson and the $\Dz$ meson which decays from the $\Dstz$ state, in the $\X$
decay rest frame\footnote{This is not the traditional helicity angle, which is
the angle between the $K$ meson and the $\Dzb$ meson, in the $\X$ decay rest
frame.}.
Of course, 
the $\X$ decays occur in a superposition of $\X \to \Dstz\Dzb$ and
$\X \to \Dstzb\Dz$. For the latter, the interesting variable would be
${\overline{\theta}_K}^{\X}$, the angle between the $K$ meson and the $\Dzb$
meson, in the $\X$ decay rest frame.

While it is not generally possible to get into the $\X$ rest frame without
reconstructing the light neutral, it is possible to make an acceptable
approximation to the $\X$ rest frame by considering the $\DDbSys$ rest
frame.
The angle ${\theta_K}^{\DDbSys}$ in the $\DDbSys$ rest frame is
correlated to ${\theta_K}^{\X}$, and thus provides separation power between
$\XtoDDbP$ and $\XtoDDbG$ where $\X \to \Dst\D$. There is a convenient
symmetry, as ${\overline{\theta}_K}^{\X}$ will have the same correlation 
to ${\theta_K}^{\DDbSys}$ due to the decay topology. 

\begin{figure}[t]
\flushleft
\includegraphics[page=413,scale=0.169]{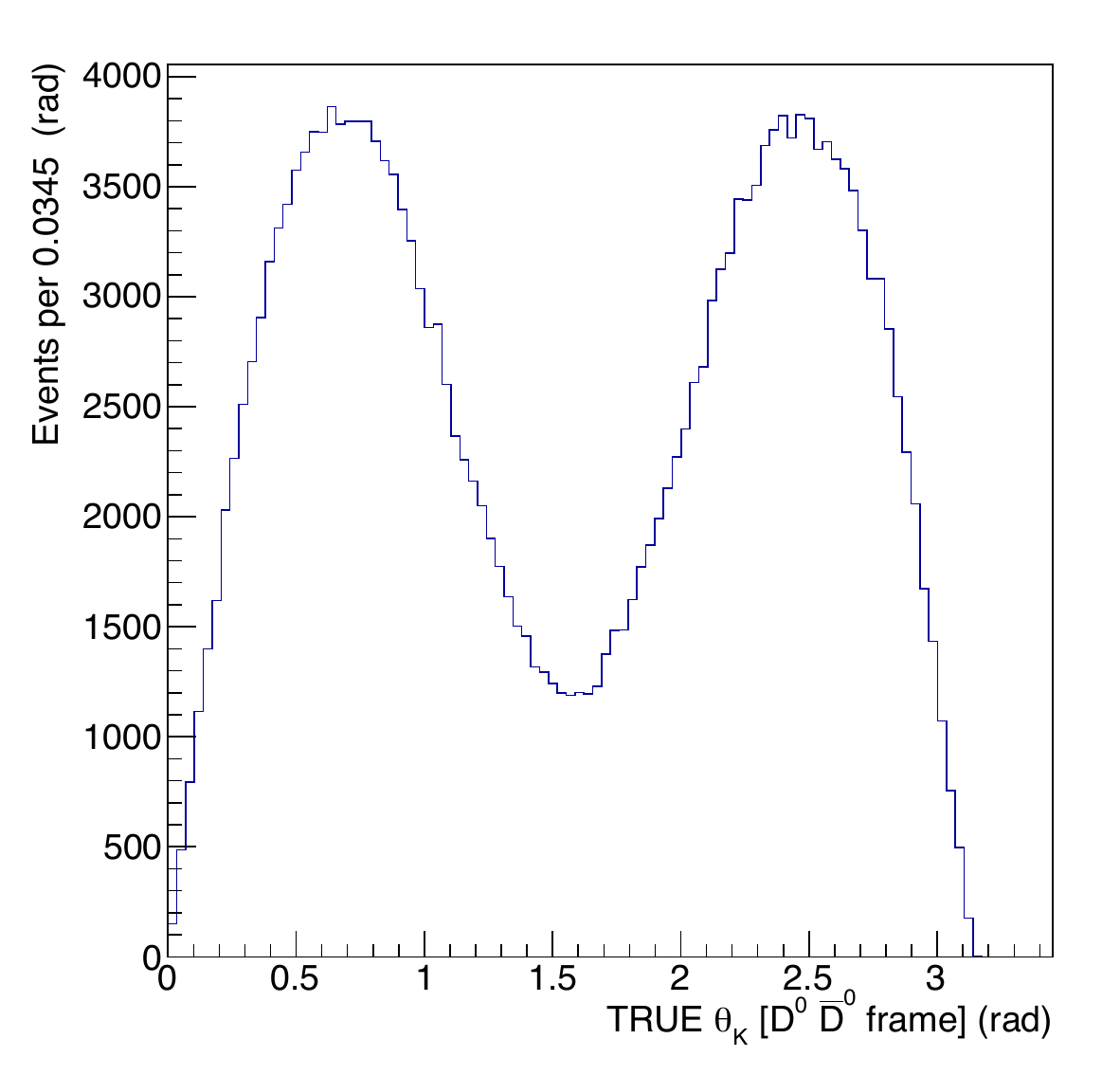}
\includegraphics[page=409,scale=0.169]{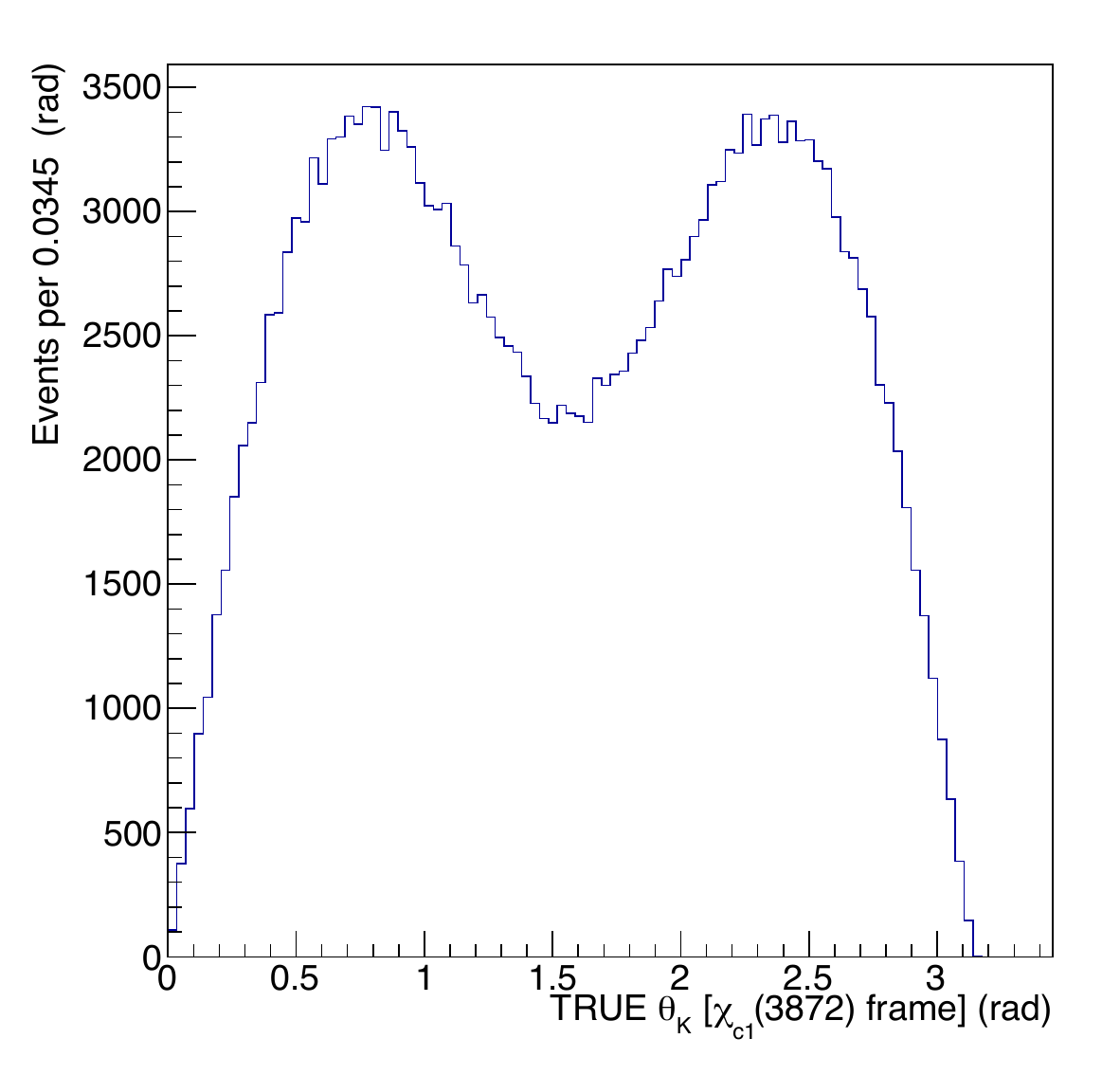}
(a)
\includegraphics[page=327,scale=0.169]{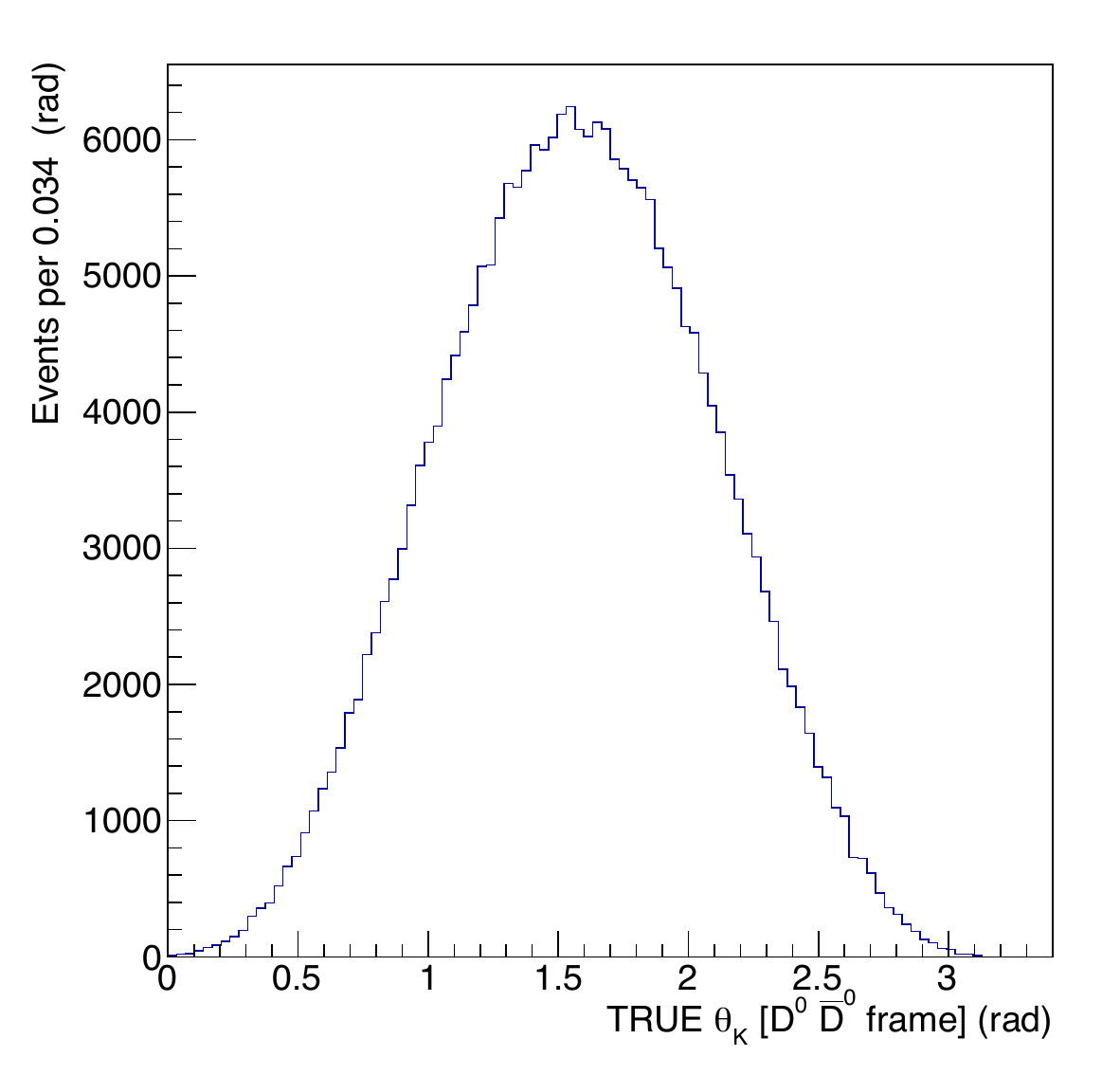}
\includegraphics[page=323,scale=0.169]{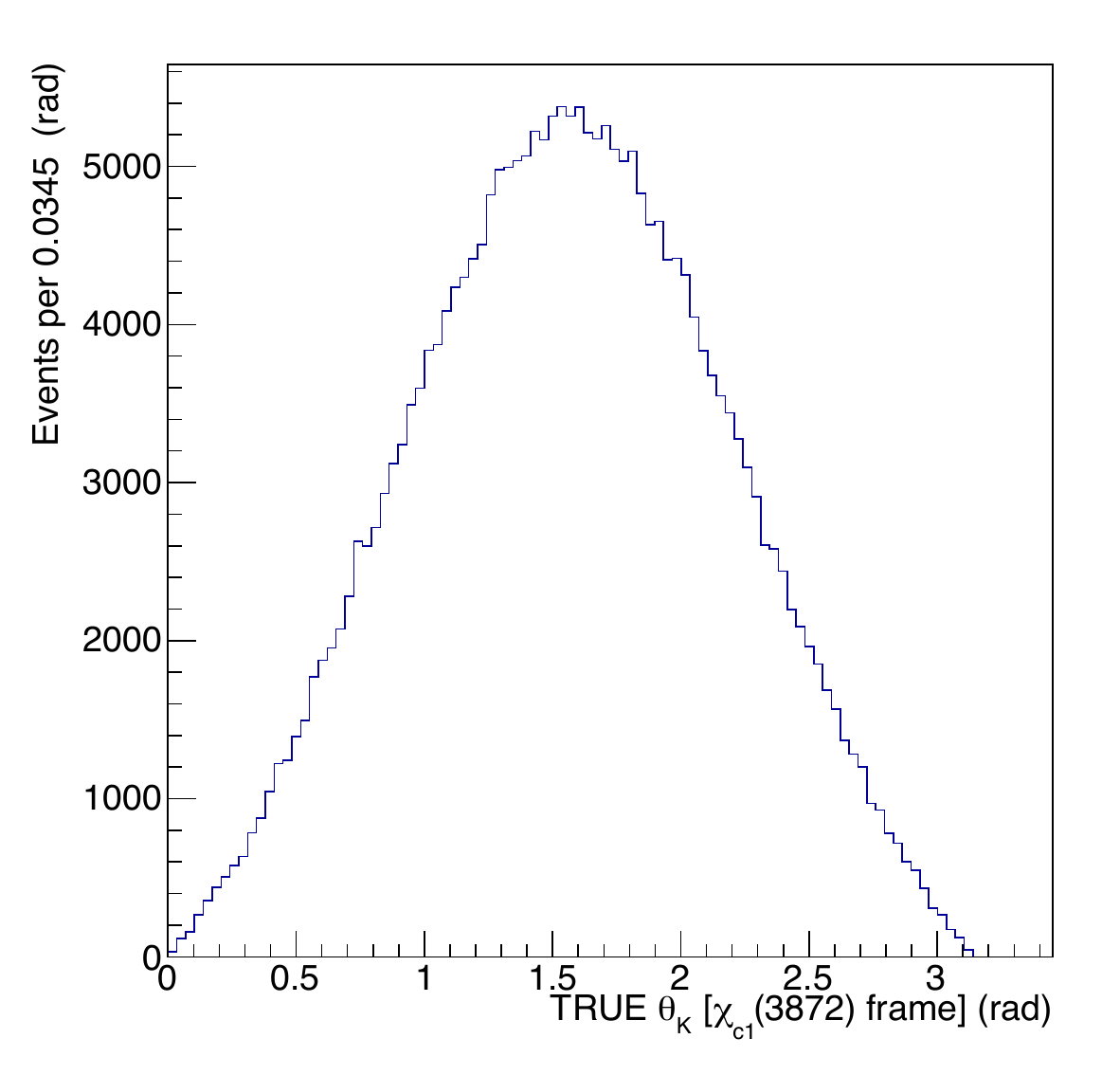}
(b)
\includegraphics[page=413,scale=0.169]{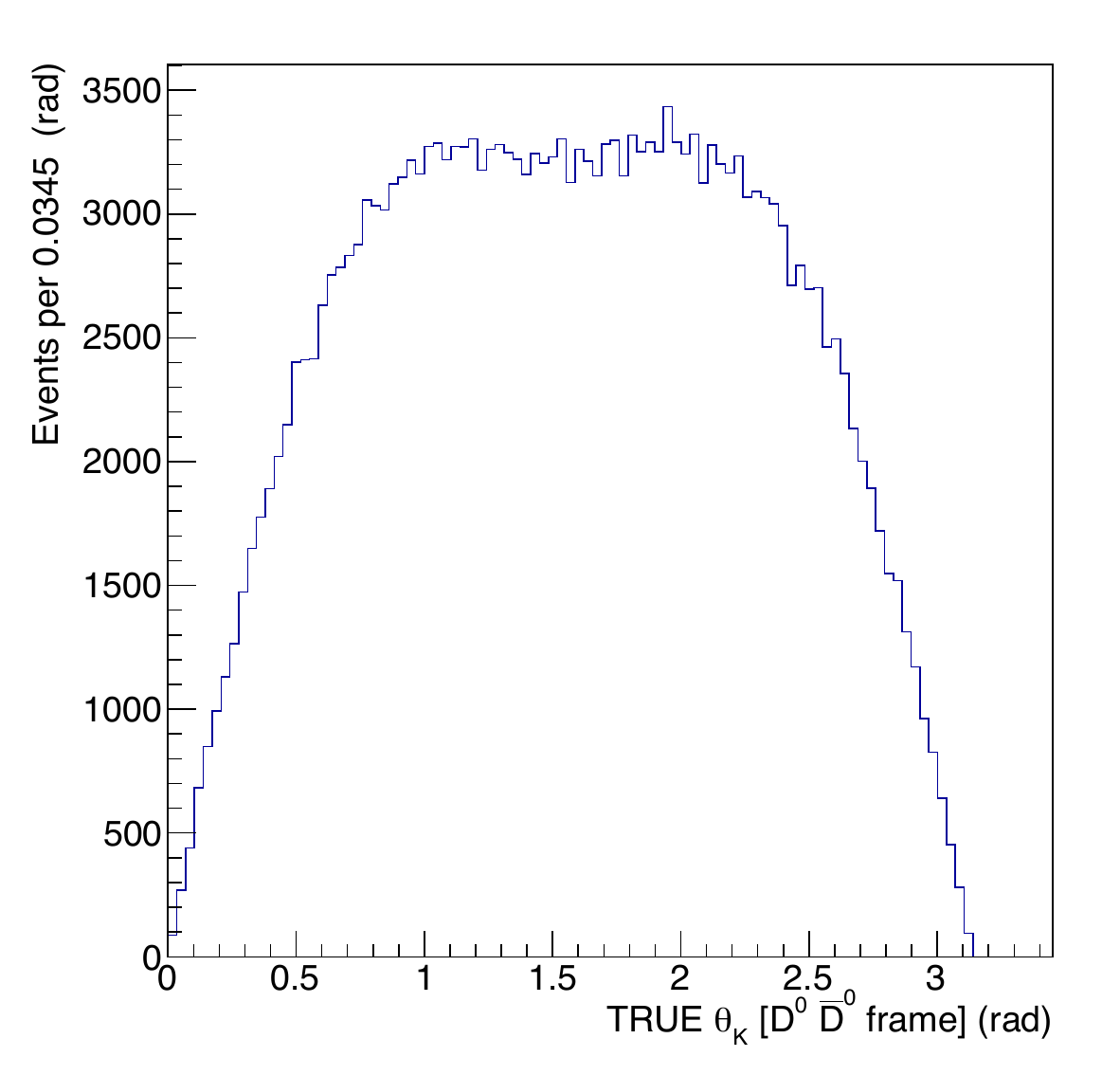}
\includegraphics[page=409,scale=0.169]{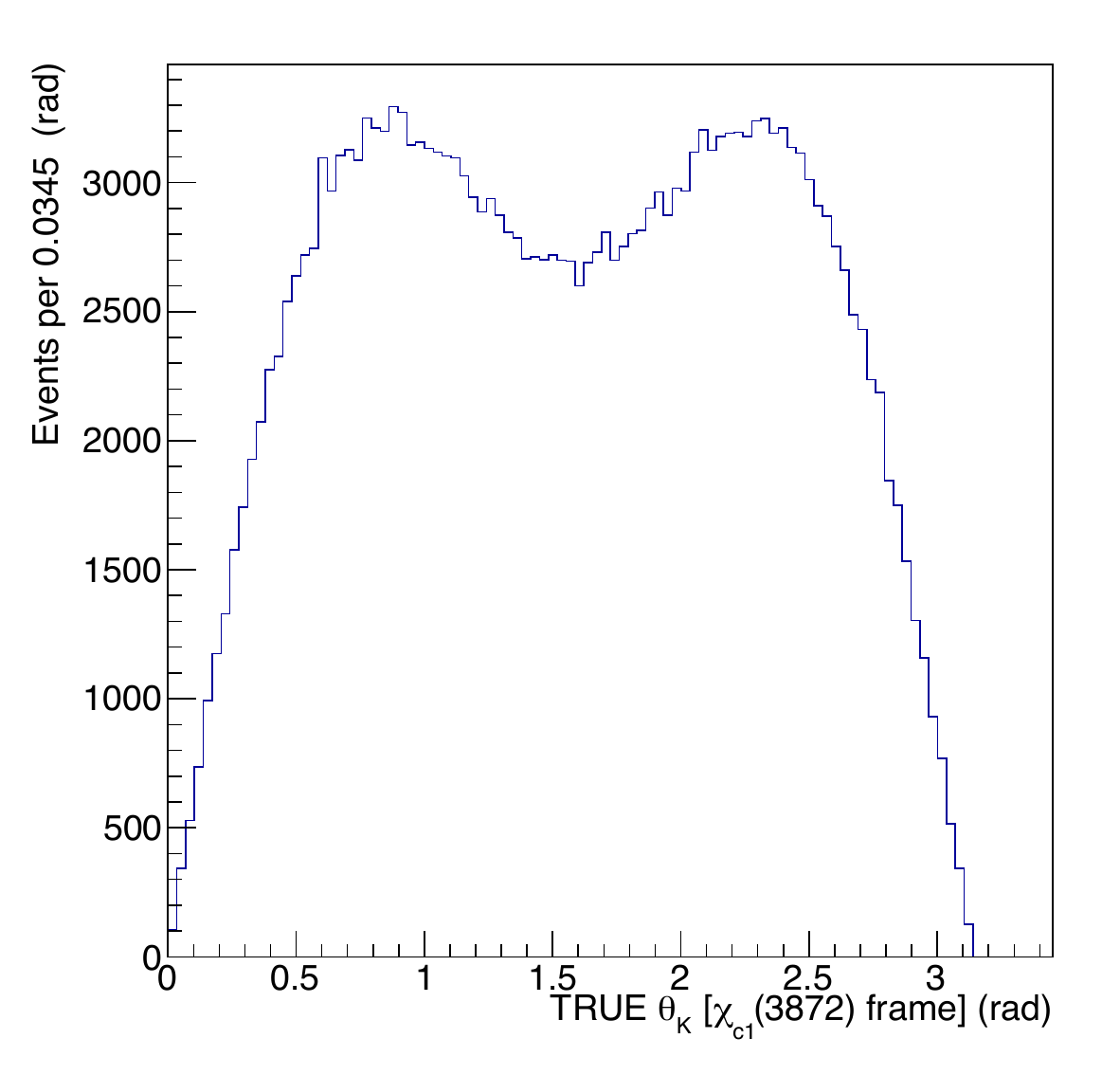}
(c)
\includegraphics[page=327,scale=0.169]{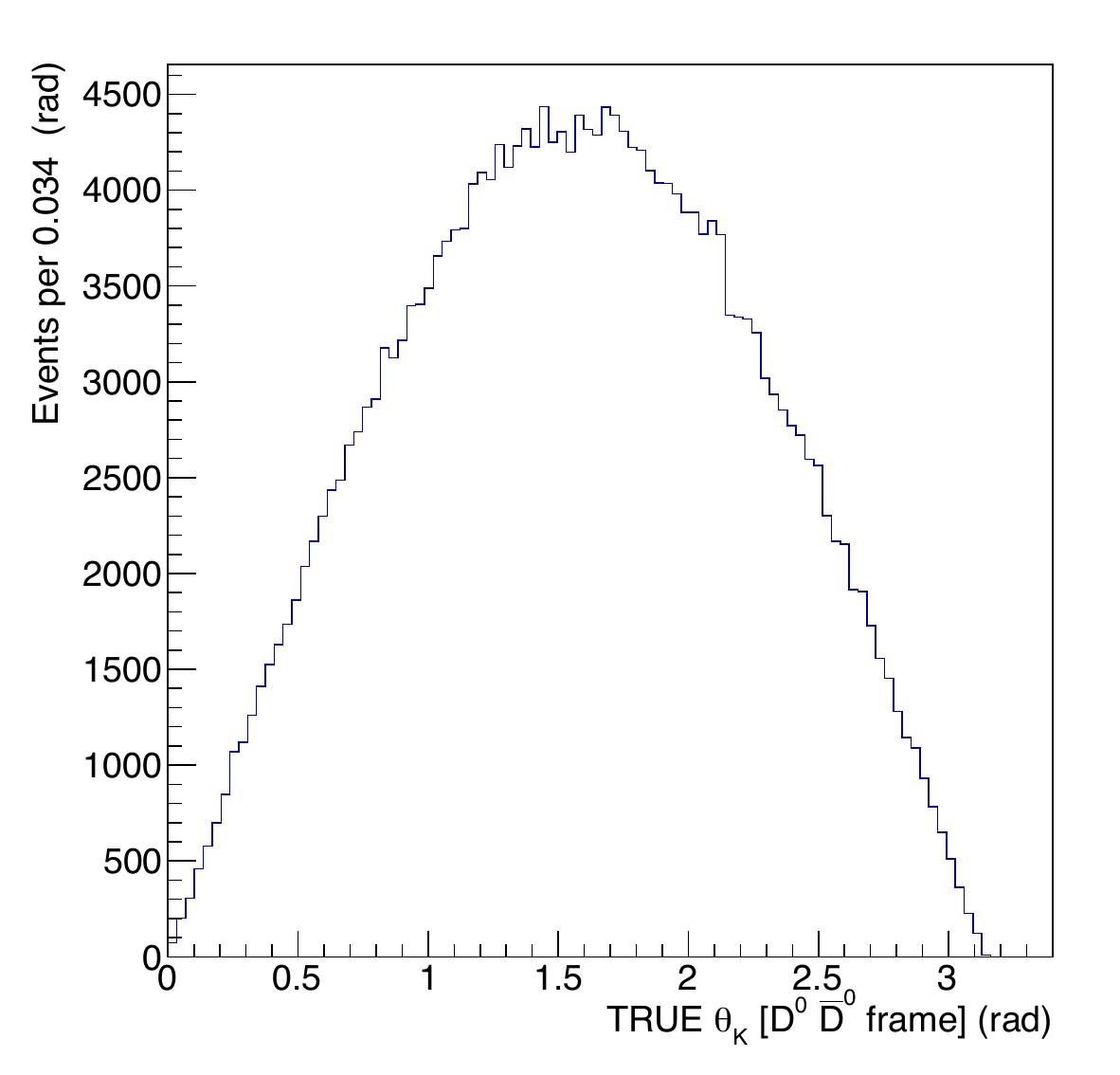}
\includegraphics[page=323,scale=0.169]{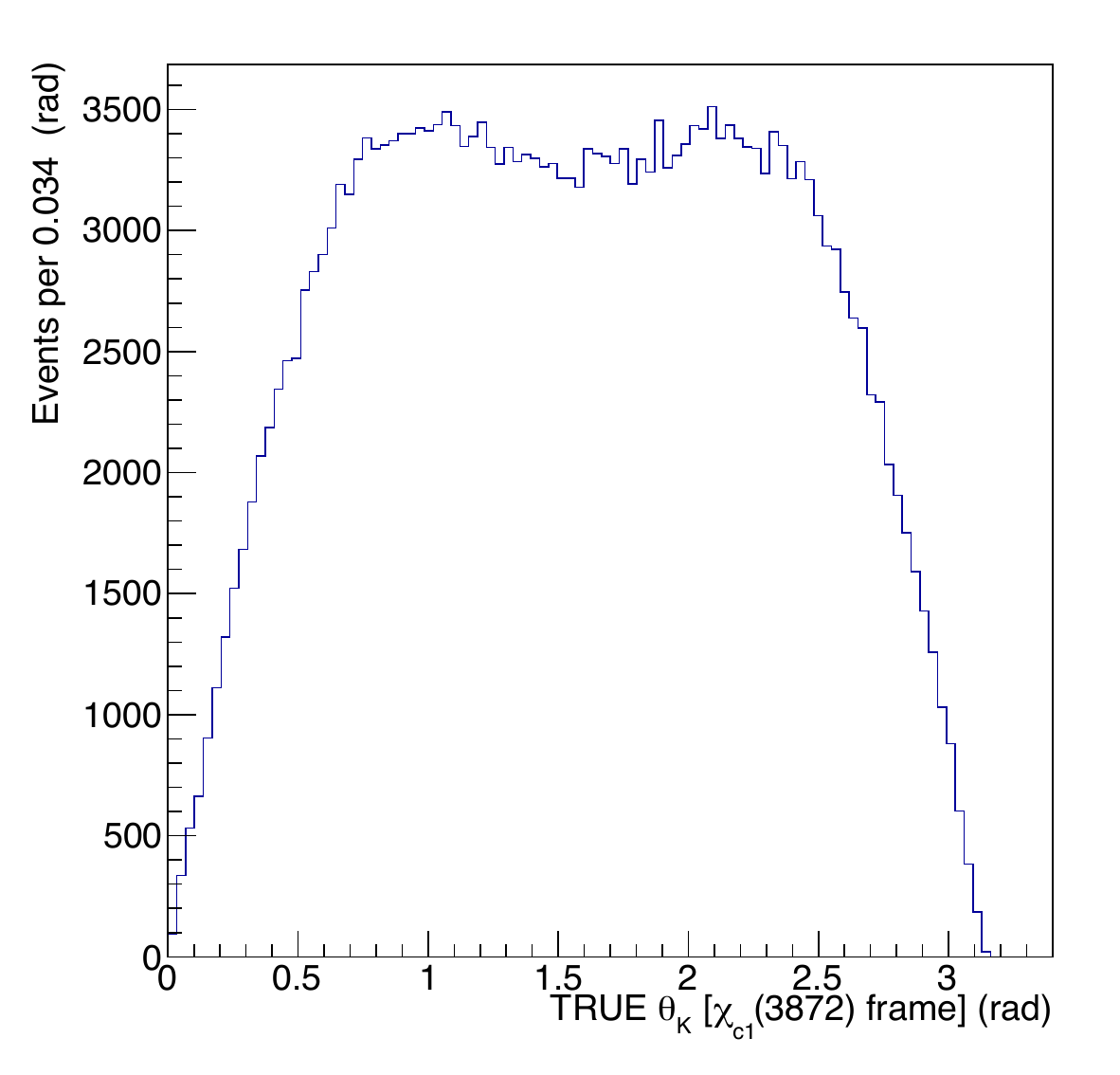}
(d)
\includegraphics[page=647,scale=0.169]{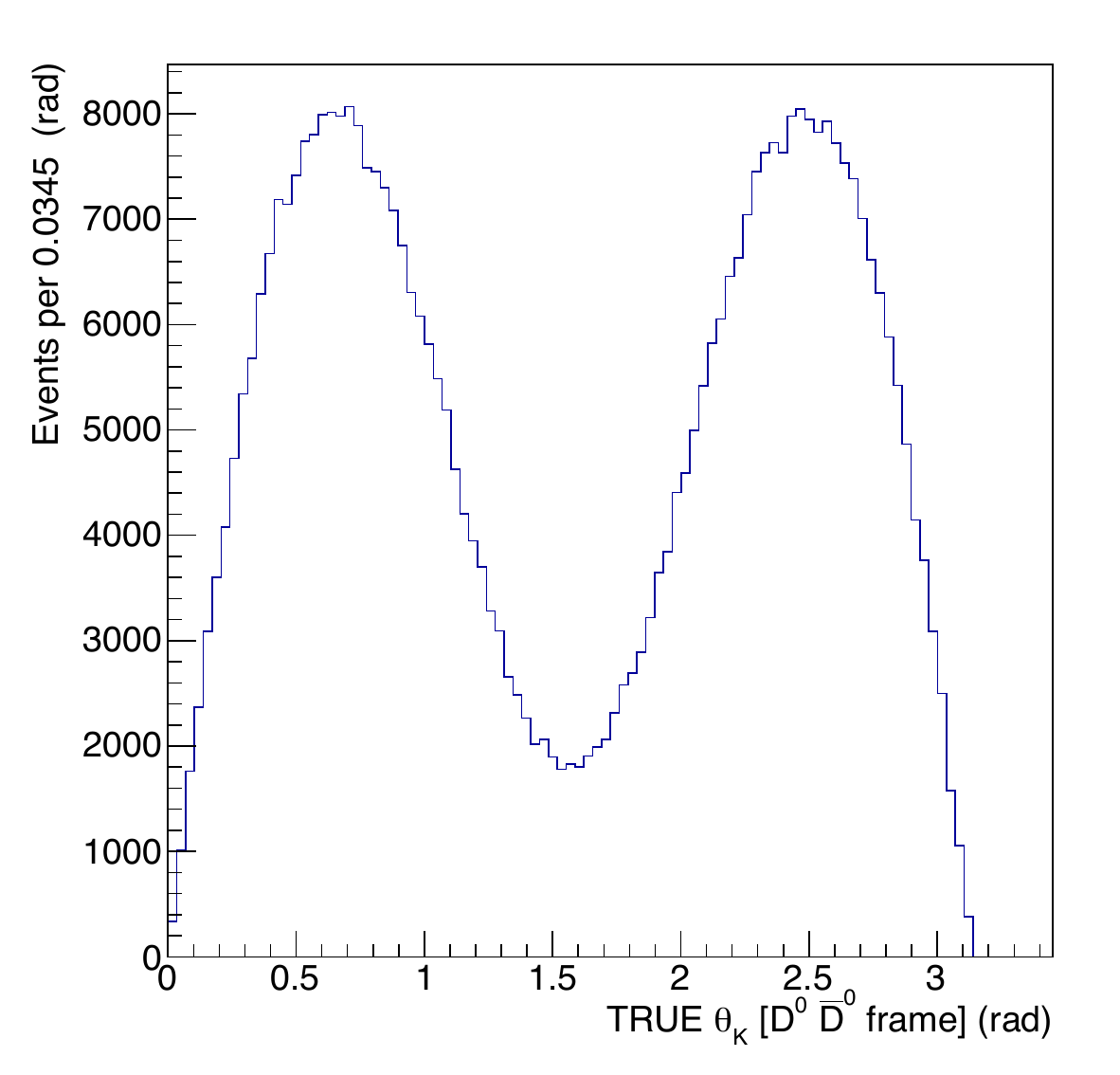}
\includegraphics[page=643,scale=0.169]{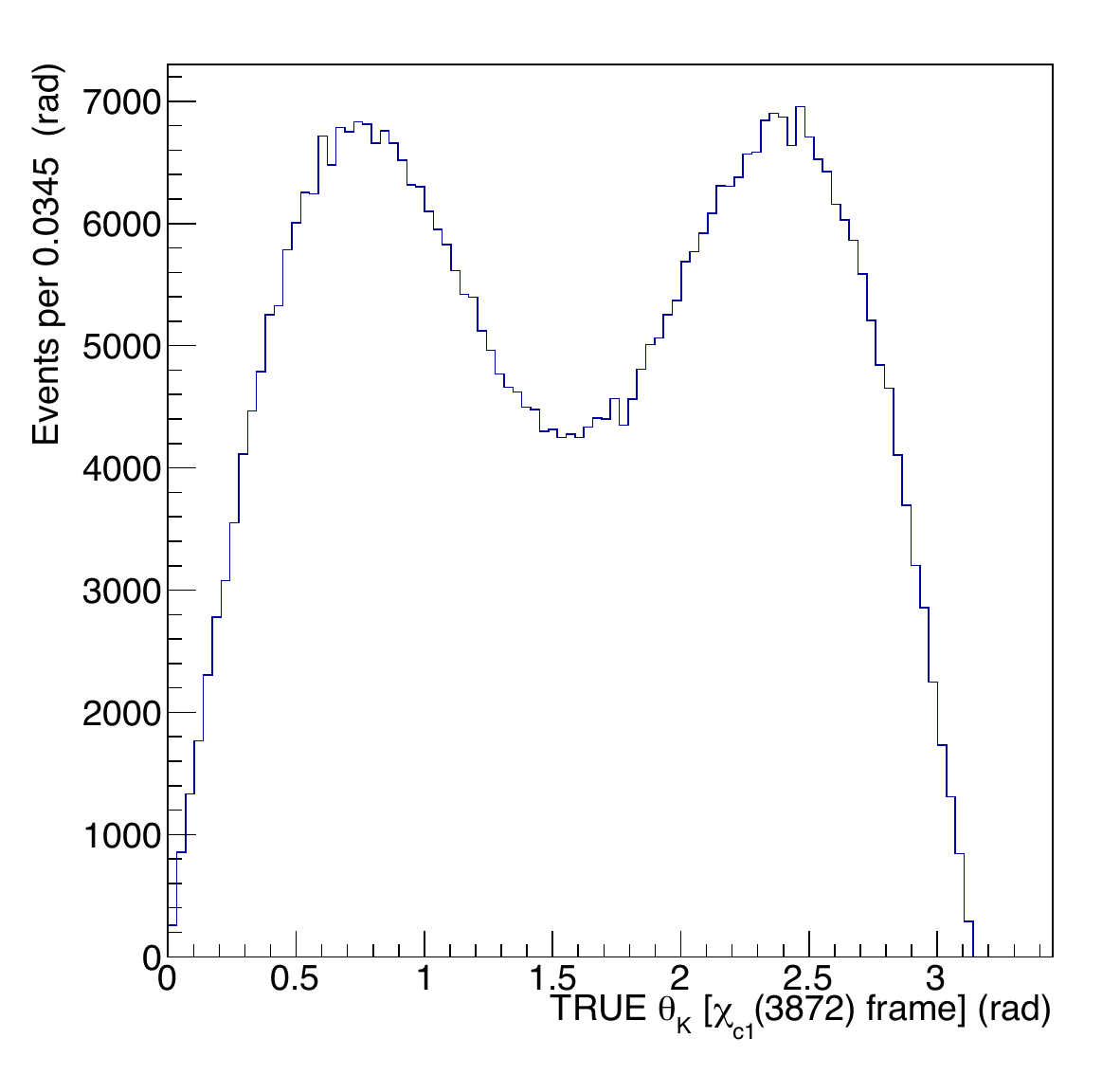}
\centering(e)
\caption{
TRUE ${\theta_K}^{\DDbSys}$ (left) and TRUE
${\theta_K}^{\X}$ (right), shown for secondary (a) $\XtoDDbP$ 50/50 S-Wave
only (correlation = 0.65), (b) $\XtoDDbG$ 50/50 S-Wave only (correlation = 0.51),
(c) $\XtoDDbP$ 50/50 D-Wave only (correlation = 0.65), 
(d) $\XtoDDbG$ 50/50 D-Wave only (correlation = 0.52), and
(e) $\XtoDDbP$ Model (correlation = 0.78)
\label{fig:thetaK_1D}}
\end{figure}

\Cref{fig:thetaK_1D} 
shows comparisons
between the TRUE ${\theta_K}^{\X}$ and the TRUE ${\theta_K}^{\DDbSys}$ for simulated
events, with correlation factors between the variables shown in the figure
caption.
Note that the separation is better for the ${\theta_K}^{\DDbSys}$ variable, as it has access
to the separation power of ${\theta_K}^{\X}$ and
${\overline{\theta}_K}^{\X}$ simultaneously.

\begin{figure}[t]
\flushright
\includegraphics[page=8,scale=0.35]{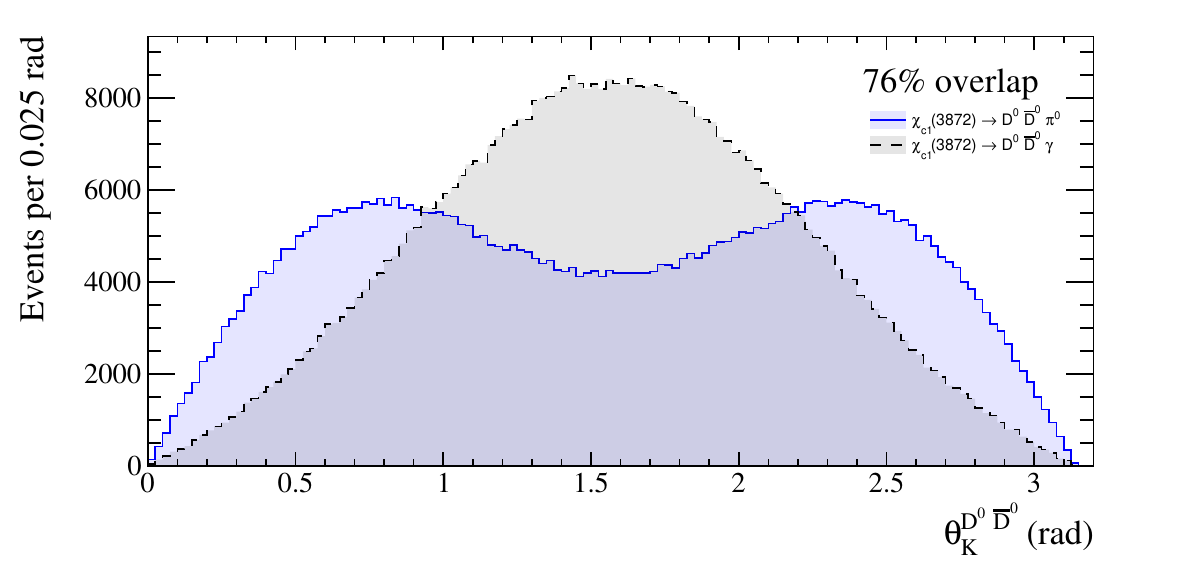}(a)
\includegraphics[page=8,scale=0.35]{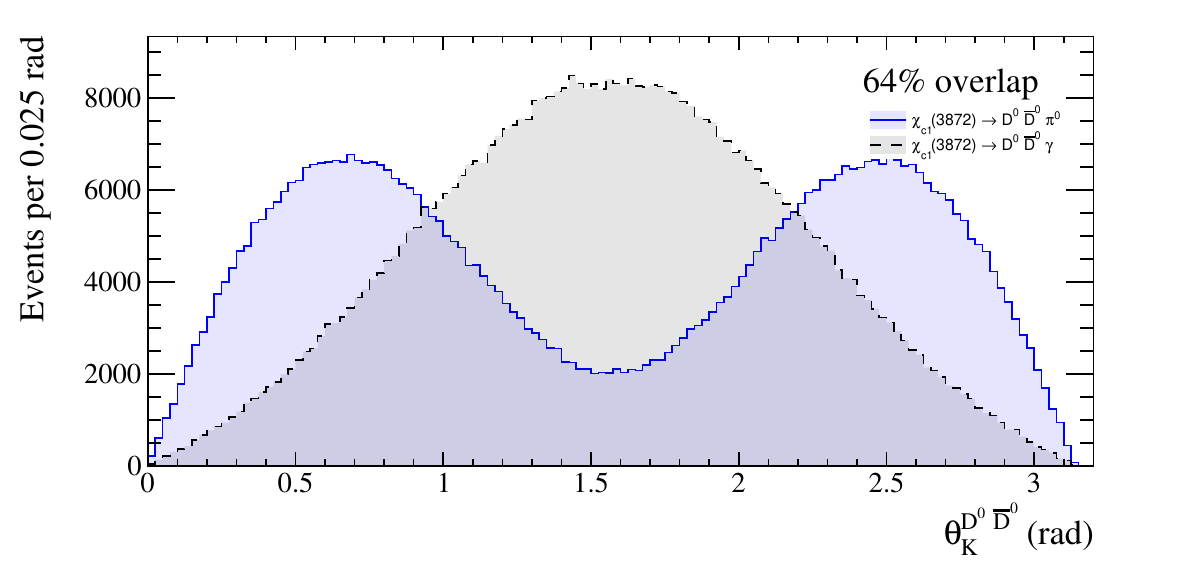}(b)
\includegraphics[page=8,scale=0.35]{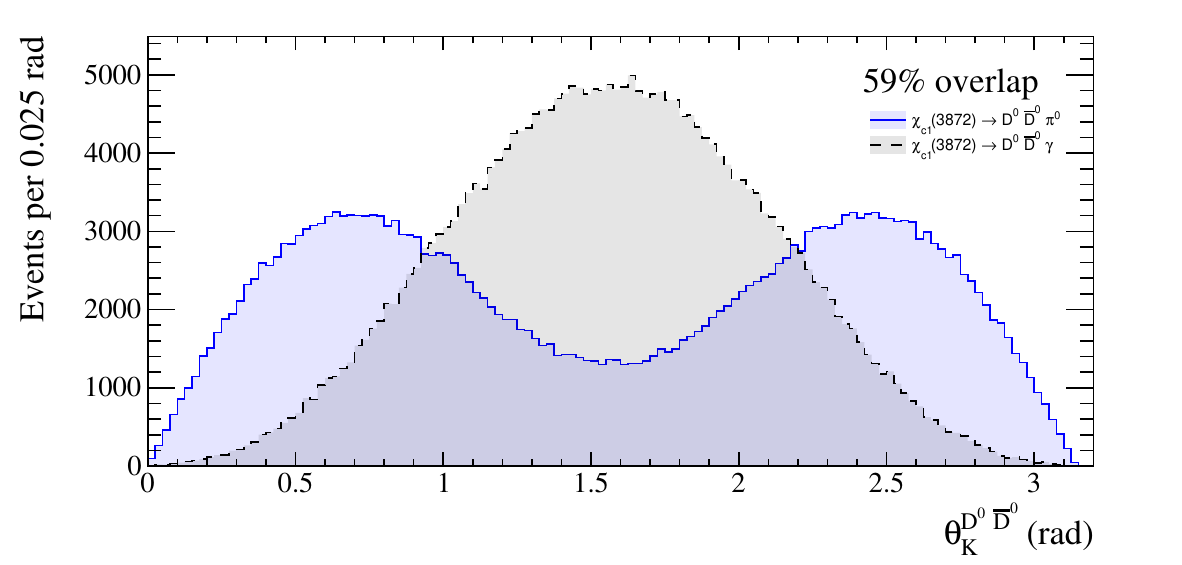}(c)
\includegraphics[page=8,scale=0.35]{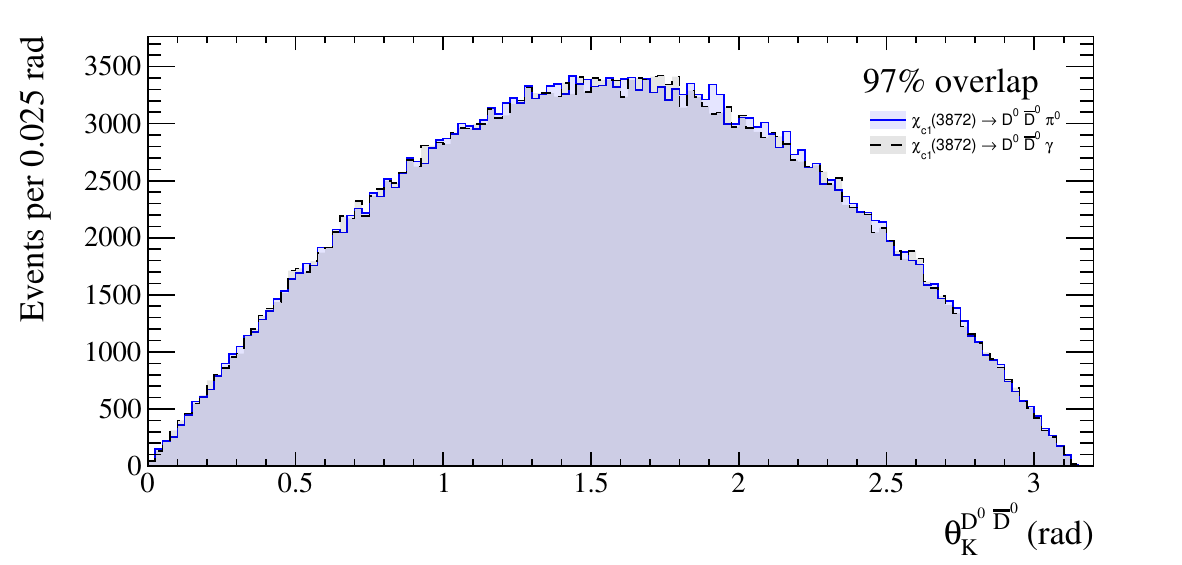}(d)
\caption{
${\theta_K}^{\DDbSys}$, compared for
secondary 
(a) $\XtoDDbP$ 50/50 (light blue) vs. $\XtoDDbG$ 50/50 (grey, dashed line), 
(b) $\XtoDDbP$ Model (light blue) vs.  $\XtoDDbG$ 50/50 (grey, dashed line), 
(c) $\XtoDDbP$ 50/50 S-Wave only (light blue) vs. $\XtoDDbG$ 50/50 S-Wave
only (grey, dashed line), and (d) $\XtoDDbP$ PHSP (light blue) vs $\XtoDDbG$ PHSP (grey, dashed line). 
The fraction of $\XtoDDbP$ (or equivalently $\XtoDDbG$) events in the
overlap region is displayed on each plot.
\label{fig:B-compare-thetaK}}
\end{figure}

\Cref{fig:B-compare-thetaK} 
shows comparisons of 
${\theta_K}^{\DDbSys}$ 
for both $\XtoDDbP$ and $\XtoDDbG$ decays, in simulated 
secondary
events. Unlike in the previous comparisons shown, here the $\XtoDDbP$ Model
decays separate better than $\XtoDDbP$ 50/50 decays, from $\XtoDDbG$ 50/50
decays.

\FloatBarrier\subsection{Summary 
}
\label{subsec:separationSummary}

The level of separations demonstrated in the {\sc RapidSim} simulations should
be representative, but  {\sc RapidSim}  is not
a full \lhcb detector simulation. Thus in 
data resolutions should not
be expected to be as good as shown here. 
Mass, vertex, and other constraints on the $\X$ exotic meson and $D$ (and $B$,
where applicable) meson decays may
modify the level of separation. Also, real backgrounds are 
more complex
than those presented here, and of course real 
$\XtoDDbP$ and $\XtoDDbG$ decays may not be perfectly described by the
decay models used to generate decays for this study.

Nevertheless, 
the
variables
presented 
here have strong 
potential
to help distinguish between
$\XtoDDbP$ and $\XtoDDbG$ decays, in experiments without sufficient calorimeter
resolution to resolve the light neutral particle (of course, information from
the light neutral itself could only be helpful).
This step is
crucial to performing quantum-correlated analyses with the resulting
$\DDbSys$ systems. Separation of these decays should also be able to help
considerably in other situations where separation is desired --- for example,
the determination of $\XtoDDbP$ and $\XtoDDbG$ branching fractions.

%% file: inputs/4-timeReversal.tex
\FloatBarrier\section{\boldmath Tests of 
\texorpdfstring{$(CP)T$}{(CP)T} violation with 
\texorpdfstring{$C = +1$ $\DDbSys$}{C = +1 D0 anti-D0} 
systems} 
\label{TV}

In addition to the possibility of adding data to existing charm
quantum-correlated analyses, the possibility to also
perform tests of $CP(T)$ and time-reversal ($T$) conservation with  
$C = +1$ $\DDbSys$ systems is proposed here.
A reconstructibility advantage in the required final states gives 
$C = +1$ $\DDbSys$ systems
the potential to provide the first experimental constraints on
time-reversal violation in the charm system. The same reconstructibility
advantage extends to tests of $CPT$ conservation in the charm system --- here 
considerable theoretical work on what may manifest $CPT$ violation in
charm (e.g. SM-extension, quantum-gravity models) and how to study it has been
performed~\cite{Colladay:1995qb,Kostelecky:1994rn,Kostelecky:2001ff,Bernabeu:2003ym,Bernabeu:2006av,Shi:2011aa,Shi:2013lua,Huang:2013iaa,Roberts:2017tmo,Edwards:2019lfb};
experimental constraints on $CPT$ conservation in the charm system have
been obtained~\cite{Link:2002fg}.

$CP$ 
violation has been found to be 
small in the charm system~\cite{Aaij:2019kcg}, but it is unclear if the amount
found is entirely due to Standard Model processes; unexpected deviations from
the expectation of small $CP$ violation could be a sign of new physics. 
The $CPT$ theorem~\cite{Blum:2022eol} requires conservation of the combined $CP$
and $T$ symmetries. 
If a test of $T$ symmetry
is performed 
and $T$ violation is observed to be different
than $CP$ violation in the charm system, it would be a clear indication of
physics beyond the Standard Model; perhaps it would mean that $T$ violation 
in the charm system arises from different mechanisms
than $CP$ violation in the charm system~\cite{Bevan:2013rpr}. 
New physics can enter these decays through quantum loops, 
as they
do in the beauty system. 
It is possible to search for 
$CPT$, 
$CP$, and $T$
violation in 
entangled charm systems, by using a similar method
to that used 
to make the first observation of time-reversal violation in the beauty quark
system~\cite{Lees:2012uka, Bernabeu:2012ab}.

For the charm quark system, this method has been generally discussed in the
context of collisions at potential asymmetric $e^+ e^-$ flavor factories 
operating at
the $\psix$ resonance~\cite{Bevan:2013wwa, Bevan:2013rpr, Bevan:2015ena,
Bevan:2015nra, Rama:2015pmr, Shi:2016bvo}, and exploiting the time evolution of
the $C = -1$ $\DDbSys$ quantum superposition; each $\D$ meson may decay at different times to a flavor-specific or a
$CP$-definite 
filter basis 
(i.e.
final state), which identifies (or ``tags'') the $\D$
meson as having decayed to a specific flavor basis (implying the parent is in a
$\Dz$ or $\Dzb$ flavor state) or being in a superposition of flavor states 
that decay to a definite $CP$ eigenstate 
(the superpositions are denoted as $\DCPp$ or $\DCPm$),
respectively. 

Specifically, $\DCPp$ and $\DCPm$ are considered to be states forming an
alternative orthogonal basis to the flavor states~\cite{Bernabeu:2012ab,Banuls:2000ki}; in
this case \cref{minus} can then be written as~\cite{PhysRevD.15.1254}:
\begin{equation}
\frac{\ket{\Dz\Dzb}-\ket{\Dzb\Dz}}{\sqrt{2}}
=
\frac{\ket{\DCPm\DCPp}-\ket{\DCPp\DCPm}}{\sqrt{2}}
~{\rm when}~ C_{{\DDbSys}} 
= -1,
\label{minus2}
\end{equation}
and thus the $\D$-decay final states are expected to be
$CP$-anticorrelated\footnote{There can be a very small $CP$-correlated
signal due to $CP$ violation in the charm system, but this is of second-order
in the $CP$-violating parameters~\cite{Petrov:2004rf}.}.

Here it is proposed to expand this formalism, compared to previous
discussions in the literature, to allow all  
decays
discussed in \cref{subsec:C} to be included; thus $C =
+1$ quantum-correlated $\DDbSys$ can also be exploited for potential analysis,
which has some key advantages. 
\Cref{plus} can be written as~\cite{PhysRevD.15.1254}: 
\begin{equation}
\frac{\ket{\Dz\Dzb}+\ket{\Dzb\Dz}}{\sqrt{2}}
=
\frac{\ket{\DCPp\DCPp}-\ket{\DCPm\DCPm}}{\sqrt{2}}
~{\rm when}~ C_{{\DDbSys}}  
= +1,
\label{plus2}
\end{equation}
indicating that the $\D$-decay final states are expected to be $CP$-correlated.

Refs.~\cite{Bevan:2013wwa, Bevan:2013rpr, Bevan:2015ena, Bevan:2015nra} also
suggest using semi-leptonic final states to cleanly identify the flavor-tagged
$\D$ meson.
Here it can be proposed to instead use hadronic pseudo-flavor tags,
e.g. $\Dz \to K^- \pi^+$. The existing formalism can be maintained, but
with the addition of a nuisance parameter to account for the $\Dzb \to
K^- \pi^+$ background of 3.5 per mille~\cite{Amhis:2019ckw}. Hadronic tags have
the advantage that they can be completely reconstructed, and also add more data
for potential analyses. Of course, experiments that can reconstruct semileptonic
decays of flavor-specific modes at hadron colliders (or any other experiments)
would certainly boost analysis prospects.

\begin{figure}[t]
\centering
\includegraphics[width=0.9925\textwidth]{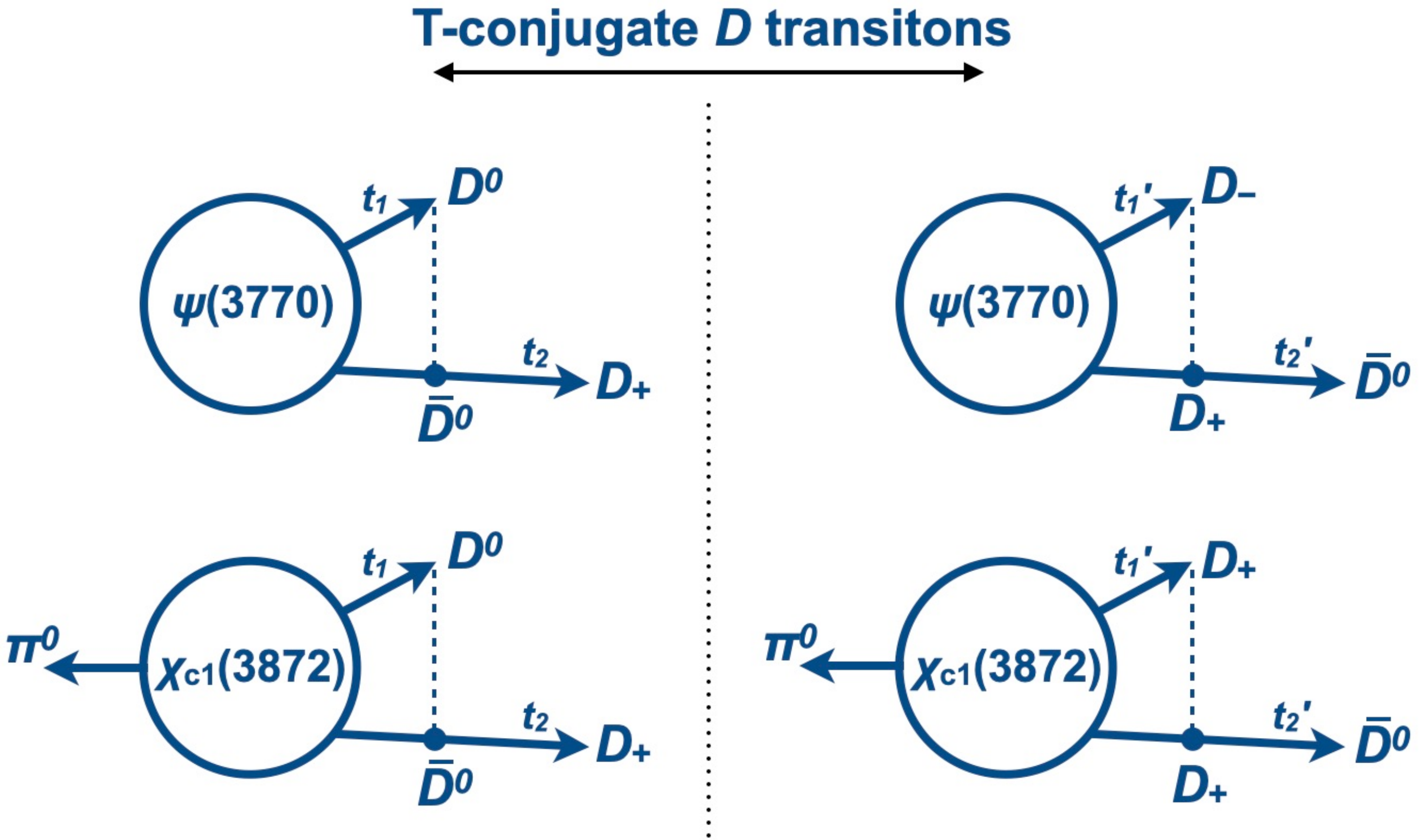}
\caption{Examples of $T$-conjugate decay processes involving transitions between
$\D$ eigenstates of flavor and $\D$ eigenstates of $CP$.
\Cref{table:compareC} lists other symmetry-conjugate decay processes of
interest, involving different transitions.
}
\label{fig:psiT} 
\end{figure}

As discussed in refs.~\cite{Bevan:2013wwa, Bevan:2013rpr,
Bevan:2015ena, Bevan:2015nra}, and as seen in
\cref{fig:psiT}, 
it is
possible to construct $T$ conjugate decay processes from the quantum-correlated
decays of the $\psix$, and compare their rates directly to search for
symmetry violations. 
In this example,
if at time\footnote{Proper time is henceforth implied.} $t_1$ the $\Dz$
meson decays to a flavor state (e.g. $K^- \pi^+$), and at a later time
$t_2$ the other $\D$ meson is found in a $CP+$ eigenstate
($\ket{\DCPp}$), 
for the latter
decay this represents the process $\Dzb|_{t_1} \to \DCPp|_{t_2}$.
The $T$-conjugate process can be then obtained as follows. 
If at time $t_1'$ a neutral 
$\D$ meson from another $\psix$ meson decays to a
$CP-$ eigenstate ($\ket{\DCPm}$), 
and at a later time $t_2'$ the other $\D$ meson is found in 
an anti-flavor state (e.g. $K^+ \pi^-$),
for the latter decay this represents the process $\DCPp|_{t_1'} \to
\overline{D}{}^{0}|_{t_2'}$. The processes can then be studied in terms of
proper time difference $\Delta t = t_2' - t_1' = t_2 - t_1$.

We can also construct pairs of processes to be compared that are $CP$ and
$CPT$ (applying both $CP$ and $T$) conjugates. For example, the processes
$\Dzb|_{t_1} \to \DCPp|_{t_2}$ and $\Dz|_{t_1'} \to \DCPp|_{t_2'}$ can be
compared to look for $CP$ violation; the processes $\Dz|_{t_1} \to
\DCPm|_{t_2}$ and $\DCPm|_{t_1'} \to \Dzb|_{t_2'}$ can be compared
to search for $CPT$ violation.

Using the concepts discussed in the previous sections, this idea can be extended
beyond $e^+ e^- \to \psix$ to all decays that could produce
quantum-correlated $\DDbSys$. A key example is $\X \to \DDbP$, also
shown in \cref{fig:psiT}. Here the resulting neutral $\D$ mesons have
$C=+1$ and are thus $CP$-correlated.
For $\X$ decay the experiment can be set up as with $\psix$ decay, except
for the spectator $\pi^0$ meson and the change required to obtain a proper $T$
conjugate. As for $\psix$ decay, if at time $t_1$ the $\Dz$ meson decays to a
flavor state, and at a later time $t_2$ the other $\D$ meson
is found in a $CP+$ eigenstate ($\ket{\DCPp}$), for the
latter decay this represents the process $\Dzb|_{t_1} \to \DCPp|_{t_2}$.
The $T$-conjugate process in the case of $\X$ decay can be then obtained as
follows.
If at time $t_1'$ a neutral $
{D}$ meson from another $\X$ decays to a
$CP+$ eigenstate ($\ket{\DCPp}$), 
and at a later time $t_2'$ the other $\D$ meson is found in 
an anti-flavor state (e.g. $K^+ \pi^-$),
for the latter decay this represents the process $\DCPp|_{t_1'} \to
\overline{D}{}^{0}|_{t_2'}$.
Pairs of processes to be compared that are $CP$ and
$CPT$ conjugates can also be constructed.

\Cref{table:compareC} lists the testable symmetries and possible pairings of
reference and conjugate transitions as discussed in
refs.~\cite{Bevan:2015ena,Bevan:2015nra}, and also shows the
representations of the required $\D$ decay final states that would need to be
reconstructed at time ($t_1$, $t_2$, $t_1'$, $t_2'$) to test each pairing. A
reconstructibility advantage occurs when using $C = +1$ correlated decays, as
$T$ and $CPT$ symmetries can be tested without the use of more
difficult-to-reconstruct $\DCPm$ states (that typically involve reconstructing
neutral $\piz$, $\eta$, $\omega$, and $K_S^0$) --- this is not possible for $C
= -1$ correlated charm decays.
\backrefsetup{disable}
{
\begin{table}[t]
\begin{center}
\begin{tabular}{ccc||c|c} \hline\hline
Testable       & Reference        & Conjugate        &     $(\DDbSys)_{C = -1}$    &     $(\DDbSys)_{C = +1}$     \\
Symmetry       & Transition       & Transition       &        Detection Modes      &        Detection Modes       \\ 
   $S$         & $a \to b$        & $a' \to b'$      & at ($t_1$, $t_2$, $t_1'$, $t_2'$) &  at ($t_1$, $t_2$, $t_1'$, $t_2'$) \\  
\hline\hline 
$CP$ and $T$   & $\Dz\to \Dzb$    & $\Dzb\to \Dz$    & (\mb$\Dzb,\Dzb,\Dz,\Dz$)    & (\mb$\Dzb,\Dzb,\Dz,\Dz$)     \\
$CP$ and $CPT$ & $\Dz\to \Dz$     & $\Dzb\to \Dzb$   & (\mb$\Dzb,\Dz,\Dz,\Dzb$)    & (\mb$\Dzb,\Dz,\Dz,\Dzb$)     \\ 
\hline 
$T$ and $CPT$  & $\DCPp\to \DCPm$ & $\DCPm\to \DCPp$ & ($\DCPm,\DCPm,\DCPp,\DCPp$) & ($\DCPp,\DCPm,\DCPm,\DCPp$)  \\ 
\hline 
$CP$           & $\Dzb\to \DCPm$  & $\Dz \to \DCPm$  & ($\Dz,\DCPm,\Dzb,\DCPm$)    & ($\Dz,\DCPm,\Dzb,\DCPm$)     \\
               & $\DCPp\to \Dz$   & $\DCPp \to \Dzb$ & ($\DCPm,\Dz,\DCPm,\Dzb$)    & (\mb$\DCPp,\Dz,\DCPp,\Dzb$)  \\
               & $\Dzb\to \DCPp$  & $\Dz\to \DCPp$   & (\mb$\Dz,\DCPp,\Dzb,\DCPp$) & (\mb$\Dz,\DCPp,\Dzb,\DCPp$)  \\
               & $\DCPm\to \Dz$   & $\DCPm\to \Dzb$  & (\mb$\DCPp,\Dz,\DCPp,\Dzb$) & ($\DCPm,\Dz,\DCPm,\Dzb$)     \\ 
\hline
\T             & $\Dzb\to \DCPm$  & $\DCPm \to \Dzb$ & ($\Dz,\DCPm,\DCPp,\Dzb$)    & ($\Dz,\DCPm,\DCPm,\Dzb$)     \\
               & $\DCPp\to \Dz$   & $\Dz \to \DCPp$  & ($\DCPm,\Dz,\Dzb,\DCPp$)    & (\mb$\DCPp,\Dz,\Dzb,\DCPp$)  \\
               & $\Dzb\to \DCPp$  & $\DCPp\to \Dzb$  & ($\Dz,\DCPp,\DCPm,\Dzb$)    & (\mb$\Dz,\DCPp,\DCPp,\Dzb$)  \\
               & $\DCPm\to \Dz$   & $\Dz\to \DCPm$   & ($\DCPp,\Dz,\Dzb,\DCPm$)    & ($\DCPm,\Dz,\Dzb,\DCPm$)     \\ 
\hline 
\CPT           & $\Dzb\to \DCPm$  & $\DCPm \to \Dz$  & ($\Dz,\DCPm,\DCPp,\Dz$)     & ($\Dz,\DCPm,\DCPm,\Dz$)      \\
               & $\DCPp\to \Dz$   & $\Dzb \to \DCPp$ & ($\DCPm,\Dz,\Dz,\DCPp$)     & (\mb$\DCPp,\Dz,\Dz,\DCPp$)   \\
               & $\Dz\to \DCPm$   & $\DCPm\to \Dzb$  & ($\Dzb,\DCPm,\DCPp,\Dzb$)   & ($\Dzb,\DCPm,\DCPm,\Dzb$)    \\
               & $\DCPp\to \Dzb$  & $\Dz\to \DCPp$   & ($\DCPm,\Dzb,\Dzb,\DCPp$)   & (\mb$\DCPp,\Dzb,\Dzb,\DCPp$) \\ 
\hline\hline
\end{tabular} 
\caption{The fifteen possible pairings of reference and symmetry conjugated 
transitions used to study $CP$, \T and $CPT$ for
pairs of neutral $\D$ mesons, as demonstrated by Bevan~\cite{Bevan:2015ena,Bevan:2015nra}.
In four of these pairings, both $a$ and $a'$ can be 
established without the use of $C$-correlated charm (e.g. via $\D^{*+} \to \Dz
\pi^+$ flavor tags) and thus the symmetry can also be tested elsewhere.
Listed next to these pairings are the states that must be measured at ($t_1$,
$t_2$, $t_1'$, $t_2'$) for $C = -1$ and $C = +1$ correlated $\DDbSys$ systems,
to establish the conjugated-transitions pair (see \cref{fig:psiT}).
Sets of states that do not require $\DCPm$ to be reconstructed are highlighted
in {\bf bold}; only $C = +1$ correlated $\DDbSys$ allow tests of $T$ and $CPT$
without the use of the more difficult-to-reconstruct $\DCPm$ states.}
\label{table:compareC}
\end{center}
\end{table}
}
\backrefsetup{enable}

Note that in the weak Hamiltonian all $CP$ violation may be a combination of $T$
and $CPT$ violations; it is proposed that results obtained from
samples outlined in \cref{table:compareC} could be combined to extract the
individual contributions of $T$ and $CPT$ violation 
under this model~\cite{Fidecaro:2013gsa}.
Since
mixing effects in the charm system are small, time-integrated measurements of
asymmetries could be considered first; small corrections could then be
applied to account for any mixing~\cite{Bevan:2015nra}.
In addition to $CP(T)$ violation, $C$ and $P$ violation, known to exist in
the charm system, may also be tested within multi-body $\D$ decays, by performing
a simultaneous triple product analysis 
on correlated
four-body $\Dz$ and $\Dzb$ decays, as discussed 
in 
ref.~\cite{Bevan:2015ena}. 
 
LHCb and Belle II are currently the only
operating experiments where it is possible
to reasonably study the flight distances of 
$\D$ mesons from $\X$ and $\psix$
decays in both direct production and from 
$B \to \X K$ and $B \to \psix K$ decays, the latter having
the added benefit that one could search for $CP$ violation directly by taking
advantage of Bose symmetry in the decay~\cite{Sahoo:2013mqa}. 
The ability to study time dependence in $\psix$ decays could also help
motivate a future asymmetric charm factory~\cite{Asner:2006sd}.

%% file: inputs/5-conclusion.tex
\FloatBarrier\section{Conclusion}
\label{Conclusion}

There remains an
opportunity to go beyond traditional 
quantum correlated analyses of charm, by
exploring $C = +1$ $\DDbSys$ systems 
for 
studies of $CP(T)$
violation, time-reversal violation, mixing, and 
relative strong phase measurements. 
Amplitude analysis of $\B \to \DDbSys X$ decays may also benefit from
targeted reconstruction; for example, reconstruction of 
$\B \to (\DDbSys)_{C=+1} X$ via $\D$ decay filter bases, such that $C=-1$
$\DDbSys$ resonances will not be present in the $\B \to \DDbSys X$ decay. 

Several
variables of separation exist to help identify the $\XtoDDbP$ and $\XtoDDbG$ components of
$\X \to \Dstz\Dzb$ decays, even if the final state photon or $\piz$
cannot be reconstructed. Separating these components would allow the
extraction of $C = +1$ $\DDbSys$ systems from these decays, 
and also improve measurements of the $\XtoDDbP$ and $\XtoDDbG$ branching fractions.

Also outlined are advantages for studying time-reversal violation and $CPT$
violation with $C = +1$ $\DDbSys$ systems, due to the ability to collect useful
data in reconstruction modes suitable at both annihilation experiments and hadron
colliders. 

A survey of the various production mechanisms and potential
laboratories for studying quantum-correlated $\DDbSys$ systems is given
in
\cref{Laboratories}.
A future tau-charm factory~\cite{Asner:2006sd} 
or the proposed \panda experiment~\cite{Peters:2017kop, PANDA:2021ozp} 
offer great potential to obtain world-leading 
amounts of charmonium decays to double open charm. 

In the meantime, 
several current experiments have the ability to collect the relevant
charmonia(-like) decays.
The planned
\besiii $\DDbSys$ sample~\cite{BESIII:2020nme} can be augmented with correlated
$\DDbSys$ systems collected above $\psix$ threshold at
\besiii and at \belleII, particularly if there are dedicated \belleII runs
at the $\Upsilon(1S)$ resonance~\cite{Li:2006wx}. \lhcb will also produce many 
correlated $\DDbSys$ systems, including 
large samples of prompt $\X$ decays; further experimental work to 
isolate these decays for quantum-correlated analyses is encouraged. 
A
quantum-correlated analysis of any sample of $\X$ decays would provide 
interesting (and possibly first) measurements, and provide a valuable blueprint 
for analysis of $C = +1$ $\DDbSys$ systems at upgraded and future experiments.

%% file: inputs/A-appendices.tex

\FloatBarrier\section{Potential laboratories} 
\label{Laboratories}

A subset of promising production 
mechanisms for quantum-correlated $\DDbSys$ systems, and possible experiments
where these systems could be studied, are overviewed here. 
Decays of the $\X$ exotic meson
 have  been recently observed in photo(-muo)production, lead-lead
collisions, and two-photon interactions~\cite{Guskov:2017nzr,
CMS:2021znk, Belle:2020ndp}, but will not be discussed further here.
Electron-hadron facilities could also compliment existing experiments that produce 
$\X$ and other exotic states~\cite{Albaladejo:2020tzt}.

\FloatBarrier\subsection{Quantum correlated states from
\texorpdfstring{$e^+ e^-$}{positron -- electron} annihilation}
\label{subsec:QCfrom-ee}

This section discusses the production of quantum-correlated $\DDbSys$
systems via resonant electron-positron annihilation.  In this process, quantum
correlated $\DDbSys$ can be created at any energy above the open charm threshold, but are
particularly prominent in $1^{--}$ charmonium(-type) resonances. Further
investigations have been encouraged to search for states of
charmonia that decay to other charmonia and one or more $\piz$ or
$\gamma$~\cite{Guo:2013nza}. Production of a $\X$ exotic meson with an
associated photon is also of interest. Active and future experiments that could exploit this 
production mechanism
include \besiii, \belleII, 
and a 
future tau-charm factory~\cite{Asner:2006sd}.

\FloatBarrier\subsubsection{\texorpdfstring{The process $e^+ e^- \xrightarrow[]{\psix}
\DDbSys$}{Positron -- electron annihilation into D0 anti-D0 at the psi(3770)
resonance} }
\label{subsec:QCfrom-psi}

The entirety of quantum-correlated $\DDbSys$ samples 
collected and analyzed thus
far have come from this production mechanism. The cross 
section for this specific process
has been measured by \besiii as:
\begin{equation}
\sigma[e^+ e^-
\to \DDbSys] =  
(3.615\pm 0.010 \pm 0.038)~{\rm nb},
\end{equation}
near the $\psix$ resonance peak
($\sqrt{s}=3.773$~GeV)~\cite{Ablikim:2018tay}.

\cleoc collected an integrated luminosity of 818 pb${}^{-1}$ 
at this energy; $3.0 \times 10^6$ $\DDbSys$ systems were
produced~\cite{Bonvicini:2013vxi}. 
 \besiii has collected 2.93 fb${}^{-1}$ so
far, corresponding to $1.06 \times 10^7$ $\DDbSys$ systems~\cite{Ablikim:2018tay}.
To date, several quantum correlated $\DDbSys$ analyses have been performed on
these data
samples~\cite{Rosner:2008fq,Asner:2008ft,Asner:2012xb,Insler:2012pm,Libby:2010nu,Ablikim:2014gvw,Nayak:2014tea,Malde:2015mha,Evans:2016tlp,Harnew:2017tlp,Resmi:2017fuo,Ablikim:2020lpk,Ablikim:2020cfp,BESIII:2020hlg,BESIII:2021eud,BESIII:2022qkh,BESIII:2022wqs,BESIII:2022ebh}.
A recent call for further collection and study of quantum-correlated $\DDbSys$ 
focuses primarily on further data-taking at the $\psix$ resonance
peak~\cite{Wilkinson:2021tby,BESIII:2020nme}; 20 fb${}^{-1}$ of
additional integrated luminosity is expected from \besiii by the year 2025.

A recently proposed future tau-charm factory, with instantaneous luminosity
$\mathcal{L} = 0.5 \times 10^{35}$~cm$^{-2}$~s$^{-1}$, could be expected 
to produce 1 ab$^{-1}$ of
integrated luminosity, or $3.6 \times 10^9$ correlated $\DDbSys$ systems, per
year~\cite{Zhou:2021rgi}.
\cleoc with 818 pb${}^{-1}$ of integrated luminosity reconstructed $1731 \pm 42
\pm 11$ $e^+e^- \to (\Km \pip)_\Dz (\Kp \pim)_\Dzb$ decays with an efficiency of
($40.0\pm0.2$)\%\footnote{Decays in many other final states were reconstructed
from this $\DDbSys$ sample, including approximately 550 
$e^+ e^- \to (K^\pm \pi^\mp)_{D} (h^+ h^-){}_{\overline{D}},~h \in \{\pi, K\}$ 
at similar efficiency levels.}~\cite{Asner:2012xb}.
Presuming a 50\% efficiency could be reached, this would correspond to
$\mathcal{O}(2.6 \times 10^6)$ reconstructed $e^+e^- \to (\Km \pip)_\Dz (\Kp
\pim)_\Dzb$ decays per year at a future tau-charm factory. This can be compared
to the $\mathcal{O}(5 \times 10^4)$ $e^+e^- \to (\Km \pip)_\Dz (\Kp
\pim)_\Dzb$ that could be reconstructed from 
20 fb${}^{-1}$ of \besiii data.

\FloatBarrier\subsubsection{\texorpdfstring{The process $e^+ e^- \xrightarrow[]{}
\DDbSys + m \gamma + n \piz$}{Positron -- electron annihilation into D0 anti-D0,
m photons, and n neutral pions} }
\label{subsec:QCfrom-ee-4040-4140-and-Upsilon}

As described in \cref{subsec:C}, both $C=+1$ and $C=-1$ $\DDbSys$ systems can
be produced from $e^+ e^-$ collisions above $\DDbSys$ threshold. The charm
factories have yet to perform quantum correlated analyses with any $C=+1$
$\DDbSys$. The cross section and potential inefficiencies lead to an order of
magnitude fewer $C=+1$ $\DDbSys$ than the $C=-1$ $\DDbSys$ that would be
collected from the same integrated luminosity of data taken at $\psix$
threshold; however, it has been shown that the combination
of such decays would give significant improvement over
certain quantum correlated measurements made  
with $C=-1$ $\DDbSys$
alone~\cite{Asner:2005wf}. Production is more plentiful at $1^{--}$
charmonium resonances of course, such as through $e^+ e^-
\xrightarrow[]{\psi(4040)} \DDbSys + m \gamma + n \piz$~\cite{Bondar:2010qs} or
$e^+ e^-
\xrightarrow[]{\psi(4160)} \DDbSys + m \gamma + n \piz$~\cite{Xing:1996pn}.
Production may also be plentiful at $1^{--}$
bottomonium resonances, such as through $e^+ e^-
\xrightarrow[]{\Upsilon(1S)} \DDbSys + m \gamma + n \piz$~\cite{Li:2006wx};
yearly production rates of correlated $\DDbSys$ are expected to be
very competitive with \besiii, if the 
luminosity delivered to \belleII 
approaches $10^{36}~{\rm cm}^{-2}{\rm s}^{-1}$ as expected. 
There is a possibility of collecting $e^+ e^-
\xrightarrow[]{\Upsilon(4S)} \DDbSys + m \gamma + n \piz$ decays within the
planned \belleII 50 ab$^{-1}$ sample, but the relevant decay branching
fractions are unknown; future work should focus on determining the real or
theoretical decay rates.

\FloatBarrier\subsubsection{\texorpdfstring{The process $e^+ e^- \xrightarrow[]{\Yx} \X
\gamma$}{Positron -- electron annihilation into X(3872) and a photon at the
psi(4230) resonance}  }
\label{subsec:QCfrom-ee-4230}

It is not possible to produce resonant $\X$ exotic meson production 
in electron-positron
annihilation due to its quantum numbers. However, 
the $\X$ exotic meson has been
observed in radiative production from the $\Yx$ state~\cite{Ablikim:2013dyn}.
The cross section $\sigma[e^+ e^-
\to \gamma \X]\times\mathcal{B}[\X\to J/\psi \pi^+ \pi^-]$ was measured by
\besiii at center-of-mass energies from 4.0 to 4.6 \GeVm~\cite{Ablikim:2019zio}.
This allowed the following maximum cross section to be calculated in
ref.~\cite{Li:2019kpj}:
\begin{equation}
\sigma[e^+ e^-
\to \gamma \X] = 
(5.5^{+2.8}_{-3.6})~{\rm pb},
\end{equation}
reaching this value at $\sqrt{s}=4.226$~GeV. 

Using the branching fraction of $\X \to \Dz
\Dstzb + \Dzb \Dstz$, as determined 
in ref.~\cite{Li:2019kpj}, the following cross section 
can be estimated:
\begin{equation}
\sigma[e^+ e^-
\to \gamma \X]\times \mathcal{B}(\X \to \Dz
\Dstzb + \Dzb \Dstz) 
\approx {2.9}~{\rm pb}.
\end{equation}
While this cross section is comparatively small compared to $\psix$ decays, it
could comprise a useful source of correlated $\DDbSys$ systems, particularly
because the separation techniques discussed in \cref{simulationSeparation} could be
used to more cleanly identify the $C = +1$ $\DDbSys$ components of the $\X$
decays, if at least the photon produced along with the $\X$ exotic meson can be
reconstructed.
In the case where all light neutral particles are reconstructed, \besiii has
recently observed approximately 50 $\X \to \Dz \Dstzb + \Dzb \Dstz$ decays, in 9 fb$^{-1}$ of $e^+e^-$ collision data at
center-of-mass energies ranging from 4.178 to 4.278 \GeVE~\cite{BESIII:2020nbj}.
The neutral $D$ mesons were reconstructed in $K\pi$, $K\pi\pi^0$, and
$K\pi\pi\pi$ final states.

Of course, the $\Yx$ state or $e^+ e^-$
decays directly to $\DDbSys \pi^0 \gamma$ and $\DDbSys \gamma \gamma$
certainly also provide quantum
correlated $\DDbSys$ systems --- depending on the relative cross sections
determining the $C$ eigenstate of the $\DDbSys$ may be a challenge, though could
be overcome by studying the final states in the admixture~\cite{Asner:2005wf}. 
Further searches for and collection of radiatively produced $\X$ exotic mesons
from other $1^{--}$ states could help build a further sample.

Initial state
radiation (ISR) production of $\Yx$ state decays at \belleII should be
competitive with the collection of these decays in direct production at
\besiii~\cite{Belle:2007qxm,Kou:2018nap}; for example during the full \belleII
running period, over 2 fb$^{-1}$ of integrated luminosity from ISR events will be available in every
10~\MeVE energy region from 4 to 5~\GeVE in center-of-mass energy.

\FloatBarrier\subsection{Charmonium(-like) states from
\texorpdfstring{$p\overline{p}$}{proton -- anti-proton} annihilation}
\label{subsec:QCfrom-ppbar}

Proton-antiproton annihilation allows the generation of 
several $J^{PC}$ states, including
$1^{++}$ and
$1^{--}$, through gluon-rich annihilation processes~\cite{Kaplan:2011hv}. 
The future
\panda experiment~\cite{Peters:2017kop, PANDA:2021ozp}, colliding antiprotons
onto a proton fixed target, potentially offers a laboratory for resonant
production of these decays with the possibility of very large cross sections.

\FloatBarrier\subsubsection{\texorpdfstring{The process $p\overline{p} \to
\psix$}{proton-antiproton to psi(3770)}}
\label{subsec:QCfrom-ppbar-psi}

The annihliation cross section for $p\overline{p} \to
\psix$ may be determined using the Breit-Wigner formula~\cite{PDG2020} if
the branching fraction of the time-reversed process $\psix \to
p\overline{p}$ is known;
this branching fraction has been extracted from a 
sample of $e^+ e^-\xrightarrow[]{\psix}p\overline{p}$ decays collected at
\besiii~\cite{Ablikim:2014jrz}.

At $\sqrt{s}=3.774$~GeV, this cross section has been estimated to
have the following two solutions~\cite{Ablikim:2014jrz}:
\begin{equation}
\sigma[p\overline{p} \to \psix] = (9.8^{+11.8}_{-3.9})~{\rm nb}~( < 27.5
{\rm~nb~at~90\%~C.L.})~{\rm or}~(425.6^{+42.9}_{-43.7})~{\rm nb}.
\end{equation}

This result indicates that proton--antiproton collisions
at future experiments such as \panda~\cite{IkegamiAndersson:2016wxs} can produce
potentially large boosted $\psix$ samples from $p\overline{p}$
annihilations.
In just one day of
dedicated proton--antiproton collisions at the design specifications
of \panda, $\mathcal{O}(8 \times 10^4)$ or
$\mathcal{O}(3 \times 10^6)$ boosted $\psix$ mesons could be produced in
\panda's high luminosity mode; in \panda's high resolution mode the event yields
are decreased by a factor of 10.
There are potential scenarios where the background from threshold $D\overline{D}$ pair production is low enough to see a clear 
$\psix$ signal peak~\cite{Haidenbauer:2015vra}.

\FloatBarrier\subsubsection{\texorpdfstring{The process $p\overline{p} \to \psix
\pi^0$}{proton-antiproton to psi(3770) and a neutral pion}}
\label{subsec:QCfrom-ppbar-psipi}

The cross section for $p\overline{p} \to
\psix \pi^0$ may be determined by crossing relations~\cite{Lundborg:2005am}
if the branching fraction of the process $\psix \to p\overline{p} \pi^0$ is known;
this branching fraction has been extracted from a 
sample of $e^+ e^- \xrightarrow[]{\psix} p\overline{p} \pi^0$ decays
collected at \besiii~\cite{Ablikim:2014kxa}

This cross section (maximal at a center of mass energy of $5.26$~GeV) has been
estimated to have the following two solutions~\cite{Ablikim:2014kxa}:
\begin{equation}
\sigma[p\overline{p} \to \psix\piz] = ( < 0.79
{\rm~nb~at~90\%~C.L.})~{\rm or}~(122\pm10)~{\rm nb}.
\end{equation}

\FloatBarrier\subsubsection{\texorpdfstring{The process $p\overline{p} \to \X
(\pi^0)$}{proton-antiproton to X(3872) (neutral pion)}}
\label{subsubsec:pbar-p-to-X}

Measurements of $\X \to p\overline{p}(\pi^0)$ would allow 
determinations of cross sections for $p\overline{p} \to \X (\pi^0)$,
similar to those made for $p\overline{p} \to \psix (\pi^0)$.
The method of detailed balance, combined with available experimental input, 
predicts an upper limit of $\sigma(p\overline{p} \to \X) <
68$~nb~\cite{Prencipe:2015cgg}. If the cross section is close to the upper
limit, depending on the operational mode of \panda, $\mathcal{O}(5 \times
10^5)$ (high luminosity mode) or $\mathcal{O}(5 \times
10^4)$ (high resolution mode) could be produced per
day. 

There is potential destructive interference that could substantially
reduce the cross section of $p\overline{p} \xrightarrow[]{\psix} \DDbSys$ and
enhance the cross section of $p\overline{p} \xrightarrow[]{\psix}
D^+D^-$~\cite{Shyam:2015hqa}.
Thus one advantage of $\X$ decays over $\psix$ decays at \panda is that 
$p\overline{p} \xrightarrow[]{\X} \Dz\Dstzb + \Dstz\Dzb$ should not
suffer from the same type of interference since the $\X$ exotic meson is below
$D^{*+}D^-$ threshold.

Searches for $\X \to p\overline{p}(\pi^0)$ at currently running experiments
would be very useful to help determine a more precise $p\overline{p} \to
\X$ cross section.
If this cross section is determined to be comparable or larger than for $\psix$
decays, even if $p\overline{p} \xrightarrow[]{\psix} \DDbSys$ is
suppressed there could be a great opportunity to source both $C=+1$ and $C=-1$
$\DDbSys$ systems at \panda.

Production of $\Yx$ and $\Zc$ states is also expected at \panda with
upper limits on their cross sections approximately 30 and 150 times lower than
for $\X$, respectively~\cite{Prencipe:2015cgg}. Since the $\Zc$ state is also
not narrow, $C$-changing $\DDbSys$ contamination from these decays should be minimal.

\FloatBarrier\subsection{Charmonium(-like) states from \texorpdfstring{$b$}{b} hadron decays}
\label{subsec:QCfrom-B}

Active experiments that could exploit this 
production mechanism
include \lhcb and \belleII.

The cross section for prompt $B^\pm$ production at 
\lhcb in 13~TeV $pp$
collisions is~\cite{Aaij:2017qml}:
\begin{eqnarray}
\sigma(pp \to B^\pm X) 
&=& 86.6 \pm 0.5 ({\rm stat.}) \pm 5.4 ({\rm syst.}) \pm 
3.4 (\mathcal{B}(B^\pm \to J/\psi K^\pm))~{\mu}{\rm b}, \nonumber\\ 
&&2.0<y<4.5,~0< p_T < 40~{\rm GeV}/c. 
\end{eqnarray}

In combination with the branching fractions
$\mathcal{B}(\BptoPsiKp) \times
\mathcal{B}(\psix \to \DDbSys) = (1.5 \pm 0.5) \times 10^{-4}$
\cite{PDG2020}, the cross-section for $\psix \to \DDbSys$ from $B^\pm$ 
decays at 13
TeV is $\approx{ 13}$ nb. 
In combination with the branching fractions
$\mathcal{B}(\BptoXKp) \times
\mathcal{B}(\X \to \Dstzb\Dz) = (8.5 \pm 2.6) \times 10^{-5}$ 
\cite{PDG2020}, the cross-section for $\X \to \Dstzb\Dz$ from $B^\pm$ 
decays at 13
TeV is $\approx{ 7}$ nb. 

In the \lhcb Upgrade and Upgrade II the samples collected should be much
higher, though efficency of obtaining the sample may be low. The \lhcb
collaboration has predicted reconstructed yields in 
the $\BptoXKp, \X \to J/\psi \pi \pi$ decay channel to be 14k,
30k, and 180k, with 23, 50, and 300/fb samples of \lhcb Upgrade II
data, respectively, and also estimates a yield of 11k for 50/ab of \belleII
data~\cite{Bediaga:2018lhg}. Determining the relative efficiency for detecting
$\BptoXKp, \X \to \Dz \Dstzb + \Dzb \Dstz$ from \lhcb without
reconstructing the light neutral $\piz$ or $\gamma$, should give an 
indication of how many quantum
correlated $\DDbSys$ decays could be collected from this channel.

Recently, an amplitude analysis of $\Bp \to \D^+ \D^- K^+$ was performed by the
\lhcb collaboration~\cite{LHCb:2020bls,Aaij:2020ypa}. A sample of 1260 candidate
events (with purity greater than $99.5\%$) were studied. The resonant decay $\BptoPsiKp$
is found to have a fit fraction of $14.5 \pm 1.2 \pm 0.8 \%$, indicating
$\mathcal{O}(180)$ $\BptoPsiKp$ decays were detected. These events are
relatively isolated from other resonant contributions; there is a small non-resonant
contribution underneath the $\BptoPsiKp$ contribution, for which the
interference effect is consistent with zero. A large charge asymmetry in $B \to
\D\Db K$ has been noticed and studied by Bondar and Milstein, who find the ratio
of the $\Bp \to \D^+ \D^- K^+$ decay probability and the $\Bp \to \DDbSys K^+$
decay probability in the $(\D\Db)_\psix$ region is $0.27 \pm
0.13$~\cite{Bondar:2020eoa}.
Thus it is possible under a similar efficiency profile $\mathcal{O}(700)$
$\BptoPsiKp$ should be observed in $\Bp \to \DDbSys K^+$ decay in the 
combined \lhcb Run 1 and Run 2 data samples;
lower branching fractions in neutral $D$ reconstruction modes may be balanced by
reconstructing two fewer particles in the overall final state (presuming the
neutral $D$ mesons are only reconstructed in $K \pi$ final states).
The yields should be reasonably expected to increase by one 
to two orders of magnitude during the \lhcb Upgrade and Upgrade
II eras --- this would correspond to $\mathcal{O}(10^4)$
and $\mathcal{O}(6 \times 10^4)$ reconstructed $\psix \to (\Km
\pip)_\Dz(\Kp \pim)_\Dzb$ from $\Bp \to \psix K^+$ decays after \lhcb collects
50 and 300 fb$^{-1}$ of integrated 
luminosity, respectively~\cite{Craik:2017dpc,Bediaga:2018lhg}. The number of
reconstructed decays from this channel alone, in a 300 fb$^{-1}$ \lhcb data set,
would likely be comparable to the number of decays available from the \psix
resonance in \besiii data, as discussed in \cref{subsec:QCfrom-psi}.

Substantial $\Bp \to \chicJ K^+ |_{J\in{\rm even}}$ components are
also observed in the aforementioned \lhcb analysis, and also peak strongly in
the $D^+ D^-$ spectrum. However, Bondar and Milstein notice this component is
suppressed in $\Bp \to \DDbSys K^+$, as no such component was observed by
\babar~\cite{Lees:2014abp}; this suppression may be related to the
aforementioned charge asymmetry. 
An interesting consequence of this suppression is
that by applying the selection suggested in \cref{subsec:CdefiniteBdecays},
the $C=-1$ $\DDbSys$ resonances will be completely suppressed, leaving only
the $DK$ resonant amplitudes as dominant components of the total $\Bp \to
\DDbSys K^+$ decay amplitude.
It is possible that the decay $B^0 \to \DDbSys K^0$ has a substructure more favorable to $\Bp \to \chicJ K^+ |_{J\in{\rm even}}$ components, so this decay mode is worth further investigation, as these particular $\chi_{cJ}$ states could also be a source of $C = +1$
$\DDbSys$~\cite{DanAndTimG}. 

There is opportunity to collect $\psix$ and $\X$
decays from other $b$ hadron decays as well. The cross section for $b$ hadron
production at \lhcb in 13~TeV pp
collisions is $144 \pm 1 ({\rm stat.}) \pm 21
({\rm syst.})$~${\mu}$b, for $2 < y < 5$ \cite{Aaij:2016avz}.
An incomplete list of decay modes from which these decays could be collected
include:
$B^0 \to \psix K^0$, $\BptoPsiKp$, $B^0 \to \X K^0$, $B^0 \to \X K^0$,
$\Bp \to \psix K^0\pi^+$, $B^0 \to \psix K^+\pi^-$, $\Bp \to \X K^0\pi^+$,
$B^0 \to \X K^+\pi^-$, $\Lambda_b^0 \to \X p K^-$, $\Lambda_b^0 \to \psix p
K^-$.
Some of the branching fractions have not yet been measured, 
or could be measured more accurately~\cite{PDG2020}. 
Triangle singularities may contribute observably to the branching fractions
involving $\X K \pi$ final states~\cite{Braaten:2019yua}. Decays of the $B_s^0$
such as $B_s^0 \to \X \phi$ may also be worth exploring~\cite{Sirunyan:2020qir}.
The $\X$ decay separation techniques discussed in
\cref{simulationSeparation} would apply to
these decays; also, the $B \to \X K$-specific techniques 
could be
applied analogously for $B \to \X K^*$ and $\Lambda_b^0 \to \X \Delta(1520)$.
Note that \CPV originating from $B$ decays should not need to be considered in
most cases, if the primary interest is in the decays of the
resulting correlated $\DDbSys$ systems.

As noted in~\cref{subsec:C}, $\DDbSys$ systems
from rarer weak decays such as $B_{(s)}^0 \to \DDbSys$ must 
exist in fixed states of $C=+1$ due to angular
momentum conservation.
The branching fraction for $B_{(s)}^0 \to \DDbSys$ is
small~\cite{Aaij:2013fha}, but the possibility to obtain these systems without
interference backgrounds is also worth exploring. 
LHCb observed $13 \pm 6$~($45 \pm 8$) $B_{(s)}^0 \to \DDbSys$ events in
1~fb$^{-1}$ of their Run 1 data; the yields should be reasonably expected to
increase by two orders of magnitude during the \lhcb Upgrade era~\cite{Bediaga:2018lhg}.
Quantum correlations in
$\DDbSys$ formed from weak decays of the $B_{(s)}^0$ meson could be
experimentally confirmed, by demonstrating that the branching fractions 
of the
reconstructed final states shown in \cref{table:Cdefinite} are 
proportional to those expected by quantum
correlations (rather than the naive expectation).

There is also the possibility that the $X(3720)$, a theorized $J^{PC} = 0^{++}$
state~\cite{Gamermann:2006nm,Gamermann:2007mu}, could produce $C=+1$ correlated
$\DDbSys$ systems near charm threshold. It has been suggested that the existence
of this state could be inferred from studies of $B \to \DDbSys K$
systems~\cite{Dai:2015bcc} or 
$\Lambda_b \to \DDbSys \Lambda$ \cite{Wei:2021usz}.

\FloatBarrier\subsection{Charmonium(-like) states from proton-proton collisions}
\label{subsec:QCfrom-pp}

Large cross sections for $pp \to \X {X}$ have been observed
by \cms~\cite{Chatrchyan:2013cld}, \atlas~\cite{Aaboud:2016vzw}, and
\lhcb~\cite{Aaij:2011sn,LHCb:2021ten}.

For example, at $\sqrt{s} = 7$~TeV \lhcb observed:
\begin{eqnarray}
\sigma(pp \to \X {X}) \times && \nonumber \\ \mathcal{B}(\X \to
J/\psi\pi^+\pi^-) 
&=& 5.4 \pm 1.3 ({\rm stat.}) \pm 0.8 ({\rm syst.})~{\rm nb}, \nonumber \\
&&2.5<y<4.5,~5 < p_T < 20~{\rm GeV}/c. 
\end{eqnarray}

Using the branching fractions of $\X \to
J/\psi\pi^+\pi^-$ and $\X \to \Dz
\Dstzb + \Dzb \Dstz$, as determined 
in ref.~\cite{Li:2019kpj}, the following cross section 
can be estimated:
\begin{eqnarray}
\sigma(pp \to \X {X}) \times && \nonumber \\ \mathcal{B}(\X \to \Dz
\Dstzb + \Dzb \Dstz) 
&\approx& {69}~{\rm nb}, \nonumber \\ && 2.5<y<4.5,~5
< p_T < 20~{\rm GeV}/c. 
\end{eqnarray}

Significant numbers of prompt $\psix \to
\DDbSys$ and $\X \to \DDbSys X$ decays, where $\Dz \to K^- \pi^+$ and $\Dzb \to
K^+ \pi^-$, have recently been observed at \lhcb~\cite{Aaij:2019evc} with the
full \lhcb Run 1 and Run 2 datasets. Both samples sit on considerable backgrounds, 
however the cleaner $pp \to \X {X}$ samples are
easier to obtain given that the $\X$ decay width is relatively narrow. In these
samples, \lhcb observed $(5.1 \pm 0.5 ({\rm stat.})) \times 10^3$ $\psix$
decays.
A precise number of $\X \to \DDbSys {X}$ decays 
was not determined, but $\mathcal{O}(10^4)$ decays are visible in the
$m_{\DDbSys}$ spectra. Other charmonia
decaying to $\DDbSys X$ final states were also observed, including
$\chi_{c2}(3930)$, and the $X(3842)$ (presumed to be the $\psi_3(3842)$).

Ref.~\cite{Bediaga:2018lhg} describes how prompt $\Dz$ meson yields will
increase from the existing \lhcb runs to the \lhcb Upgrade and Upgrade II eras. 
Prompt $\X$ exotic meson yields
will scale similarly,
thus we can expect reconstructed yields in the prompt
$\X \to \DDbSys {X}$ decay channel, where $\Dz \to K^- \pi^+$ and $\Dzb \to
K^+ \pi^-$, could be
$\mathcal{O}(5 \times 10^4)$, $\mathcal{O}(2 \times 10^5)$, and
$\mathcal{O}(10^6)$, with 23, 50, and 300/fb samples of \lhcb Upgrade and
Upgrade II data, respectively --- these yields compare favorably to the number
of decays available from the \psix resonance in \besiii data, as discussed in
\cref{subsec:QCfrom-psi}.
 
Very recently, an 
exotic tetraquark
$T_{cc}^+$ was observed decaying to the $\Dz \Dz \pi^+$ final state just below 
$\DstarPlus\Dz$ threshold~\cite{LHCb:2021vvq,LHCb:2021auc}. The resulting
$\Dz \Dz$ systems are not entangled, but when reconstructing $D$ decays to \CP~eigenstates these could form a background to similarly reconstructed
entangled $\DDbSys$ systems. Such potential contamination can be estimated from
ref.~\cite{LHCb:2021auc} to be only on the order of 5\%, and perhaps can be
abated experimentally by rejecting $\DDbSys$ sytems sharing a vertex with an
associated charged pion.

\FloatBarrier\section{\boldmath Amplitude model for \texorpdfstring{$\X \to
\Dz\Dzb\pi^0$}{X(3872) decays to D0 anti-D0 pi0}}
\label{sec:Amplitude}

\FloatBarrier\subsection{Formalism}
\label{subsec:AmpFormalism}

The probability distribution $\mathcal{I}$ for the $\X \to
\Dz\Dzb\pi^0$ decay, as a function of a point $\mathbf x$ in the $\X$ decay
phase space, is proportional to the sum of the
spin-projection dependent decay probabilities, taken over all 
unobservable initial-state spin projections ${s_z}_{X}$: 
\begin{equation}
    \mathcal{I(\mathbf x)} 
    \propto 
    \sum_{ 
    {s_z}_{X}  
    } 
    \left\vert A_{\Xs}({\mathbf x}, 
    {s_z}_{X}  
    )\right\vert^2 ,
	\label{eq:intense}
\end{equation}
where $A_{\Xs}({\mathbf x},  {s_z}_{X}  )$ consists of the coherent sum 
over all intermediate state amplitudes $A_i({\mathbf x},   {s_z}_{X}  )$
representing the complete set of decay chains $\mathbb{P}$:
\begin{equation}
A_{\Xs}({\mathbf x},  {s_z}_{X}  ) = \sum_{i \in \mathbb{P}} A_i({\mathbf x}, 
{s_z}_{X}  )
\end{equation}

The intermediate state
decay amplitudes are parameterized as a product of 
form factors ${B_i}_{L}$ for the initial particle $\Xs$ and resonance $R$ decay
vertices, 
Breit-Wigner propagators ${T_i}_{R}$ included for each resonance,
and an overall angular distribution represented by a spin factor $S_i$:
\begin{equation}
	A_{i}({\bf x},  {s_z}_{X}  ) =  {B_{i}}_{L_{\Xs}}({\bf x}) \,
	{B_{i}}_{L_{R}}({\bf x})  \, {T_{i}}_{R}({\bf x}) \,  S_{i}({\bf x},  {s_z}_{X}
	)  \, .
	\label{eq:ampX}
\end{equation}

The Blatt-Weisskopf penetration factors, 
derived in ref.~\cite{Bl2}
by assuming a square well interaction potential with radius $r_{\rm BW}$, account 
for the finite size of the decaying resonances; these are chosen for the
form factors $B_L$.
They depend on the breakup momentum $q$,
and the orbital angular momentum $L$, between the resonance daughters.
Their explicit expressions are:
\begin{align}
         \nonumber
	B_{0}(q)  &= 1 ,  \\ \nonumber
	B_{1}(q)  &= 1 / \sqrt{{1+ (q \, r_{\rm BW})^{2}}} ,  \\
	B_{2}(q)  &= 1 / \sqrt{9+3\,(q \, r_{\rm BW})^{2}+(q \, r_{\rm BW})^{4}} . 
\end{align}
Here, $r_{BW} = 1.5~{\rm GeV}^{-1}$ is chosen for both $\X$ and $D^*$ decays.
Resonance lineshapes are described as function of the energy-squared, $s$, by Breit-Wigner propagators
\begin{equation}
	T(s) = \frac{1}
	{M^{2}(s) - s - i\,m_{0}\,\Gamma(s)}   \, ,
	\label{eq:BW}
\end{equation}
featuring the energy-dependent mass-squared $M^{2}(s)$, and total width,
$\Gamma(s)$.
The latter is normalized to give the nominal width, $\Gamma_{0}$, when evaluated at the nominal mass $m_{0}$, 
i.e.
$\Gamma_{0} = \Gamma(s = m_{0}^{2})$. While $M^{2}(s)$ typically
follows the Kramers-Kronig dispersion
relation~\cite{PhysRevD.39.1357,Vojik:2010ua}, in practice 
the approximation $M^{2}(s) = {m_0}^2$ is used, which is justified since for the
relatively narrow resonances included this quantity is approximately constant
near the on-shell mass. The nominal masses and widths of the resonances are taken from 
ref.~\cite{PDG2020}, with the exceptions described below.

For a decay into two stable particles $R \to AB$, the energy dependence of the decay width can be described by 
\begin{equation}
	\Gamma_{R \to AB}(s) = \Gamma_{0} \, \frac{m_{0}}{\sqrt s} \, \left(\frac{q}{q_{0}}\right)^{2L+1} \, \frac{B_{L}(q)^{2}}{B_{L}(q_{0})^{2}}  \, ,
	\label{eq:gamma2}
\end{equation}
where $q_{0}$ is the value of the breakup momentum at the resonance pole \cite{BW}.

The spin factors in the covariant Zemach (Rarita-Schwinger) tensor
formalism~\cite{Zemach,Rarita,helicity3} are constructed here, applied in the
same manner as in ref.~\cite{dArgent:2017gzv}.
The fundamental objects of the covariant tensor formalism 
are spin projection operators and angular momentum tensors, 
which connect the particle's four-momenta to the spin dynamics of the
reaction~\cite{Zou,Filippini}.

A spin-$S$ particle with four-momentum $p$, and spin projection ${s_z}$, 
is represented 
by the polarization tensor $\epsilon_{(S)}(p,{s_z})$, which is symmetric, traceless and orthogonal to $p$.
The Rarita-Schwinger conditions reduce the a priori $4^{S}$ 
 elements of the rank-$S$ tensor to $2S +1$ independent  elements in accordance with the number of degrees of freedom of a spin-$S$ state\cite{Rarita,Zhu}.

The spin projection operator $P^{\mu_{1} \dots \mu_{S_R} \nu_{1} \dots
\nu_{S_R}}_{(S_R)}(p_{R})$, for a resonance $R$, with spin $S_{R} = \{0,1,2\}$,
and four-momentum $p_{R}$, is given by \cite{Filippini}:
\begin{align}
	\nonumber
	P_{(0)}^{\mu \nu}(p_{R}) &= 1 \\
	P_{(1)}^{\mu \nu}(p_{R}) &= \sum_{s_z} \epsilon_{(1)}^\mu(p,s_z)
	\epsilon_{(1)}^{*\nu}(p,s_z) =  
	- \, g^{\mu \nu} + \frac{p_{R}^{\mu} \, p_{R}^{\nu}}{m_{R}^{2}} 
	\label{eq:defP} \\
    \nonumber
	P_{(2)}^{\mu\nu\alpha\beta}(p_{R})  &=
	 \frac{1}{2} \,  \left[ P_{(1)}^{\mu \alpha}(p_{R})  \, P_{(1)}^{\nu \beta}(p_{R})  + P_{(1)}^{\mu \beta}(p_{R})  \, P_{(1)}^{\nu \alpha}(p_{R}) \right] - \frac{1}{3} \, P_{(1)}^{\mu \nu}(p_{R}) 
	 \, P_{(1)}^{\alpha\beta}(p_{R})    \,,
\end{align} 
where $g^{\mu \nu}$ 
is the Minkowski metric.
Contracted with an arbitrary tensor, the projection operator selects 
the part of the tensor that satisfies the Rarita-Schwinger conditions.

For a decay process $R \to A B$, with relative orbital angular momentum
$L_{AB}$, between particle $A$ and $B$, the angular momentum tensor
$L_{(L_{AB})}$ is obtained by projecting 
the rank-$L_{AB}$ tensor $q_R^{\nu_{1}}   \,  q_R^{\nu_{2}}  \dots  \, 
q_R^{\nu_{L_{AB}}} $, constructed from the relative momenta $q_{R} = p_{A} -
p_{B}$, onto the spin-$L_{AB}$ subspace,
\begin{equation}
	L_{(L_{AB})\mu_{1}  \dots \mu_{L_{AB}}}(p_{R},q_{R}) = (-1)^{L_{AB}}  \,
	P_{(L_{AB})\mu_{1} \dots \mu_{L_{AB}} \nu_{1}  \dots \nu_{L_{AB}}}(p_R)  \,
	q_R^{\nu_{1}} \dots  \,  q_R^{\nu_{L_{AB}}}  .
\label{eq:defL}\end{equation}
Their $\vert \vec q_{R} \vert^{L_{AB}} $ dependence accounts 
for the influence
of the centrifugal barrier on the transition amplitudes. 
For the sake of brevity, the following 
notation is introduced, 
\begin{align}
\nonumber
\epsilon_{(S_R)}(R) & \equiv \epsilon_{(S_R)}(p_{R},{s_z}_{R}) , \\ \nonumber
P_{(S_R)}(R) & \equiv P_{(S_R)}(p_{R}), \\
L_{(L_{AB})}(R) & \equiv L_{(L_{AB})}(p_{R},q_{R})  .
\end{align}

Following the isobar approach, a three-body decay spin amplitude is described as
a product of two-body decay spin amplitudes.
Each sequential two-body decay $R \to A \, B$, 
with relative orbital angular momentum $L_{AB}$, total intrinsic spin $S_{AB}$,
and total final state momentum $p_{AB} \equiv  p_{R}$, contributes a term to the
overall spin factor given by
\begin{align}
	S_{R \to A B}(\mathbf x) 
	=
	\epsilon_{(S_{R})}(R) \, \Xi(S_{R},L_{AB},S_{AB}) \, 
L_{(L_{AB})}(R) \,    
\Phi_{AB}(\mathbf x ),
          \label{eq:spin1}
\end{align}
where
\begin{align}
	 \Phi_{AB}(\mathbf x)  &= 
	 P_{(S_{AB})}(R) \, \Xi(S_{AB},S_{A},S_{B})  \, \epsilon^{*}_{(S_{A})}(A)  \,
	 \epsilon^{*}_{(S_{B})}(B) \,   ,
          \label{eq:spin2}
\end{align}
and
\begin{equation}
	  \Xi(J_{1},J_{2},J_{3}) = 		
	  \begin{cases}
			  1 & \mbox{if } J_{1} + J_{2} + J_{3} \; {\rm even} \\
			   \epsilon_{\alpha \beta \gamma \delta} \, 
			   p_{R}^{\delta} & \mbox{if } J_{1}
			   + J_{2} + J_{3} \; {\rm odd} \end{cases} \, ,
\label{eq:Xi}\end{equation}
where $\epsilon_{\alpha\beta\gamma\delta}$ is the Levi-Civita symbol and $J$
refers to the arguments of $\Xi$ defined in
\cref{eq:spin1,eq:spin2}.
Its antisymmetric nature ensures the correct parity 
transformation behavior of the amplitude.

Here, a polarization vector is assigned to the decaying particle 
and the complex conjugate vectors for each decay product.
The spin and orbital angular momentum couplings are described by the tensors $P_{(S_{AB})}(R)$
and $L_{(L_{AB})}(R)$, respectively.
Firstly, the two spins $S_{A}$ and $S_{B}$, are coupled 
to a total spin-$S_{AB}$ state, $\Phi_{AB}(\mathbf x)$,
by projecting the corresponding polarization vectors  onto the spin-$S_{AB}$
subspace transverse to the momentum of the decaying particle.
Afterwards, the spin and orbital angular momentum tensors are properly contracted with the
polarization vector of the decaying particle to give a Lorentz scalar.

The spin factor for a whole decay chain, is obtained by combining the two-body
terms. For example, the decay $X \to (R \to AB)$ would have the following
overall spin factor:
\begin{equation}
	S_{X \to (R \to AB)}({\bf x})%
	=
	S_{X \to R C}(\mathbf x)%
	\, 
	S_{R \to AB}(\mathbf x)%
	.
\end{equation}

\FloatBarrier\subsection{Model for \texorpdfstring{$\X \to
\Dz\Dzb\pi^0$}{X(3872) decays to D0 anti-D0 pi0}}
\label{subsec:ModelFormalism}

For $\X \to \DDbSys\pi^0$, a decay model consisting
of only the intermediate states $\Dstz\Dzb$ and $\Dz\Dstzb$ is considered.
Note that in a strong decay of a pseudovector particle to a vector and a
pseudoscalar, conservation of parity dictates the possible orbital angular
momentum amplitudes; only even orbital angular momenta $L$ are possible. 
In this case, the Levi-Civita symbol of \cref{eq:Xi} is never required.

The spin factors can be explicitly constructed
for each possible
two-body decay stage within $\Xs \to \Dstz\Dzb \to \DDbSys\pi^0$:
\begin{eqnarray}
S_{\Xs \to (\Dstz\Dzb)_{L = 0}}(\mathbf x) 
&=&  
{\epsilon_{(1)}}^a(\Xs) %
{\epsilon_{(1)}}_a^*(\Dstz) %
\nonumber \\
S_{\Xs \to (\Dstz\Dzb)_{L = 2}}(\mathbf x)
&=& 
{\epsilon_{(1)}}^a(\Xs) %
L_{(2) ab}(\Xs) %
{\epsilon_{(1)}}^{b*}(\Dstz) %
\nonumber \\
S_{\Dstz \to (\Dz\pi^0)_{L = 1}}(\mathbf x) 
&=&
{\epsilon_{(1)}}^c(\Dstz) %
{L_{(1)}}_{c}(\Dstz) %
\end{eqnarray}

Using the identities in \cref{eq:defP},
the overall spin factors are determined
for $\Xs \to (\Dstz\Dzb)_{L = 0} \to \DDbSys\pi^0$:
\begin{equation}S_{\Xs \to (\Dstz\Dzb)_{L = 0} \to \DDbSys\pi^0}(\mathbf x) = 
{\epsilon_{(1)}}_a(\Xs) %
{L_{(1)}}^{a}(\Dstz) %
.
\end{equation}
and $\Xs \to (\Dstz\Dzb)_{L = 2} \to \DDbSys\pi^0$:
\begin{equation}S_{\Xs \to (\Dstz\Dzb)_{L = 2} \to \DDbSys\pi^0}(\mathbf x) = 
{\epsilon_{(1)}}^a(\Xs) %
{L_{(2)}}_{ab}(\Xs) %
{L_{(1)}}^{b}(\Dstz) %
. %
\end{equation}

Spin factors can be determined similarly for the decay chains originating
from $\Xs \to \Dz\Dstzb$. 
The tensor
implementation of the spin factors in {\sc EvtGen} was cross-checked with 
those provided by {\sc qft++}~\cite{Williams:2008wu}.

\FloatBarrier\section{\boldmath The
\texorpdfstring{$\DmDb$}{D0 - anti-D0} momentum in the 
\texorpdfstring{$\X$}{X(3872)} and \texorpdfstring{$\DDbSys$}{D0
anti-D0} frames}
\label{sec:DmDbPreservation}

%
Consider a $\XtoDDbP$ or $\XtoDDbG$ decay in the $\X$ rest frame.
If we align the light neutral momentum to the negative $x$-axis, the $\DDbSys$ 
momentum points in the positive $x$-direction. The transverse components of
the $\DmDb$ momentum are conserved in a boost to the $\DDbSys$ rest
frame. This leaves the longitudinal component, which under a Lorentz
transformation to the $\DDbSys$ rest frame is the following:
\begin{eqnarray}
{p_\Dz}_x^{\D\Db} - {p_\Dzb}_x^{\D\Db} &= &\left(1 - \left(\frac{(p_\Dz +
p_\Dzb)c}{E_\Dz + E_\Dzb}\right)^2\right)^{-\frac{1}{2}} \nonumber \\ 
&  &\times 
\left({p_\Dz}_x -
{p_\Dzb}_x - \frac{p_\Dz + p_\Dzb}{E_\Dz +
E_\Dzb} ( E_\Dz -
E_\Dzb )
\right).
\label{booster}
\end{eqnarray}

\begin{figure}[th]
\flushleft
\includegraphics[page=33,scale=0.35]{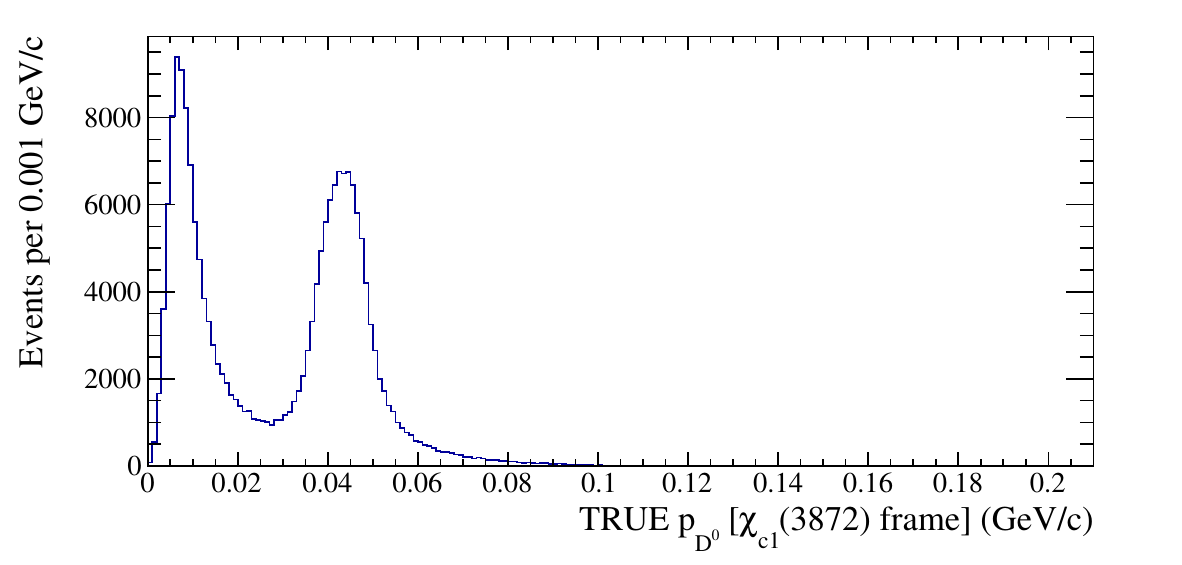}(a)
\includegraphics[page=33,scale=0.35]{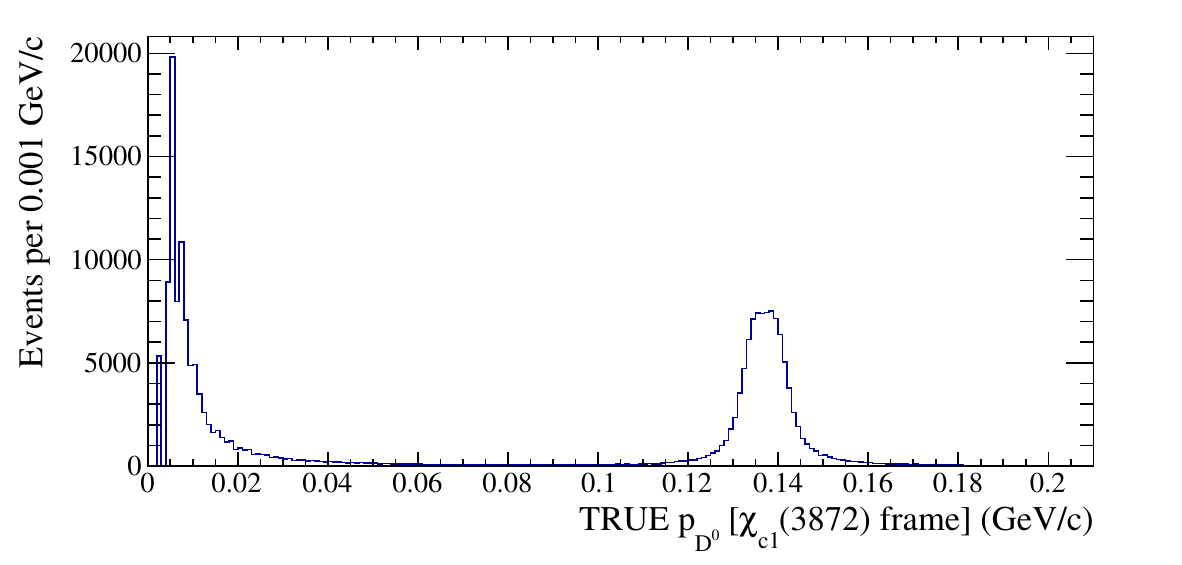}(e)
\includegraphics[page=33,scale=0.35]{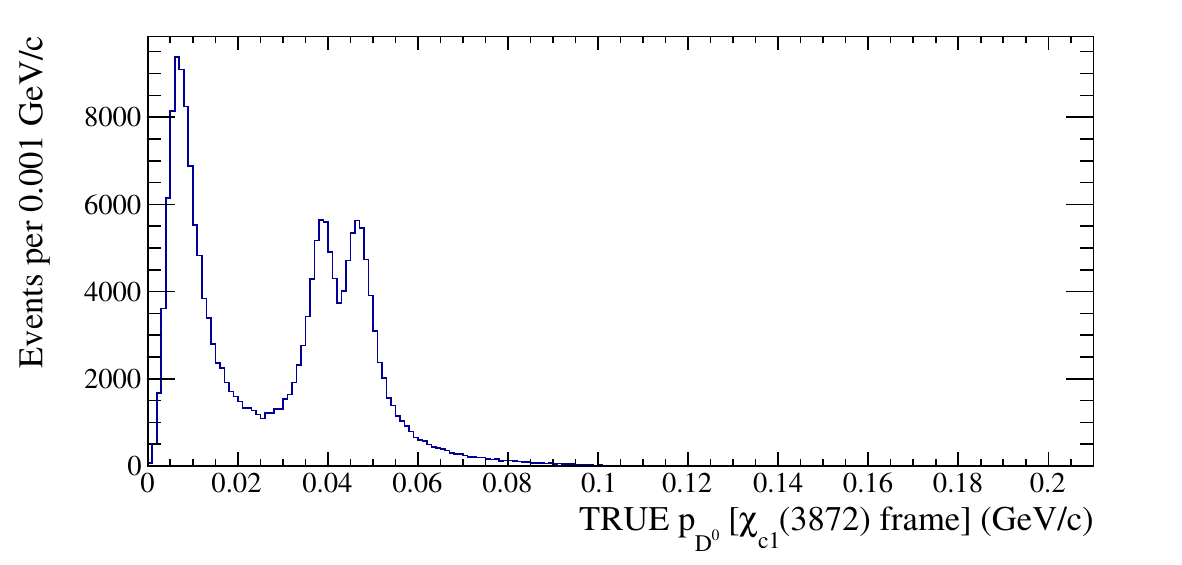}(b)
\includegraphics[page=33,scale=0.35]{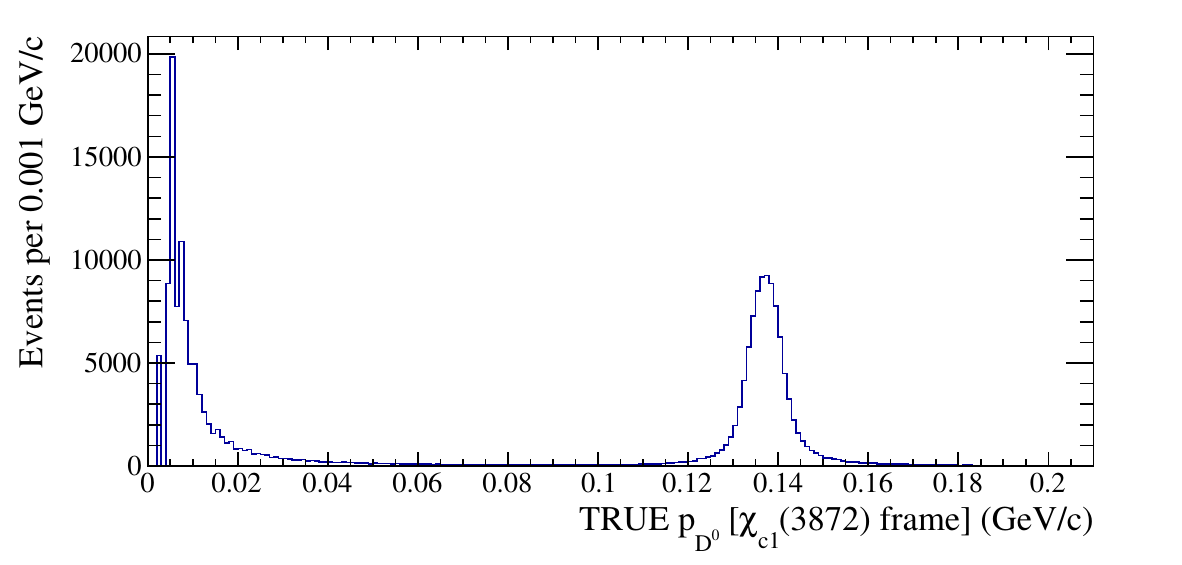}(f)
\includegraphics[page=33,scale=0.35]{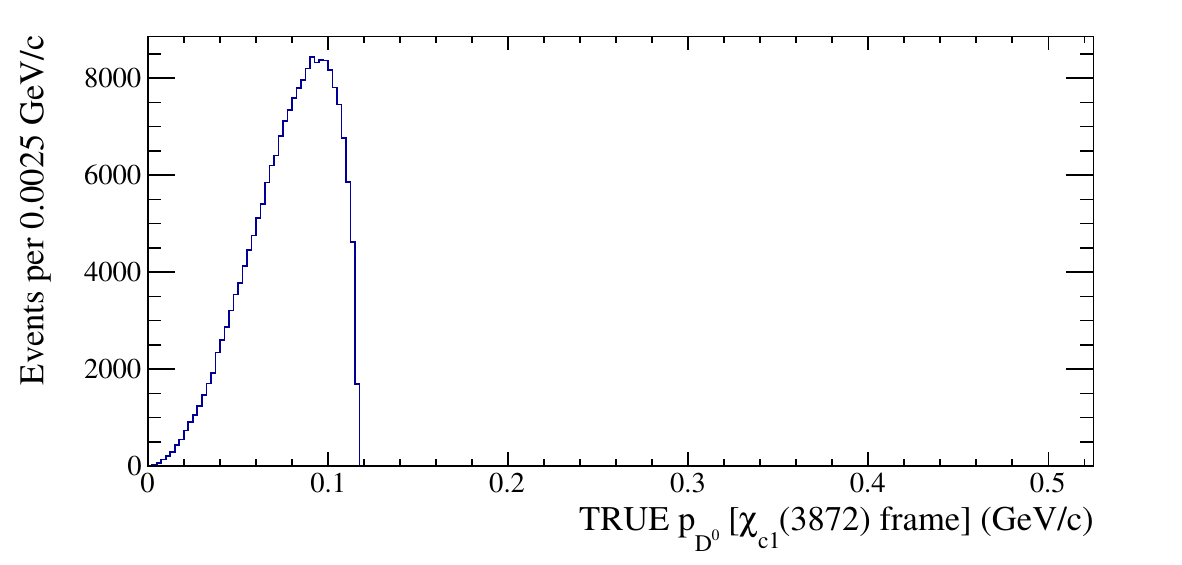}(c)
\includegraphics[page=33,scale=0.35]{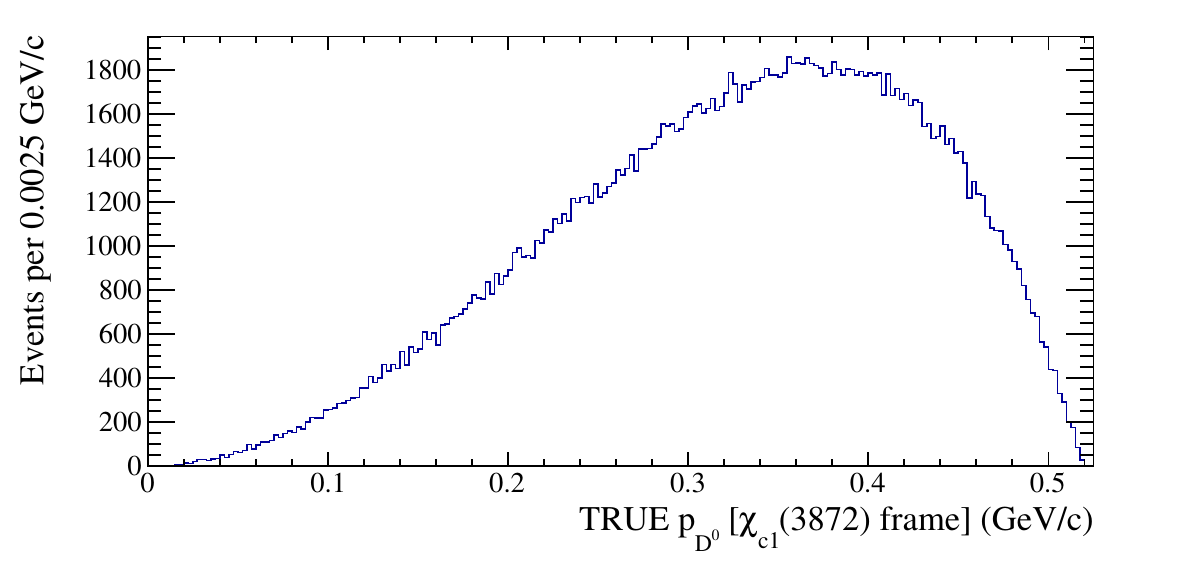}(g)
\includegraphics[page=33,scale=0.35]{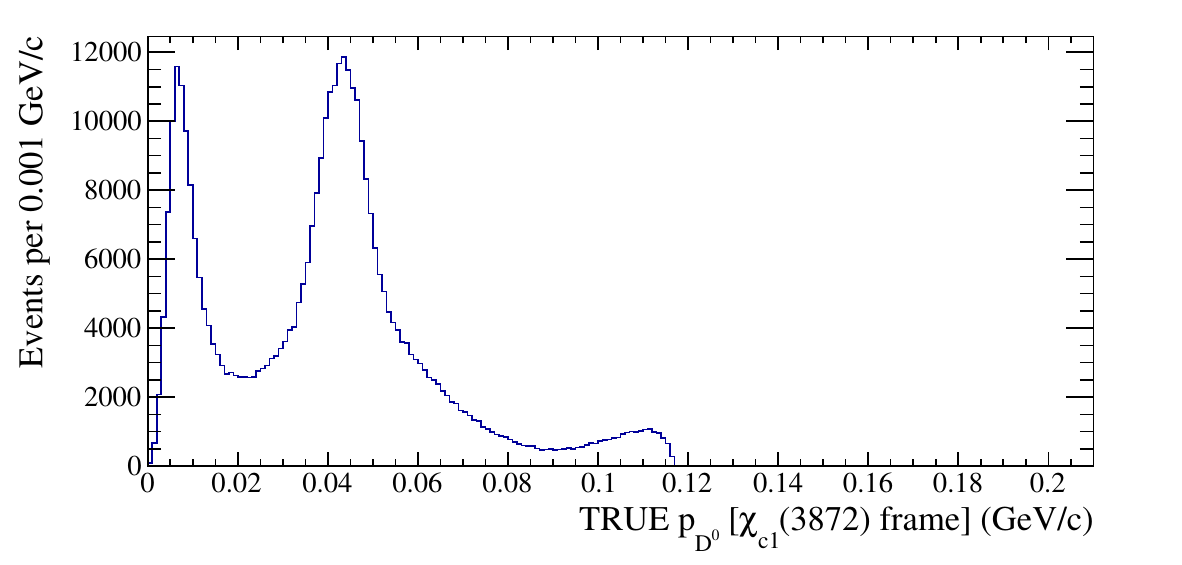}(d)
\caption{TRUE $p_\Dz$ in the $\X$ rest frame for prompt (a) $\XtoDDbP$ 50/50
S-Wave, (b) $\XtoDDbP$ 50/50 D-Wave, (c) $\XtoDDbP$ PHSP, (d) $\XtoDDbP$ Model
(e) $\XtoDDbG$ 50/50 S-Wave, (f) $\XtoDDbG$ 50/50 D-Wave, and (g)
$\XtoDDbP$ PHSP. Recall that a $\D$ meson not associated with a $\Dst$ is
near rest in the $\X$ rest frame, and a $\D$ meson associated with the $\Dst$
meson moves with momentum approximately opposite to the light neutral, hence the
two peaks in the 50/50 samples' distributions.
No detector simulation is applied.
The $p_\Dzb$ distributions appear identical.
}
\label{fig:D0-mom}
\end{figure}

The momentum of each $\D$ meson in the $\X$ decay rest frame is a
small fraction of the $\D$ rest mass.
In the case of $\X \to \Dstz\Dzb \to \DDbP$ the $\D$ momenta are negligible 
compared to the $\D$ mass, and in the
case of $\X \to \Dstz\Dzb \to \DDbG$ it is a small fraction. The $\D$ momentum is
still a small fraction in $\XtoDDbP$ PHSP decays; in $\XtoDDbG$ PHSP
decays the $\D$ momentum can reach about 27\% of the total $\D$ energy. Plots of
the $\D$ momentum in the $\X$ rest frame are shown in \cref{fig:D0-mom} 
for TRUE sample prompt decays; plots of this variable for TRUE sample secondary
decays are qualitatively identical. 

In the cases where the $\D$ momenta are small, we can presume $\left(\left(p_\Dz + p_\Dzb\right)c\right)^2
<< \left(E_\Dz + E_\Dzb\right)^2$ and $E_\Dz
= E_\Dzb \simeq m_\Dz$, and thus the $\DmDb$ momentum in the
$\DDbSys$ rest frame is essentially the same as it is in the $\X$ rest frame,
since:
\begin{equation}
{p_\Dz}_x^{\D\Db} - {p_\Dzb}_x^{\D\Db} \simeq  {p_\Dz}_x - {p_\Dzb}_x.
\label{equals}
\end{equation}
Simulations confirm that, on the scales of interest, the $\DmDb$ momentum is
practically identical in both frames.
%